\documentclass[12pt,a4paper]{article}
\usepackage{amsfonts}
\usepackage{amsmath}
\usepackage{array}
\usepackage{arydshln}
\usepackage{rotating}
\usepackage{amssymb}
\usepackage{bm}
\usepackage{geometry}
\usepackage{setspace}
\usepackage{graphicx}
\usepackage{lscape}
\usepackage{verbatim}
\usepackage{natbib}
\usepackage{xcolor}
\usepackage{calrsfs}
\usepackage{mathrsfs}
\usepackage{algorithm2e}
\usepackage{algpseudocode}
\usepackage{tablefootnote}
\usepackage{tabularx}
\usepackage{ragged2e}
\definecolor{winered}{rgb}{0.5,0,0}
\usepackage[bookmarks=true, bookmarksnumbered=true, allbordercolors={1 1 1}]{hyperref}
\hypersetup{
  colorlinks   = true, 
  urlcolor     = blue, 
  linkcolor    = blue, 
  citecolor    = winered,
}
\usepackage[open=true, numbered=true]{bookmark}
\usepackage{float}
\usepackage{fancyhdr}
\usepackage[toc,page]{appendix}
\usepackage{scalefnt}
\usepackage{afterpage}
\usepackage[left]{lineno}
\usepackage{caption}
\usepackage{subcaption}
\usepackage{varioref}
\usepackage{authblk}
\usepackage{cleveref}

\DeclareFontFamily{OT1}{pzc}{}
\DeclareFontShape{OT1}{pzc}{m}{it}{<-> s * [0.900] pzcmi7t}{}
\DeclareMathAlphabet{\mathpzc}{OT1}{pzc}{m}{it}

\makeatletter
\def\blfootnote{\xdef\@thefnmark{}\@footnotetext}
\makeatother

\emergencystretch=\maxdimen
\begin{document}

\title{\LARGE{Monitoring multicountry macroeconomic risk}\blfootnote{We would like to thank Raffaella Giacomini, Sylvia Kaufmann, Massimiliano Marcellino, Christian Matthes, Mirco Rubin, Neil Shephard, Leif Anders Thorsrud and participants at the following conferences, for useful discussions and comments: 12th European Seminar on Bayesian Econometrics in Salzburg; ``Advances in alternative data and machine learning for macroeconomics and finance'' in Paris; Barcelona Workshop on Financial Econometrics; 27th International Conference on Macroeconomic Analysis and International Finance in Rethymno; 2023 Finance and Business Analytics Conference in Lefkada; 10th IAAE Annual Conference in Oslo. We would also like to thank seminar participants at the following institutions: BI Norwegian Business School, European Central Bank, University of Lancaster. \\
The views expressed are those of the authors and do not necessarily reflect those of Norges Bank or any of the affiliated institutions.}}
\author[1]{Dimitris Korobilis}
\author[2]{Maximilian Schr\"{o}der} 
\affil[1]{{\footnotesize University of Glasgow, and Rimini Center for Economic Analysis}}
\affil[2]{{\footnotesize BI Norwegian Business School, and Norges Bank}}

\date{\today}

\maketitle
\begin{abstract}
\noindent We propose a multicountry quantile factor augmeneted vector autoregression (QFAVAR) to model heterogeneities both across countries and across characteristics of the distributions of macroeconomic time series. The presence of quantile factors allows for summarizing these two heterogeneities in a parsimonious way. We develop two algorithms for posterior inference that feature varying level of trade-off between estimation precision and computational speed. Using monthly data for the euro area, we establish the good empirical properties of the QFAVAR as a tool for assessing the effects of global shocks on country-level macroeconomic risks. In particular, QFAVAR short-run tail forecasts are more accurate compared to a FAVAR with symmetric Gaussian errors, as well as univariate quantile autoregressions that ignore comovements among quantiles of macroeconomic variables. We also illustrate how quantile impulse response functions and quantile connectedness measures, resulting from the new model, can be used to implement joint risk scenario analysis.

\bigskip

\bigskip \medskip

\end{abstract}
\thispagestyle{empty} 

\newpage
\onehalfspace
\setcounter{page}{1}

\section{Introduction}
The so-called Great Recession of 2008-2009 marked the beginning of an era in which global shocks are more pervasive, are able to generate domino effects, and can increase macroeconomic risks to unprecedented levels. Recent evidence is provided by \cite{Adrianetal2019} who find that the expected distribution of GDP growth skews left during recessions. This skewness can change over time \citep{Jensenetal2020} and is positively related to macroeconomic volatility \citep{BekaertPopov2019}. Similar distributional asymmetries occur for inflation \citep{Korobilis2017,LopezLoria2019}, as for numerous other economic and financial time series. Although some global risks -- such as global inflation and oil crises -- are eerily familiar to economists, the current unsustainable levels of debt, low growth, and the climate emergency create new challenges for macroeconomic policy-makers. In the euro area, which shares common monetary, regulatory, and other policies, measuring and monitoring the heterogeneous performance of countries facing modern global risks is a particularly challenging quantitative exercise. As an example, following the first response to the coronavirus (Covid-19) health crisis, euro-area GDP was 4.9\% below its prepandemic level in the first quarter of 2021. Nevertheless, country-level performance was remarkably heterogeneous, with Ireland reporting growth of 13.2\% and Spain experiencing a contraction of -9.3\% \citep{Muggenthaleretal2021}.

In this paper we develop a novel quantile regression methodology to capture, in a parsimonious and interpretable way, the heterogeneous responses of euro-area and country-specific macroeconomic aggregates to global risks. Our approach builds on combining the strengths of vector autoregressions (VARs) for structural inference, with the flexibility of modeling individual percentiles of the data distribution of macro data using quantile regression methods. The main challenge with modeling multivariate quantiles is that different percentiles of different variables might be correlated, which is not typically an issue in univariate quantile models (where each quantile level is estimated independently).\footnote{From another perspective, \cite{Adrianetal2021} argue that during crises the joint distribution of economic and financial conditions becomes multimodal. Such multimodalities support the argument that percentiles of joint distributions are not symmetrically correlated.} With just two macroeconomic variables $(y_{1},y_{2})$ and three quantile levels $(q_{1},q_{2},q_{3})$, there are numerous ways any two quantiles of these variables could be correlated. As a result, despite the fact that VAR modeling of the mean of $y_{1},y_{2}$ is a bivariate system, a quantile VAR would require specifying all six variables $y_{1(q_{1})},y_{1(q_{2})},y_{1(q_{3})},y_{2(q_{1})},y_{2(q_{2})},y_{2(q_{3})}$ as endogenous. When the interest is in modeling quantiles of macroeconomic variables for many countries, VAR inference on quantiles can quickly become high-dimensional, over-parameterized and, potentially, computationally cumbersome.

We solve this modeling conundrum by specifying a quantile factor-augmented vector autoregression (QFAVAR) that extends the popular factor-augmented VAR approach outlined in \cite{Bernankeetal2005} and \cite{StockWatson2005} to the quantile regression setting. In a setting with several macro variables for several countries, we extract variable-specific quantile factors for three percentiles of interest (10th, 50th, and 90th percentiles). As a result, the factors not only capture common dynamics across the cross-section of the data (which is a key feature of multi-country factor studies such as \citealt{Koseetal2003}), but also they are a parsimonious way of modeling cross-quantile dependence among the original variables. The benefit is that, similar to the recent methodologies in \cite{Chenetal2021} and \cite{KorobilisSchroeder2022}, the macroeconomic variables load onto quantile factors independently for each quantile level, allowing computational convenience and numerical stability. However, the QFAVAR allows all quantile factors to be dynamically correlated for all percentile levels by means of a joint VAR-state evolution. 

Methodologically, the new QFAVAR adds novel features to various interconnected literatures in macroeconometrics. We first build on established literature that uses dynamic factor models (DFMs) to characterize comovements and heterogeneities among different countries. \cite{Koseetal2003} use multicountry dynamic factor models to measure the degree of synchronization of business cycles; \cite{CiccarelliMojon2010} and \cite{MumtazSurico2012} use common factors to measure global inflation. A common limitation of the traditional DFM/FAVAR approach is the reliance on the normality assumption of common and idiosyncratic disturbances, which does not sufficiently capture any asymmetries in higher moments of the empirical distribution of macroeconomic data.\footnote{Papers such as \cite{Korobilis2013} and \cite{KoopKorobilis2014} specify flexible FAVARs with time-varying parameters and stochastic volatility; \cite{GorodnichenkoNg2017} explicitly estimate mean and volatility factors. Nevertheless, such factor model approaches are restricted to modeling flexibly only the first two moments of the data distribution.} Next, our proposal to estimate quantile factors and combine these with VAR dynamics, adds a new tool to a recent literature that is otherwise limited to specifying static quantile factors. Key papers include \cite{AndoBai2020}, \cite{Chenetal2021}, \cite{Clarketal2021}, \cite{KorobilisSchroeder2022}, and \cite{Maetal2021}. Finally, the QFAVAR adds to another emerging literature that combines quantile regressions and VAR methods in order to identify the asymmetric effects of various macroeconomic shocks. See for example \cite{CastelnuovoMori2022}, \cite{Fornietal2021} and \cite{Loriaetal2019}. Although existing methodologies for estimating quantile VARs (QVARs) can be empirically useful and relevant, they may rely on simplifications that are context-specific.\footnote{For example, the Bayesian approaches in \cite{MumtazSurico2015} and \cite{Schuler2020} might become computationally cumbersome using the dimensions we consider in this paper. The contribution by \cite{Andoetal2022} is restricted in that the covariance matrix is based on observed data, such that QVAR estimation is implemented using univariate quantile regressions. \cite{Manganelli2020} present a bivariate quantile VAR that does not allow different quantiles of the two series to interact. Finally, the setting in \cite{Whiteeal2015}, while useful for value-at-risk applications in finance, has several limitations for structural macro inference.} In contrast, the proposed QFAVAR approach is general enough to be used in a multitude of other empirical problems where asymmetric shocks are evident, for example, issues involving climate risks, macroeconomic uncertainty, or financial shocks.

Our first contribution is to establish the workings of such a novel specification and to show how it allows for parsimonious VAR inference for quantiles without sacrificing flexibility and generality. The second contribution is to derive and test numerically likelihood-based estimators for inference in the QFAVAR. In particular, we adopt Bayesian methods and modern priors from the statistics and machine learning literature that lead to tuning-free penalized estimation in high dimensions, and we derive two algorithms for posterior inference. Our benchmark estimation is based on a Markov chain Monte Carlo (MCMC) algorithm that is a generalization of the Gibbs samplers proposed in \cite{Bernankeetal2005} and \cite{Koseetal2003}. The second algorithm is based on variational Bayes (VB) inference that extends the static quantile factor estimator proposed in \cite{KorobilisSchroeder2022}; it provides a fast and convenient approximation to the joint parameter posterior. By leveraging machine learning methods, the second algorithm is particularly useful for applications where computational time is especially important, as are frequent in policy work.  

Our third contribution is empirical, as we apply the new tool to the problem of assessing the effects of global risks to a series of euro-area macroeconomic variables. We collect five macroeconomic variables for nine countries, namely inflation, industrial production, the long-term interest rate, an index of economic sentiment, and an index country-level financial stress. We extract quantile factors from each macro variable, by aggregating over all nine countries, such that each quantile factor approximates the distribution of the respective euro area aggregate. For example, the quantile factor extracted from the nine country-level industrial production (IP) series captures the quantiles of aggregate IP for the euro area. From a measurement perspective these quantile factors are superior in capturing cross-sectional and distributional heterogeneities, compared to fitting univariate quantile regressions directly to the euro area aggregates, but allow insights into country-level heterogeneity at the same time.\footnote{A univariate quantile regression fitted to euro-area IP will not capture correctly the distribution of IP, especially in the tails. This is due to the lack of observations in the tails as well as the fact that IP is measured with error. By combining the information in the IP series of multiple countries, we expect the quantile factor to suffer less from these issues and provide superior estimates of the distributional characteristics of aggregate IP.} By definition the QFAVAR allows us to augment the quantile factors with (observed) global factors, and we choose to include global measures of inflation, supply chain pressures, financial conditions, and economic policy uncertainty.

The empirical facts can be summarized as follows. Quantile factor estimates of 10th, 50th and 90th percentiles are characterized by evident heterogeneity, implying asymmetries in the distribution of unobserved factors. The QFAVAR is significantly better to a FAVAR with symmetric Gaussian errors in forecasting the left and right tails of the distribution of inflation and IP in the euro-area countries. This observation is true, in particular, for the short, one-month-ahead horizon of both variables. Additionally, we find that the QFAVAR is superior to univariate quantile autoregressive models with or without exogenous global predictors, and the quantile dynamic factor model (that is, a special case of the QFAVAR without any global variables). These numerical results clearly show not only the benefits of adopting a multivariate approach to quantile regression forecasting, but also that there are benefits from augmenting the multi-country model with relevant global predictors. We illustrate, by means of quantile impulse response functions, quantile forecast error variance decompositions, and quantile connectedness graphs, that the QFAVAR captures a large variety of heterogeneities across different quantiles of macroeconomic variables of interest. In doing so, we also add to an open debate. \cite{Adrianetal2019} argue that financial conditions are an important predictor of downside risks to GDP. \cite{plagborgmoller2020} question the validity of this argument by providing detailed empirical evidence that no financial time series is persistently informative in univariate quantile regressions. In contrast, we find that (global) financial and economic conditions can be very informative for country-level macro risks, once the potential linkages are modelled as a multivariate system. 

The next section describes the exact specification of the QFAVAR and how estimation and inference can be tackled using Bayesian methods. As the QFAVAR is a new model in the literature, Section 3 undertakes a battery of numerical exercises that establish its usefulness for monitoring multi-country macroeconomic risks. Section 4 concludes the paper.

\section{Econometric Methodology}
Our starting point is the factor augmented vector autoregressive model from \cite{Bernankeetal2005} and \cite{StockWatson2005}, adopted for a panel of several macroeconomic time series for several countries. This modeling approach involves extracting a smaller vector of latent factors from the large panel of macroeconomic data. The latent factors evolve jointly with observed variables as a lower-dimensional vector autoregression (VAR). This setting is established in macroeconomics, and the reader should consult \cite{StockWatson2016} for a thorough review. Here we explain a conceptually straightforward extension of the FAVAR to the quantile setting, show how this extension results in an inherently high-dimensional model, outline how Bayesian inference can help tackle estimation challenges, and, finally, we show how the proposed quantile specification can be deployed for traditional structural VAR analysis.

\subsection{A multi-country quantile FAVAR (QFAVAR)}
Let $y_{ijt}$ denote macroeconomic/financial indicator $i$ for country $j$ observed at time $t$, for $i=1,...,m$, $j=1,...,n$ and $t=1,...,T$. We characterize the distribution of the $mn \times 1$ vector $\bm y_{t} = \left[ y_{11t},...,y_{1nt},...,y_{m1t},...,y_{mnt} \right]^{\prime}$ by grouping its elements into unobserved, indicator-specific factors $f^{i}_{t,(q)}$ for each quantile level $q = q_{1},...,q_{R}$, where $q_{r} \in (0,1)$ and $q_{r-1}<q_{r}$. We also assume global-level factors summarized in the $k \times 1$ vector of observed variables $\bm g_{t}$. The quantile factor model strategy begins by specifying the $q$th conditional quantile of $y_{ijt}$ as a linear function of the global indicators and the indicator-specific factor, which is of the form
\begin{equation}
Q_{q} \left( y_{ijt} \vert \bm g_{t} \right) = c_{ij(q)} +  \bm \gamma_{ij(q)} \bm g_{t} +  \lambda_{ij(q)}f^{i}_{t,(q)},
\end{equation}
where $c_{ij(q)}$ is a scalar intercept, $\bm \gamma_{ij(q)}$ is a $1 \times k$ vector of loadings on the observed global factors $\bm g_{t}$, and $\lambda_{ij(q)}$ is the scalar loading (weight) of the scalar, unobserved, indicator-specific quantile factor $f^{i}_{t,(q)}$.\footnote{The fact that we extract indicator-specific factors, and we don't just allow all variables to load on all factors, helps with identification of the factor model. We only impose normalization restrictions, where for the factor corresponding to quantile $q_{r}$, we normalize the loading of the $r^{th}$ series to be one.} Following the probabilistic approach in \cite{KorobilisSchroeder2022} this quantile factor model can be represented as a parametric regression of the form
\begin{equation}
y_{ijt}  = c_{ij(q)} + \gamma_{ij(q)} \bm g_{t} +  \lambda_{ij(q)}f^{i}_{t,(q)} +  u_{ijt(q)}, \label{measurement_ij}
\end{equation}
where $u_{ijt(q)} \sim AL\left(0,\sigma_{ij(q)}^{2},q \right)$ is an asymmetric Laplace disturbance term; that is, it has the functional form
\begin{equation}
f\left(u_{ijt(q)}\right) = \frac{q(1-q)}{\sigma_{ij(q)}^{2}} \left\lbrace e^{\left[(1-q)\frac{u_{ijt(q)}}{\sigma_{ij(q)}^{2}}\right]} \mathbb{I}(u_{ijt(q)}\leq 0) + e^{\left[(-q)\frac{u_{ijt(q)}}{\sigma_{ij(q)}^{2}} \right]}\mathbb{I}(u_{ijt(q)}>0)  \right\rbrace,
\end{equation}
with $\mathbb{I}$ denoting the indicator function. Similar to a Bayesian linear regression, where the Gaussian residual is centred around zero, the asymmetric Laplace residual has the $q$-th quantile equal to zero.

An important modeling feature of our approach is that the indicator-specific quantile factors and the global factors are contemporaneously and dynamically correlated with each other via a vector autoregressive (VAR) model with $p$ lags. We define the $mr \times 1$ vector $\bm F_{t} = \left[f^{1}_{t(q_{1})},..., f^{m}_{t(q_{1})}, f^{1}_{t(q_{2})}, ..., f^{m}_{t(q_{2})}, ..., f^{1}_{t(q_{r})}, ..., f^{m}_{t(q_{r})} \right]^{\prime}$ that summarizes all unobserved factors at all quantile levels. The VAR($p$) that characterizes the joint dynamics of the quantile factors and the global factors is of the form
\begin{equation}
\left[
\begin{array}{c}
\bm F_{t} \\
\bm g_{t}
\end{array}
\right] = \bm v +  \bm \Phi_{1} \left[
\begin{array}{c}
\bm F_{t-1} \\
\bm g_{t-1}
\end{array}
\right] + ... + \bm \Phi_{p} \left[
\begin{array}{c}
\bm F_{t-p} \\
\bm g_{t-p}
\end{array}
\right] + \bm \varepsilon_{t}, \label{state}
\end{equation}
where $\bm v$ is an $l \times 1 $ vector of intercept terms, $\bm \Phi_{c}$ are $l \times l$ matrices of autoregressive coefficients for lagged term $c=1,...,p$, and $\bm \varepsilon_{t} \sim N\left(\bm 0, \bm \Omega \right)$ is the disturbance term with $ \bm \Omega$ an $l \times l$ full, symmetric and positive definite covariance matrix. Here, $l = mr + k $ is the joint dimension of the vectors $\bm F_{t}$ and $\bm g_{t}$. As a result, in contrast to \cite{Chenetal2021} and \cite{KorobilisSchroeder2022} who estimate factors independently for each quantile, equation \eqref{state} allows for complex patterns of dynamic correlations among the quantiles to affect the estimation outcomes of the quantile factors. At the same time, as we show in \autoref{sec:VAR_form}, this latter equation is important in order to perform structural VAR inference using the QFAVAR. Finally, the model maintains its computational simplicity, as the disturbances $u_{ijt(q)}$ are independent from each other for all $i,j,q$. Therefore, equation \eqref{measurement_ij} is a collection of univariate quantile regressions that can be estimated independently from one another. In contrast, the VAR of the state equation \eqref{state} can become quite large when either considering many indicator-specific or global factors, or many quantile levels. However, it is trivial to draw from an established literature on large Bayesian VARs in order to alleviate high-dimensionality concerns. In particular, we adopt the efficient algorithm of \cite{Carrieroetal2022}. We next discuss likelihood-based inference in the QFAVAR in detail.

\subsection{Likelihood, priors, and posterior inference} \label{sec:lik_priors}
The QFAVAR consists of seemingly disjointed equations \eqref{measurement_ij} and \eqref{state}. In order to write them as a joint system and facilitate likelihood-based inference, we simplify our notation by dropping intercepts and assuming one lag in the VAR part of the model. Under these simplifications -- and as shown in detail in appendix A -- we can combine equations \eqref{measurement_ij} and \eqref{state} into the following linear state-space system
\begin{eqnarray}
\left[
\begin{array}{c}
\bm Y_{t} \\
\bm g_{t}
\end{array}
\right]   & = &  \left[
\begin{array}{cc}
\bm \Lambda & \bm \Gamma \\
\bm 0  &\bm I
\end{array}
\right]   
\left[
\begin{array}{c}
\bm F_{t} \\
\bm g_{t}
\end{array}
\right] + 
\left[
\begin{array}{c}
\bm u_{t} \\
\bm 0
\end{array}  
\right],  \label{measurement_full}  \\
\left[
\begin{array}{c}
\bm F_{t} \\
\bm g_{t}
\end{array}
\right] & = & \bm \Phi \left[ 
\begin{array}{c}
\bm F_{t-1}\\
\bm g_{t-1}
\end{array}
\right] + \bm \varepsilon_{t}. \label{state_full} 
\end{eqnarray}
In this matrix notation $\bm Y_{t}$ is an $nmr \times 1$ vector with the vector $\bm y_{t}$ repeated $r$ times; $\bm \Lambda$ is an $nmr \times mr$ block diagonal matrix with the quantile-specific loadings $\bm \lambda_{(q)}$ on its diagonal; $\bm \lambda_{(q)}$ in turn is an $nm \times m$ block-diagonal matrix with the factor-specific loading $\bm \lambda_{i(q)} = \left[ \lambda_{i1(q)},...,\lambda_{im(q)} \right]^{\prime}$ on its diagonal; and $\bm u_{t}$ is an $nmr \times 1$ vector of disturbances with $q$-th element $\bm u_{t(q)} = \left[u_{11t(q)},...,u_{mnt(q)} \right]^{\prime}$. The above equations define a state-space model that characterizes the joint likelihood of the unobserved state variable $\left[\bm F_{t}^{\prime} , \bm g_{t}^{\prime} \right]^{\prime}$ and other latent parameters. Because this can become a high-dimensional system with many parameters, we follow \cite{Bernankeetal2005} in part and adopt Bayesian inference as our preferred likelihood-based approach. 

The first reason for addressing estimation using Bayesian inference is the vast availability of suitable prior distributions that provide automatic regularization to the joint likelihood, especially when considering estimation of extreme quantiles where not many observations are available. Following \cite{Feldkircheretal2022}, \cite{Korobilis2022} and others, we specify the Horseshoe prior for sparse signals of \cite{Carvalhoetal2010} for the elements of the matrices $\bm \Lambda$ and $\bm \Gamma$, as well as the elements of the VAR coefficients $\bm \Phi$. For a generic $b$-dimensional vector of parameters $\bm \theta$ (where $\bm \theta$ represents column vectors obtained from vectorizing the parameter matrices $\bm \Lambda, \bm \Gamma, \bm \Phi$, respectively) the Horseshoe prior takes the form
\begin{eqnarray}
\bm \theta \vert \xi,\bm \eta & \sim & \prod_{i=1}^{b} N \left( 0,\xi \eta_{i} \right), \\
\xi & \sim & C^{+} \left( 0, 1 \right), \\
\eta_{i} & \sim & C^{+} \left(0,1 \right).
\end{eqnarray}
Shrinkage estimators regularize an equivalent unrestricted estimator by means of a scalar factor $\kappa$ that determines how much the unrestricted estimator is shrunk towards zero.\footnote{In this case, the unrestricted estimator results from placing a normal prior on $\bm \theta$ with infinite variance.} In the case of the prior above, $\kappa$ is $Beta(0.5,0.5)$ distributed, i.e. has a horseshoe shape. This shape means that with smaller parameter spaces the posterior under a horseshoe prior will tend to be unrestricted, but as the parameter space increases relative to the number of observations, an increasingly larger amount of shrinkage towards zero is favored. These properties make it ideal for our large dimensional parameter space. Both theoretically and empirically the horseshoe performs well and it is a default automatic choice for many researchers; see the detailed review of this literature in \cite{KorobilisShimizu2022}. For the scalar asymmetric Laplace scale coefficients, $\sigma_{ij(q)}$, we specify conjugate inverse gamma priors. Finally, the VAR covariance matrix $\bm \Omega$ has an inverse Wishart prior.

Other than regularized estimation via prior distributions, the second reason for choosing a Bayesian approach to inference is computational convenience. The state-space model of equations \eqref{measurement_full}-\eqref{state_full} is linear but non-Gaussian because $\bm u_{t} \sim \prod_{ijq}AL(0,\sigma_{ij(q)},q)$. Additionally, the presence of the quantile common component $\bm \Lambda \bm F_{t}$, which is a product of two high-dimensional unobserved quantities, complicates state-space estimation further. However, estimation via the Gibbs sampler simplifies inference, because conditional distributions in the QFAVAR have a simple form. As discussed, the asymmetric Laplace distribution can be written as a mixture of Gaussian distributions, converting the linear state-space model into conditionally normal form. The parameters $\bm \Lambda$ and $\bm F_{t}$ can be sampled one at a time, conditional on one other.\footnote{While this approach induces high correlation of consecutive Monte Carlo samples of these parameters, this issue can be easily alleviated by storing only every $n$-th posterior sample, for an appropriate choice of $n$.} In practice, our estimation strategy combines established ergodic samplers, and can be outlined in the following steps:
\begin{enumerate}
\item Sample $\left[ \bm F_{t}^{\prime}, \bm g_{t}^{\prime} \right]^{\prime}$ conditional on values of all other system matrices from the state-space model of equations \eqref{measurement_full}-\eqref{state_full}. This step can be implemented using numerous approaches proposed in the literature, most notably the simulation smoother of \cite{CarterKohn1994}.
\item Sample $\lambda_{ij(q)},\gamma_{ij(q)},\sigma_{ij(q)}$ (and $c_{ij}$, if an intercept is present) conditional on $f_{t(q)}^{i}$, for each $i,j,q$, using equation \eqref{measurement_ij}. This is simply a univariate quantile regression, and simple conditional posteriors are provided in \cite{Khare2012}.
\item Sample $\bm \Phi_{1},...,\bm \Phi_{p},\bm \Omega$ (and $\bm v$, if an intercept is present), conditional on all quantile factors $\bm F_{t}$ using equation \eqref{state}. This is a simple Bayesian VAR, and posterior conditionals are also quite standard, see \cite{KoopKorobilis2010}.
\end{enumerate}
Of course, in steps 2 and 3 one needs to account for the use of the Horseshoe hierarchical prior, but this is also trivial to incorporate using the hierarchical formulation of this prior proposed in \cite{MakalicSchmidt2016}. The outcome is a Gibbs sampler that is not much different to the sampler in \cite{Bernankeetal2005} for the FAVAR model and, thus, has a simple and user-friendly structure. All our results using the Gibbs sampler are based on 1,000,000 iterations after discarding an initial chain of 100,000 draws. Out of the one million iterations we store every 100th draw, as consecutive draws tend to be heavily autocorrelated.\footnote{This is not surprising for factor models, where loadings are sampled conditional on the factors and vice-versa.} Therefore, all posterior inference (posterior means, variances etc) is built on a remainder of 10,000 samples from the posterior distribution of parameters and factors.

Finally, due to the fact that the dimension of the latent states $\bm F_{t}$ (quantile factors) can be quite large, Gibbs sampling estimation tends to become computationally cumbersome in certain applications of interest to policy-makers (for example, forecasting macroeconomic risks). In order to address this issue, we also propose an approximate two-step algorithm that replaces the factors with fast variational Bayes estimates from the \cite{KorobilisSchroeder2022} probabilistic quantile factor model. This use of machine learning tools avoids the need for sampling using more demanding state-space methods. Conditional on these plug-in estimates of the quantile factors, we can estimate all other parameters ($\bm \Lambda, \bm \Phi$ etc) with a variety of methods depending on one's needs. For example, one could use MCMC as in steps 2) and 3) above, or variational Bayes, or even ordinary least squares. In our implementation, for the sake of consistency, we also use variational Bayes to obtain estimates of the parameter matrices. This two-step approach ignores the dynamics of the factors and the dependence on the global variables $\bm g_{t}$ in the QFAVAR likelihood when producing estimates of $\bm F_{t}$. However, it is justified on the basis of computational simplicity, and our empirical results suggest that there is no significant information loss from this two-step procedure relative to the one-step Gibbs sampler.\footnote{Such two-step procedures are very popular in regular FAVARs. For example, both \cite{Bernankeetal2005} and \cite{StockWatson2005} suggest using principal components in a first step and then estimating all system parameters using least squares. This two-step procedure provides a popular and asymptotically consistent estimator \citep{StockWatson2016}. Proving a similar asymptotic result for our simple two-step variational Bayes procedure would be useful, but is beyond the scope of this paper. However, the probabilistic quantile factor analysis estimator in \cite{KorobilisSchroeder2022} is numerically quite similar to the quantile factor analysis estimator in \cite{Chenetal2021}, which in turn is a generalization of the principal components estimator to a quantile setting.} A detailed overview and derivation of our proposed two-step procedure is provided in the online supplement.

\subsection{VAR inference in the QFAVAR \label{sec:VAR_form}} 
The QFAVAR implies a joint VAR for the $r$ quantiles of country-level macroeconomic variables $\bm y_{t}$ and the mean (expectation) of the global variables $\bm g_{t}$. Here we follow \cite{Bernankeetal2005}, and \cite{StockWatson2005} and we assume that the idiosyncratic disturbances $\bm u_{t}$ in equation \eqref{measurement_full} are not relevant for structural inference. Indeed, as these disturbances in our model are cross-sectionally uncorrelated (they have a diagonal covariance matrix), they can be treated as nuisance shocks (e.g. due to measurement error or country-level effects). Therefore, $\bm \varepsilon_{t}$ are the true structural shocks in the system. Given this convention, one can explicitly reduce the two-equation QFAVAR into a one-equation VARMA model on $\left[ \bm y_{t}^{\prime}, \bm g_{t}^{\prime} \right]^{\prime}$ and then show that the moving average part vanishes asymptotically (thus, simplifying into a VAR model). However, for simplicity we follow \cite{Bernankeetal2005} and \cite{StockWatson2005} and pursue a two-step alternative: first, we implement all VAR exercises of interest (conditional and unconditional forecasting, impulse response analysis, historical decompositions, etc.) using the VAR in equation \eqref{state_full}, and then we project these quantities into the original variables $\bm y_{t}$ using the projection matrices $\bm \Lambda$ and $\bm \Gamma$ in equation \eqref{measurement_full}. For example, using standard formulas \citep{Helmut2005} the VAR in equation \eqref{state_full} can provide $h$-step-ahead forecasts of $\left[ \bm F_{t+h \vert t}^{\prime}, \bm g_{t+h\vert t}^{\prime} \right]^{\prime}$, for some $h>0$, and these can be projected into forecasts for $\bm y_{t}$ using the formula $\bm y_{t+h \vert t} = \left[ \bm \Lambda, \bm \Gamma \right] \left[ \bm F_{t+h \vert t}^{\prime}, \bm g_{t+h\vert t}^{\prime}\right]^{\prime}$. Similar arguments can be made about impulse response functions and other quantities of interest.

\section{Empirical evaluation of the QFAVAR}
\subsection{Euro-area macroeconomic indicators and global data}
We use five macroeconomic variables from nine euro-area countries observed over the sample 1996M1-2022M12. The countries, series, and sources are shown in the top panel of \autoref{tab:data}. Three series come from the Statistical Data Warehouse (SDW) of the European Central Bank. Industrial production is from the OECD data website, and the Economic Sentiment Index is a composite index maintained by the DG ECFIN (and accessible from the website of Eurostat, the main statistical agency in Europe). All series were accessed in March 2023. Consumer prices are not seasonally adjusted, so we convert these to year-on-year growth rates using the transformation $100 (\log P_{t+12} - \log P_{t})$; we use the same annual growth tranformation for the ESI in order to create a smoother series. IP (which is seasonally adjusted) is converted to month-on-month growth rates using the tranformation  $100 (\log P_{t+1} - \log P_{t})$. Finally, LTIR and CLIFS are left observed in levels. We neither undertake any additional seasonal adjustment, nor do we alter the series in any other way (e.g., outlier adjustment).

\begin{table}[H]
\centering 
\caption{Euro-area and global indicators}  \label{tab:data}
\resizebox{.95\textwidth}{!}{
\begin{tabular}{cllll}\hline \hline
\textsc{Countries}	&		&	\textsc{Macroeconomic Indicators}	&	\textsc{Acronym}	&	 \textsc{Source}	\\ \hline
Austria (AT)	&		&	Harmonized Index of Consumer Prices -- Overall index	&	HICP	&	SDW$^{1}$	\\
Belgium (BE)	&		&	Industrial Production Index -- Total index	&	IP	&	OECD$^{2}$	\\
Germany (DE)	&		&	Long Term Interest Rate	&	LTIR	&	SDW$^{1}$	\\
Spain (ES)	&		&	Economic Sentiment Index	&	ESI	&	Eurostat$^{3}$	\\
Finland (FI)	&		&	Country-Level Index of Financial Stress	&	CLIFS	&	SDW$^{1}$	\\
France (FR)	&		&		&		&		\\
Italy (IT)	&		&		&		&		\\
Netherlands (NL)	&		&		&		&		\\
Portugal (PT)	&		&		&		&		\\\hline
& 	&	\textsc{Global indicators}	&	\textsc{Acronym}	&	\textsc{Source}	\\ \hline
& 	&	Global inflation (OECD countries average)	&	GINF	&	OECD$^{4}$	\\
& 	&	Global Supply Chain Pressure Index	&	GSCPI	&	NY Fed$^{5}$	\\
& 	&	Chicago Fed National Financial Conditions Index	&	FCI	&	Chicago Fed$^{6}$	\\
& 	&	Global Economic Policy Uncertainty	&	GEPU	&	EPU webpage$^{7}$ \\ \hline\hline
\multicolumn{5}{p{.9\textwidth}}{ {\footnotesize
$^{1}$\href{https://sdw.ecb.europa.eu/}{https://sdw.ecb.europa.eu/} \linebreak
$^{2}$\href{https://data.oecd.org/industry/industrial-production.htm}{https://data.oecd.org/industry/industrial-production.htm}\linebreak
$^{3}$\href{https://ec.europa.eu/eurostat/databrowser/view/teibs010/default/table?lang=en}{https://ec.europa.eu/eurostat/databrowser/view/teibs010/default/table?lang=en}\linebreak
$^{4}$\href{https://data.oecd.org/price/inflation-cpi.htm}{https://data.oecd.org/price/inflation-cpi.htm}\linebreak
$^{5}$\href{https://www.newyorkfed.org/research/policy/gscpi\#/overview}{https://www.newyorkfed.org/research/policy/gscpi\#/overview}\linebreak
$^{6}$\href{https://www.chicagofed.org/research/data/nfci/current-data}{https://www.chicagofed.org/research/data/nfci/current-data}\linebreak
$^{7}$\href{https://www.policyuncertainty.com/}{https://www.policyuncertainty.com/} }}
\end{tabular}
}
\end{table}

The global series include Global Inflation (GINF), the Global Supply Chain Pressure Index (GSCPI), the Financial Conditions Index (FCI), and Global Economic Policy Uncertainty (GEPU). The FCI is the national index for the United States, produced by the Chicago Federal Reserve Bank; strictly speaking it is not a global average of multiple countries (as is the case with the other three variables). However, this FCI is a factor of over 100 financial time series, covering developments in stock, foreign exchange, bond, and other key U.S. markets at the forefront of global financial activity. Therefore, for euro-area countries in particular, the U.S. FCI can be a proxy for global financial shocks. Global inflation is the month-on-month growth rate of the price index series. All other variables are in levels. Sources and detailed definitions are in the bottom panel of \autoref{tab:data}.

In the benchmark QFAVAR specification we extract one factor per group of country-level macroeconomic indicators in \autoref{tab:data}, we focuses on quantiles $q=0.1,0.5,0.9$, and we set a maximum of $p=6$ lags in the state equation of the model.\footnote{Recall that the horseshoe prior can flexibly restrict certain lagged parameters in a flexible way.} Note, however, that one can easily obtain special and restricted cases of our model. When $\bm \Gamma=0$ all global shocks are only transmitted to country-level data via the quantile factors, making country-level responses symmetric (because each macroeconomic indicator loads on a single factor per quantile with weight $\lambda_{ij(q)}$). If the global shocks are also restricted to be absent from the state equation, the QFAVAR collapses to a quantile dynamic factor model for variables $\bm y_{t}$. Because the Bayesian QFAVAR has a conditionally normal form and the mean of a normal distribution is identical to its median, the QFAVAR can collapse into a FAVAR by considering estimation only for $q=0.5$. In practice, whenever we estimate the FAVAR as a benchmark model we do so using algorithms similar to the QFAVAR. When forecasting in \autoref{sec:fore} we estimate the QFAVAR and the FAVAR with variational Bayes, and when doing structural analysis in \autoref{sec:structural} we estimate both models with MCMC. Finally, as explained in \autoref{sec:lik_priors}, the QFAVAR priors are automatic and adjust to varying needs for shrinkage (depending on number of countries, indicators, lags, quantiles, etc.) and we use the same priors whenever estimating the FAVAR as a benchmark for comparison.

\subsection{In-sample quantile factor estimates and model fit}
\begin{figure}[H]
\centering
\includegraphics[width=0.95\textwidth, trim={7cm 3cm 7cm 1cm}]{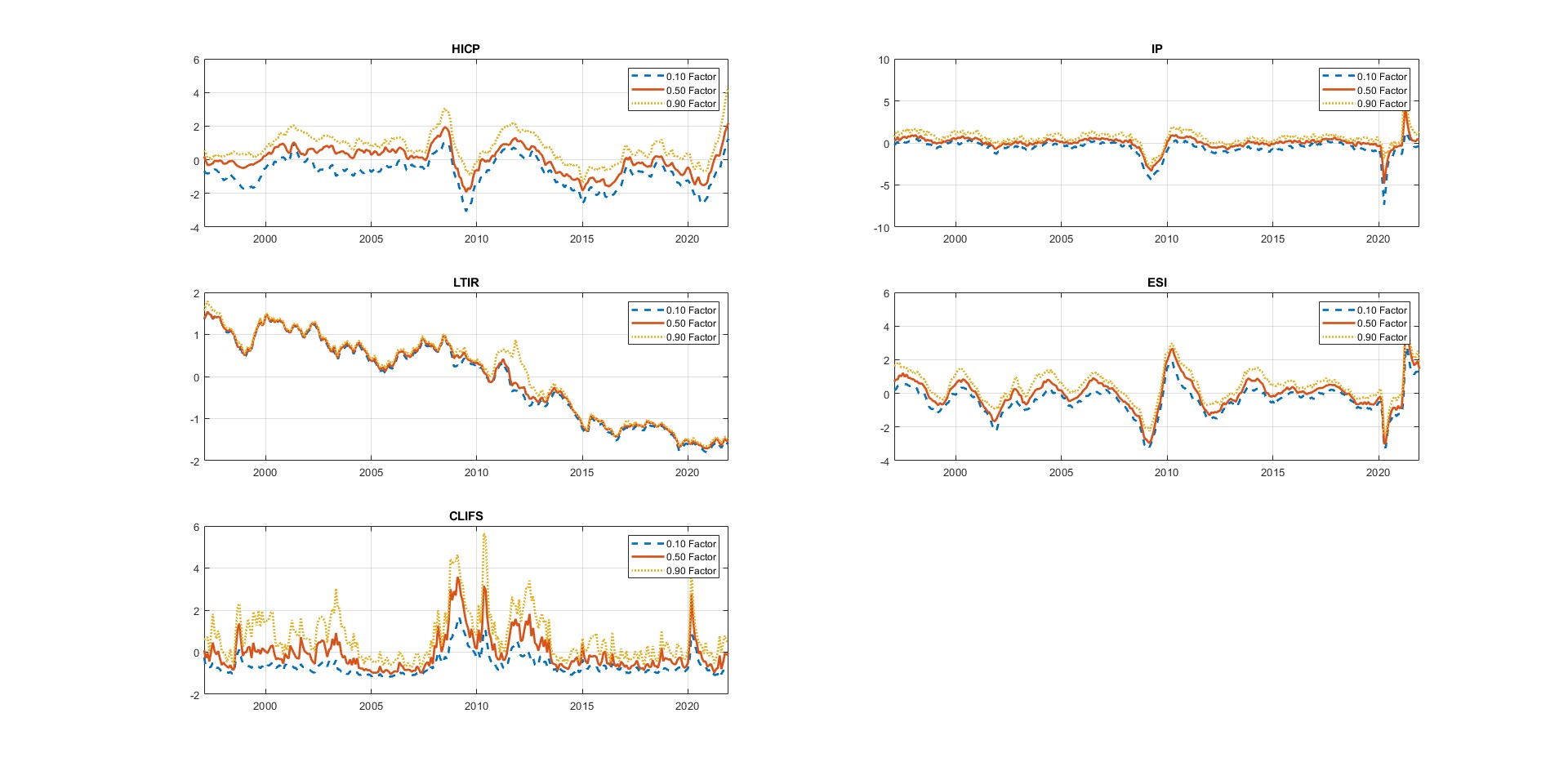}
\caption{Markov chain Monte Carlo estimates (posterior means) of the five euroarea quantile factors (10th, 50th, and 90th percentiles).} \label{fig:factor_MCMC}
\end{figure}

\begin{figure}[H]
\centering
\includegraphics[width=0.95\textwidth, trim={7cm 3cm 7cm 1cm}]{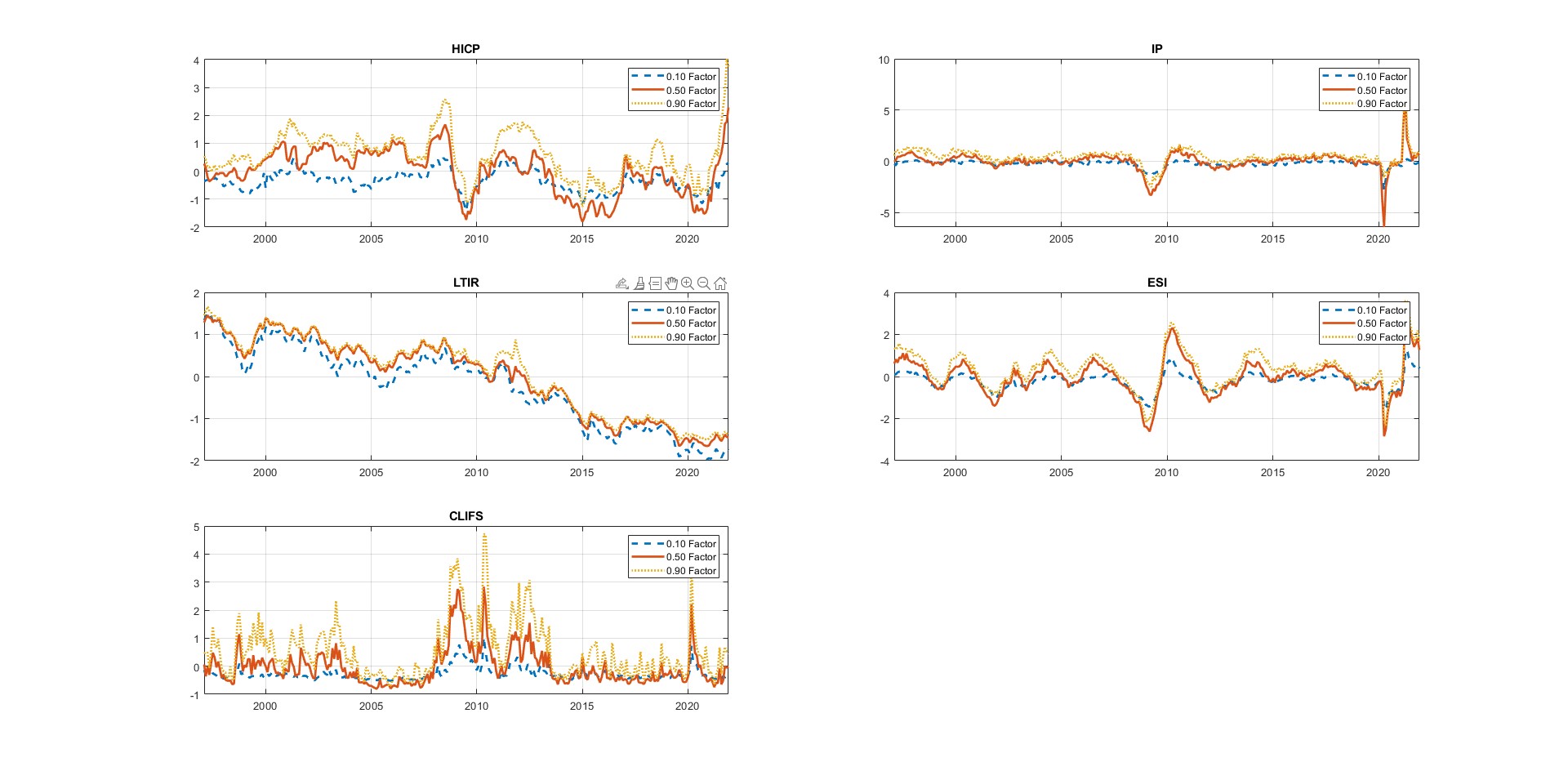}
\caption{Variational Bayes estimates (posterior means) of the five euro-area quantile factors (10th, 50th, and 90th percentiles).} \label{fig:factor_VB}
\end{figure}

It comes as no surprise that a large part of macroeconomic dynamics in the euroarea is characterized by common drivers among countries. Nevertheless, our aim is to showcase that considerable dynamics in the quantiles of macroeconomic variables remain hidden when considering in-mean factors only. In addition, these dynamics are not only interesting by themselves, but also imply strong and economically meaningful heterogeneity in the transmission of economic shocks. Figures \ref{fig:factor_MCMC} and \ref{fig:factor_VB} depict the estimated QFAVAR factors, using the one-step MCMC and two-step VB estimators, respectively. In both cases, the quantile factor estimates plotted are defined as posterior means of the 10th, 50th, and 90th percentile factors. For the sake of clarity the mean (FAVAR) factors are not plotted in this figure, as these are fairly indistinguishable from the median (50th percentile) factors. Therefore, median factors are a reference point when comparing those to tail factors.

A first inspection of the two figures reveals that there are no marked differences between the MCMC and VB estimates of the quantile factors, taking into account the fact that the two estimators are based on different modeling assumptions.\footnote{One-step MCMC estimation recovers the ``true'' dynamic factors implied by the QFAVAR specification that are also contemporaneously and dynamically correlated with the globals, $\bm g_{t}$. Two-step VB only estimates static factors as in \cite{KorobilisSchroeder2022} with no reference to $\bm g_{t}$.} The VB estimates of the 10th percentile factors sometimes cross with the median or even the 90th percentile (in the case of IP) factors, but such crossing or overlapping of quantile factor estimates is not present when using MCMC. Because latent factors combine distributional information from multiple series, some crossing might be present when using algorithms that assume complete independence between quantiles \citep[for example][]{Chenetal2021,KorobilisSchroeder2022}. Therefore, the one-step estimation performs better, possibly because it takes the correlation of the factors across quantiles into account. At the same time, the numerical results suggest that the use of one-step MCMC is essential in structural exercises where interpretation of the factors is important. When interest is in forecasting, the factors only serve the role of being reduced-form manifestations of the original data (even if they are not interpretable), in which case VB estimates can be perfectly acceptable; see discussion of this issue in next subsection.

An interesting question is whether the quantile factors capture more information in the data than the mean factors extracted with the simpler FAVAR method. To explore this question in depth, we provide an in-sample exercise in this section and a thorough out-of-sample exercise in the next section. To formally investigate the additional informational content of the quantile factors we follow \cite{DespoisDoz2022} and computing factor commonalities, defined as the $R^2$ of the multivariate regression of $y_{ij}$ onto the estimated factors
\begin{equation}
R^2_{y_{ij},\hat{F}} =  \frac{||\hat{F}(\hat{F}'\hat{F})^{-1} \hat{F}'y_{ij}||^2}{||y_{ij} ||^2}.
\end{equation}
In the above, $\hat{F}$ is the posterior mean estimate of the mean factors from the FAVAR, or the posterior mean of the $r$th quantile factor from the QFAVAR, where both models are estimated using MCMC. Table \ref{tab:commonality} presents the results for the mean factors in the first column. In the second and third column we then add the tail factors and the full set of QFAVAR factors to the mean factor. For inflation, adding the tail factors consistently increases the $R^2$, where we observe the largest gains for Finland, the Netherlands, and Portugal. Including the median factors increases the commonality further; performance improves the most for Italy and the least for Portugal. For industrial production, the general pattern is preserved. The $R^2$ for Belgium, Finland, and the Netherlands increases the most upon including the tail factors. Further small gains emerge from also including the median factors; however, the gains are overall less sizeable compared to inflation.

\begin{table}[H]
\centering
\caption{Commonality of factor estimates}  \label{tab:commonality}
\resizebox{.8\textwidth}{!}{
\begin{tabular}{lccc} \hline\hline
	&	$\phantom{+ F_{0.1} + F } F \hphantom{{0.5} + F_{0.9} }$		 &	$\hphantom{FF} F + F_{0.1} + F_{0.9} \hphantom{FF}$	&	$F + F_{0.1} + F_{0.5} + F_{0.9}$\\ \hline
	&		&		&		\\
HICP.AT	&	0.809	&	0.831	&	0.868	\\
HICP.BE	&	0.853	&	0.860   &	0.881	\\
HICP.DE	&	0.839	&	0.851	&	0.878	\\
HICP.ES	&	0.931	&	0.942	&	0.961	\\
HICP.FI	&	0.688	&	0.816	&	0.821	\\
HICP.FR	&	0.939	&	0.945	&	0.952	\\
HICP.IT	&	0.908	&	0.908	&	0.948	\\
HICP.NL	&	0.635	&	0.776	&	0.804	\\
HICP.PT	&	0.296	&	0.456	&	0.456	\\\hline
IP.AU	&	0.906	&	0.916	&	0.916	\\
IP.BE	&	0.764	&	0.809	&	0.822	\\
IP.DE	&	0.912	&	0.916	&	0.926	\\
IP.ES	&	0.951	&	0.954	&	0.955	\\
IP.FI	&	0.694	&	0.828	&	0.868	\\
IP.FR	&	0.967	&	0.971	&	0.972	\\
IP.IT	&	0.957	&	0.962	&	0.962	\\
IP.NL	&	0.655	&	0.751	&	0.763	\\
IP.PT	&	0.780   &	0.788	&	0.790   \\

 \hline \hline
\multicolumn{4}{p{.9\textwidth}}{Notes: This table contains the factor commonality estimates for inflation and industrial production for the set of the nine countries. The first columns shows the results for the mean FAVAR factor; the second column shows the results for the mean FAVAR and both QFAVAR tail factors; the final column shows the results for the mean FAVAR and all QFAVAR factors. }
\end{tabular}
}
\end{table}

\subsection{Out-of-sample evaluation of tail risks}
\subsubsection{Is the QFAVAR better than a FAVAR in capturing tail risks?}  \label{sec:fore}
In this subsection we shed more light onto the empirical fit of the QFAVAR, compared to the more established FAVAR model. Although the comparison of in-sample fit of the factors in the previous subsection (via $R^{2}$ statistics) is an informative exercise, we draw a more complete picture by comparing the out-of-sample performance of the QFAVAR versus the FAVAR. Given then importance of accurate projections and forecasts in policy work, this exercise is particularly insightful. We set up a straightforward recursive pseudo-out-of-sample (poos) exercise where we begin the estimation with $50\%$ of the total sample, forecast one to $24$ months ahead, add one more observation at the end of the sample, estimate the models, and forecast again up to $24$ horizons ahead. We compute forecasts iteratively using the VAR in state equation \eqref{state_full}, and subsequently we project them into forecasts for the conditional quantiles of the original variables $\bm y_{t}$ using the loadings matrices $\bm \Lambda$ and $\bm \Gamma$ in equation \eqref{measurement_full}.

We follow \cite{Manzan2015} and evaluate the tail forecasting performance using the following quantile score function
\begin{equation}
QS_{t\vert t-h}(q)^{M} = \left[ y_{ijt} - Q_{q}(y_{ijt}) \right] \left[ \mathbb{I}\left( y_{ijt} \leq Q_{q}(y_{ijt}) \right) - q \right]. \label{quantile_score}
\end{equation}
This score function is evaluated for the two competing models $M = \left \lbrace QFAVAR,FAVAR \right \rbrace$; the two extreme quantile levels $q=0.1, 0.9$; four forecast horizons $h=1,6,12,24$; two variables of interest $i={HICP,IP}$; and all nine euro-area countries $j$. The $QS$ function is a piecewise linear asymmetric loss function; therefore lower values signify better performance. Because we are interested in comparing the performance of only two different classes of models, we again follow \cite{Manzan2015} and calculate the $t$-statistic testing the equality of the average of the models' $QS$ functions over the whole out-of-sample period. This is defined as the spread between the $QS$ values of the QFAVAR versus the $QS$ values of the FAVAR. Values of this statistic for different variables and forecast horizons are shown in \autoref{tab:forecasting}. The visibly higher proportion of negative values implies that, overall, the QFAVAR experiences less forecast performance loss in both the left and right tails of the distribution of HICP and IP. For HICP, performance gains are statistically significant at the $5\%$ level for $h=1$ for the bottom tail and for $h=1$ and $h=6$ for the top tail of the distribution. For a subset of countries, additional significant gains also emerge for longer forecast horizons. For industrial production the picture is generally comparable with the exception of some countries; however, significant performance gains also arise for the bottom tail of the distribution for the longer horizons $h=12$ and $h=24$. Generally, the QFAVAR provides clear performance gains over the symmetric, Gaussian disturbances of the FAVAR, highlighting the importance of flexible asymmetric modeling of conditional quantiles within a multivariate setting.

\begin{table}[H]
\centering
\caption{T-statistics of equal quantile predictive accuracy}  \label{tab:forecasting}
\resizebox{.8\textwidth}{!}{
\begin{tabular}{lccccccccc} \hline\hline
	&	\multicolumn{4}{c}{$t-stat_{10}$}							&		&	\multicolumn{4}{c}{$t-stat_{90}$}							\\
	&	$h=1$	&	$h=6$	&	$h=12$	&	$h=24$	&		&	$h=1$	&	$h=6$	&	$h=12$	&	$h=24$	\\ \hline
	&		&		&		&		&		&		&		&		&		\\
HICP.AT	&	-\textbf{3.04}	&	-0.57	&	-\textbf{2.34}	&	-0.13	&		&	-\textbf{5.98}	&	-\textbf{3.71}	&	-\textbf{2.52}	&	0.12	\\
HICP.BE	&	-\textbf{2.94}	&	-0.61	&	-0.02	&	-0.81	&		&	-\textbf{5.74}	&	-\textbf{3.01}	&	-1.00	&	0.12	\\
HICP.DE	&	-\textbf{2.40}	&	-1.48	&	-\textbf{2.39}	&	-0.25	&		&	-\textbf{5.91}	&	-\textbf{3.75}	&	-1.63	&	-0.10	\\
HICP.ES	&	-\textbf{3.30}	&	-1.53	&	-1.68	&	-0.01	&		&	-\textbf{5.14}	&	-\textbf{2.10}	&	-0.63	&	0.73	\\
HICP.FI	&	-\textbf{2.15}	&	-1.54	&	-1.57	&	-\textbf{2.42}	&		&	-\textbf{4.75}	&	-\textbf{3.29}	&	-\textbf{2.95}	&	-0.71	\\
HICP.FR	&	-\textbf{3.37}	&	-1.20	&	-0.11	&	1.25	&		&	-\textbf{6.87}	&	-\textbf{2.58}	&	-1.44	&	-0.25	\\
HICP.IT	&	-\textbf{3.65}	&	-1.68	&	-0.67	&	0.35	&		&	-\textbf{5.62}	&	-\textbf{3.23}	&	-1.86	&	-0.40	\\
HICP.NL	&	-\textbf{2.68}	&	-\textbf{2.56}	&	-1.63	&	1.31	&		&	-\textbf{2.44}	&	-\textbf{2.18}	&	-1.94	&	0.11	\\
HICP.PT	&	-\textbf{4.06}	&	0.02	&	-1.07	&	-1.29	&		&	-\textbf{4.37}	&	-\textbf{2.34}	&	-0.35	&	0.76	\\
	&		&		&		&		&		&		&		&		&		\\
IP.AT	&	-\textbf{2.11}	&	-1.98	&	-1.92	&	-\textbf{4.38}	&		&	-\textbf{7.24}	&	-1.57	&	-0.63	&	-0.08	\\
IP.BE	&	-\textbf{2.19}	&	-\textbf{2.77}	&	-\textbf{2.04}	&	-\textbf{3.57}	&		&	-\textbf{3.67}	&	-\textbf{2.36}	&	-0.73	&	-1.07	\\
IP.DE	&	-1.62	&	-1.31	&	-\textbf{2.40}	&	-\textbf{3.77}	&		&	-1.88	&	-1.28	&	-0.04	&	-\textbf{2.44}	\\
IP.ES	&	-\textbf{2.06}	&	-1.90	&	-\textbf{3.67}	&	-\textbf{3.34}	&		&	-\textbf{3.66}	&	-0.77	&	1.80	&	1.08	\\
IP.FI	&	-\textbf{2.79}	&	-\textbf{2.48}	&	-\textbf{3.82}	&	-\textbf{5.70}	&		&	-\textbf{2.90}	&	-1.71	&	1.35	&	-1.87	\\
IP.FR	&	-0.93	&	-1.34	&	-1.90	&	-\textbf{2.78}	&		&	-\textbf{4.59}	&	-0.05	&	1.98	&	0.55	\\
IP.IT	&	-0.92	&	-1.82	&	-\textbf{2.31}	&	-\textbf{2.74}	&		&	-1.27	&	-0.18	&	1.25	&	-0.47	\\
IP.NL	&	-\textbf{2.24}	&	-\textbf{3.02}	&	-\textbf{2.04}	&	-\textbf{5.22}	&		&	-\textbf{2.54}	&	-0.72	&	1.04	&	0.22	\\
IP.PT	&	-0.74	&	-1.11	&	-\textbf{2.25}	&	-\textbf{2.02}	&		&	-\textbf{5.45}	&	-1.88	&	-1.69	&	-1.88	\\  \hline \hline
\multicolumn{10}{p{.9\textwidth}}{{\footnotesize Notes. Entries in this table are $t$-statistic values for the null hypothesis of equal accuracy of the quantile forecasts from the QFAVAR relative to the FAVAR benchmark. Columns 2-5 show the statistics for 10th percentile forecasts over forecast horizons $h=1,6,12,24$ months, and columns 6-9 show the statistics for the 90th percentile forecasts over the same forecast horizons. Because all values of the statistic are negative, this shows that the QFAVAR experiences less forecast loss than the benchmark FAVAR. Values lower than the critical value of -2 indicate that QFAVAR tail forecasts are significantly better than the FAVAR forecasts at the $5\%$ level.}}
\end{tabular}
}
\end{table}

Admittedly, evaluating only a certain area (left or right tail) of the distribution of inflation and output is of paramount importance to policy-makers who are interested in assessing worst-case scenarios for these two variables of interest. However, many times the full forecast distribution of the variables of interest is required, most notably when preparing fan charts similar to the ones maintained by the Bank of England \citep{Brittonetal1998}. In such cases, one has to construct the full forecast distribution by interpolating forecasts of a wide range of quantiles. This procedure would not be accurate with the benchmark specification of the QFAVAR, where for parsimony we focus only on the 10th, 50th, and 90th percentiles. For that reason, the model can be trivially extended to consider joint estimation of a larger number of quantiles. Following \cite{Chenetal2021} we re-estimate the QFAVAR model this time for $q=0.05,0.1,0.25,0.5,0.75,0.9, \text{ and } 0.95$. This implies $R=7$ and the dimension of the state now increases to $Rm+k = 7 \times 5+4 = 39$ elements for each time period $t$. Nevertheless, computational load doesn't increase noticeably when adopting the two-step variational Bayes algorithm.\footnote{This is not true for the one-step MCMC algorithm, as increasing the dimension of the state variable would increase the computational demands of the filtering problem substantially. Additionally, the one-step estimator requires writing the VAR($p$) evolution in the state equation into a VAR(1) form, which requires repeatedly sampling a state vector that is even larger; that is, a vector with $p(Rm+k)=234$ elements for $p=6$.} 

\begin{figure}[H]
\centering
\includegraphics[width=0.8\textwidth, trim={7cm 1cm 6cm 2.8cm}, clip]{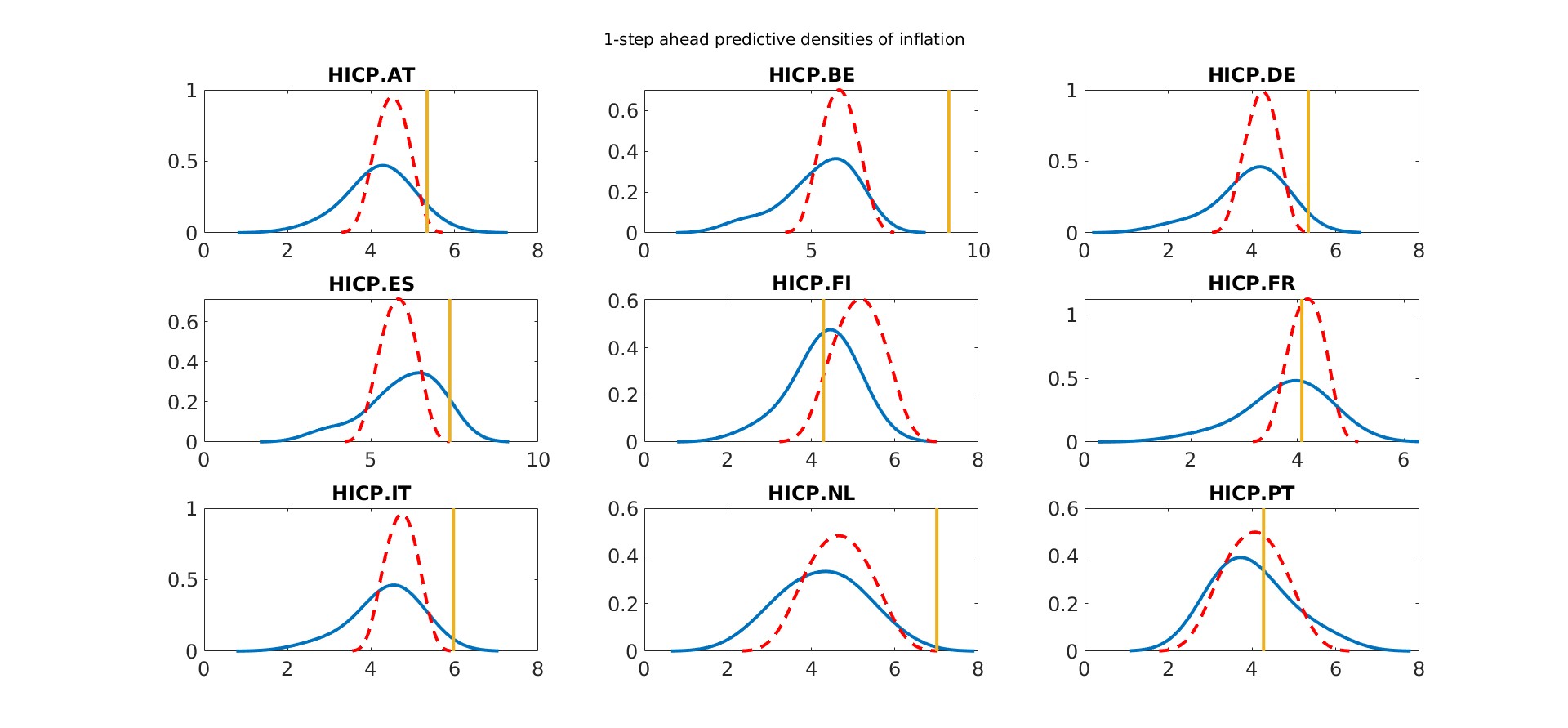}
\caption{One-step-ahead forecast distributions of inflation from QFAVAR (blue line) and the FAVAR (red line), both estimated using variational Bayes using a finer grid of quantiles ($q=0.05,0.1,0.25,0.5,0.75,0.9, \text{ and } 0.95$) compared to the benchmark specifications. The yellow vertical line indicates the realized value in January 2022.} \label{fig:HICP_fore}
\end{figure}

\begin{figure}[H]
\centering
\includegraphics[width=0.8\textwidth, trim={7cm 1cm 6cm 2.8cm}, clip]{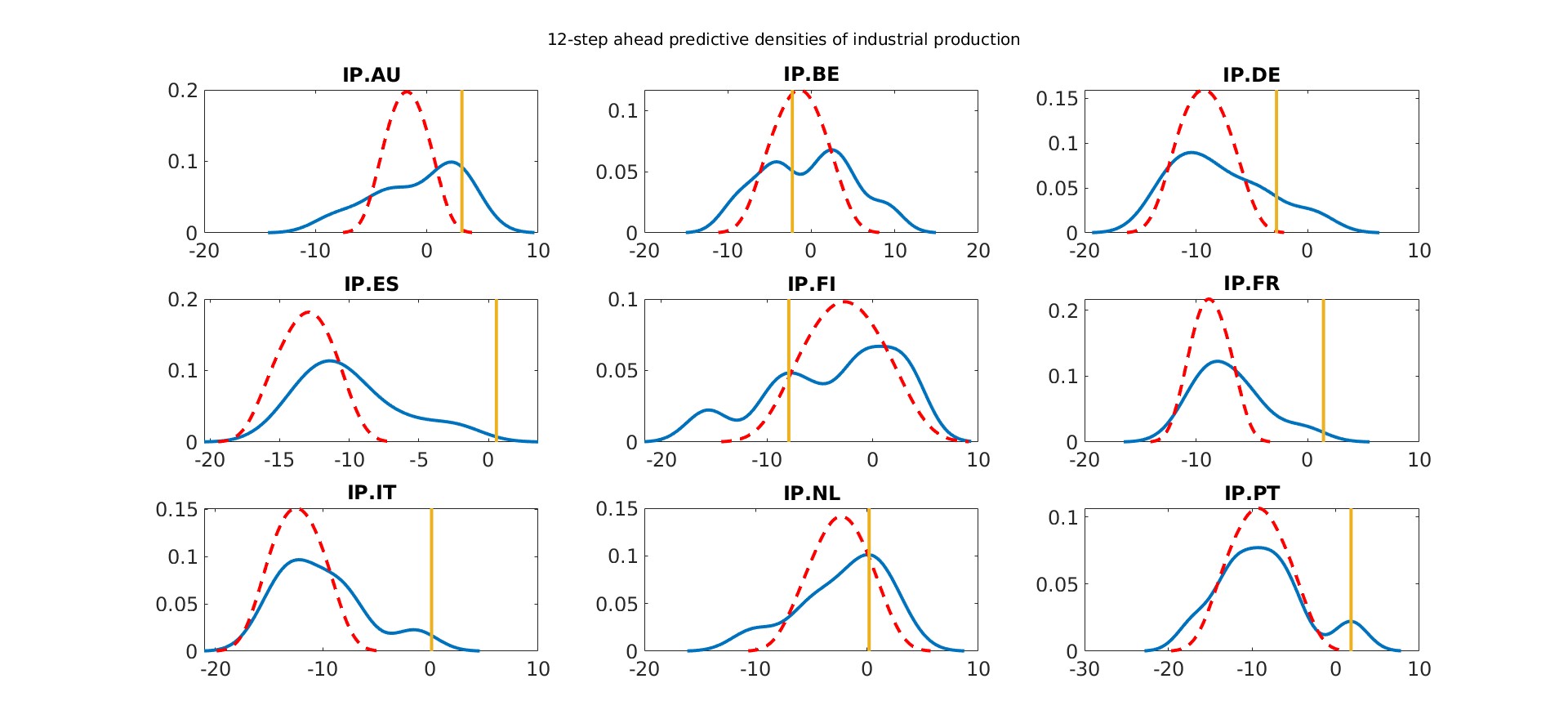}
\caption{One-step-ahead forecast distributions of industrial production from QFAVAR (blue line) and the FAVAR (red line), both estimated using variational Bayes using a finer grid of quantiles ($q=0.05,0.1,0.25,0.5,0.75,0.9, \text{ and } 0.95$) compared to the benchmark specifications. The yellow vertical line indicates the realized value in January 2022.} \label{fig:IP_fore}
\end{figure}

Figures \ref{fig:HICP_fore} and \ref{fig:IP_fore} illustrate the one-step-ahead predictive distributions of country-level inflation and industrial production, constructed from the QFAVAR (blue solid line) versus the FAVAR (red dashed line) estimated using the full sample up to December 2021. The date is chosen as it marks the beginning of an era of rapidly increasing food and energy prices, as well as shortages of raw materials and other frictions in production and global trade resulting from the Covid-19 pandemic. The yellow vertical line indicates the realized value in January 2022. We construct these densities by first predicting the one-step-ahead value of each variable for each of the $R=7$ quantiles and then fitting a Gaussian kernel smoother to these quantiles.\footnote{This is done using the MATLAB R2022b function \texttt{ksdensity} with default settings.} Based on these two figures it is clear that there are significant differences between the two modeling approaches. Compared to the symmetric FAVAR distributions, the QFAVAR features asymmetric and, in many cases, bimodal distributions. In addition, the QFAVAR distributions place more probability mass on the right tail of inflation and are left-skewed in case of industrial production. The FAVAR distributions look, by and large, like symmetric weighted averages of the more complex QFAVAR distributions, hiding the true extent of uncertainty around forecasts produced in December 2021. The last few months of 2021 signalled a period of inflationary pressures, and policy-makers faced huge uncertainty deciding whether these pressures would prove transitory or more pervasive. With the advent of the Russian invasion of Ukraine in early 2022, it was clear that food and energy inflation would persist and prospects for positive output growth could be dim due to subsequent interest rate increases and the build-up of financial risk. Given ex-post knowledge of these extreme realized macro risks, it is not surprising that forecast distributions of inflation and industrial production constructed from the QFAVAR look more plausible than the symmetric FAVAR distributions.

\subsubsection{Are global variables relevant for forecasting tail risk?} 
With multiple global shocks affecting the euro-area, it is unclear how these transmit to country-level risks of inflation and industrial production \citep{Panetta2023}. Understanding the underlying transmission channels of such global predictors is an important issue for policy-makers. A good starting point towards this direction is the related debate on whether financial predictors affect the left quantile of GDP. Despite the fact that \citep{Adrianetal2019} find that financial conditions help forecast the left tail of U.S. GDP, \cite{plagborgmoller2020} argue that no single financial predictor is persistently important for forecasting recessions. Their argument is based on the finding that the relationship between GDP and financial conditions is inherently unstable over time. This finding suggests that policy makers should not mechanically include financial conditions as predictors of tail events.

A major limitation of \cite{Adrianetal2019} and \cite{plagborgmoller2020} is that they rely on univariate quantile regressions; therefore, predictors enter in the form of right-hand-side exogenous variables for each quantile level. This is not the case with the proposed QFAVAR specification, where all estimated conditional quantiles of euro-area macroeconomic variables and the global predictors are endogenous and can interact both contemporaneously and dynamically. In order to explore the effects of global predictors on country-level macro risks, we estimate and forecast with four models: i) our benchmark QFAVAR, ii) a QDFM, which is the QFAVAR with quantile factors only and no global variables $\bm g_{t}$, iii) a quantile autoregressive (QAR) model, and iv) a quantile autoregressive model with global variables $\bm g_{t}$ as predictors (QAR-X). To eliminate the impact of the shrinkage priors in driving forecast performance, we estimate all four models with the parsimonious choice of $p=1$ lag and an intercept, deviating from the remainder of the pager where we use $p=6$ and an intercept.\footnote{In many macroeconomic forecasting problems, longer lag lengths fit very well in-sample, but when it comes to forecasting out-of-sample the choices $p=1$ or $p=2$ are hard to beat.} Finally, also for the sake of consistency we estimate the two univariate quantile models using the same variational Bayes algorithm, despite the fact that MCMC is not cumbersome in this case \citep[see for example the fast algorithm in][]{Korobilisetal2021}.

\begin{figure}[H]
\centering
\includegraphics[width=0.95\linewidth, trim={13cm 1cm 10cm 2cm}]{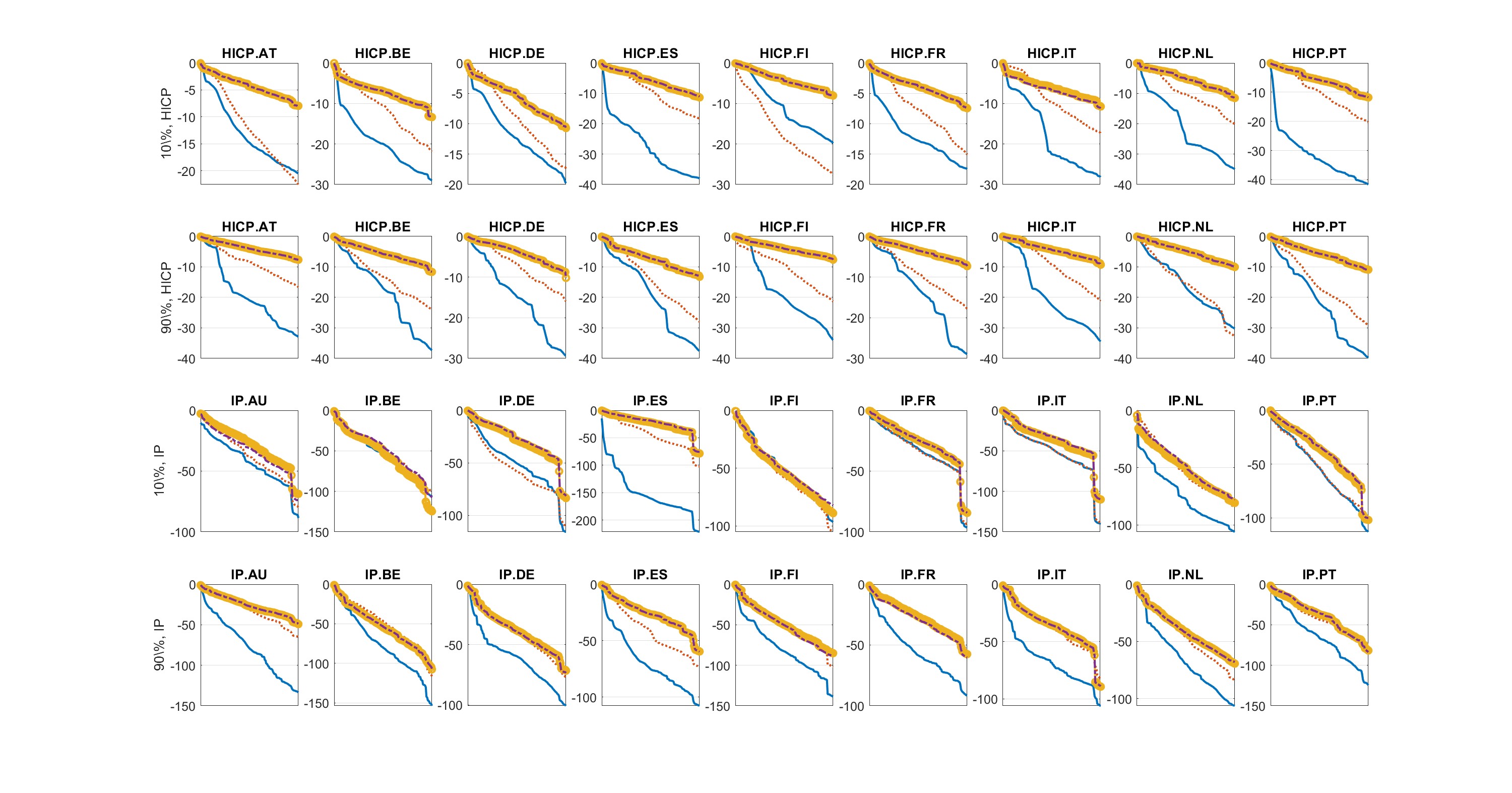}
\caption{Cumulative quantile score (QS) loss for forecast horizon $h=1$. We compare four models: QAR (yellow circled line), QAR-X (purple dashed line), QDFM (red dotted line), and QFAVAR (blue solid line). The out-of-sample evaluation period shown on the x-axis is 2011Jan to 2022Dec-$h$. The first two rows show quantile scores for the 10th and 90th percentiles of inflation, and the third and fourth rows show quantile scores for the 10th and 90th percentiles of industrial production.} \label{fig:h1QSCORES}
\end{figure}

Figure \ref{fig:h1QSCORES} plots the cumulative quantile score values for inflation and industrial production, respectively, computed using equation \eqref{quantile_score} in the previous subsection. Compared to average quantile scores, cumulative sums reflect the evolution of forecast performance of different models over the out-of-sample period. Given that the $QS_{t\vert t-h}(q)^{M}$ statistic is a loss function, the best model is the one with the lower cumulative values. The figure has four rows of nine panels, where the first two rows correspond to the QS performance at the 10th and 90th percentile of inflation at horizon $h=1$ and the columns correspond to the individual euro-area countries. The third and fourth row present the corresponding results for industrial production. Interesting patterns emerge from this graph. At first glance, it is obvious that for the short forecast horizon $h=1$, multivariate quantile models are superior to univariate quantile models in producing accurate left- and right-tail forecasts of inflation. In particular, the QFAVAR dominates all other models, and in many cases the largest reductions in loss are early in the out-of-sample period, that is, during the 2011 eurozone debt crisis.\footnote{See for example, the 10th percentile QS of HICP.IT, and the 10th percentile QS of IP.ES.} This is not a surprising result, as it is well-known that the debt crisis created strong comovements and global economic unrest. The multivariate QFAVAR is able to fit this period better than the univariate competitors and the QDFM.

Looking more closely at the results, we can infer several stylized facts regarding short-term quantile forecasting. First, global predictors seem to provide irrelevant information when considering univariate forecasting models, but this is not true when considering multivariate models. This stylized fact shows that the conclusions of \cite{plagborgmoller2020} might only apply to univariate quantile models. Second, among the multivariate models, the QFAVAR is clearly superior to the QDFM that does not consider global predictors. Third, inflation comovements in multivariate data seem to be important, because both the QFAVAR and QDFM markedly improve the performance of the univariate models. Finally, for industrial production, the QDFM provides identical forecast performance to the univariate models in most cases. The large improvements observed for the QFAVAR therefore likely come from the inclusion of global predictors as endogenous variables.

Moving to longer-term forecasts, this clear picture is lost as there is no clear winner. The only pattern that emerges is that the quantile AR(1) without global predictors is always the worst-performing model. Overall, at longer horizons both the QFAVAR and QDFM and the quantile AR with exogenous predictors performs very well. Figures supporting these results are available among the additional empirical results in the online supplement.

\subsection{Evaluating global spillovers using the QFAVAR} \label{sec:structural}
\subsubsection{Quantile factor responses} \label{sec:FEVD}
\begin{figure}[H]
\centering
\captionsetup{width=\linewidth}
\includegraphics[width=0.9\linewidth, trim={9cm 3cm 4cm 3cm}]{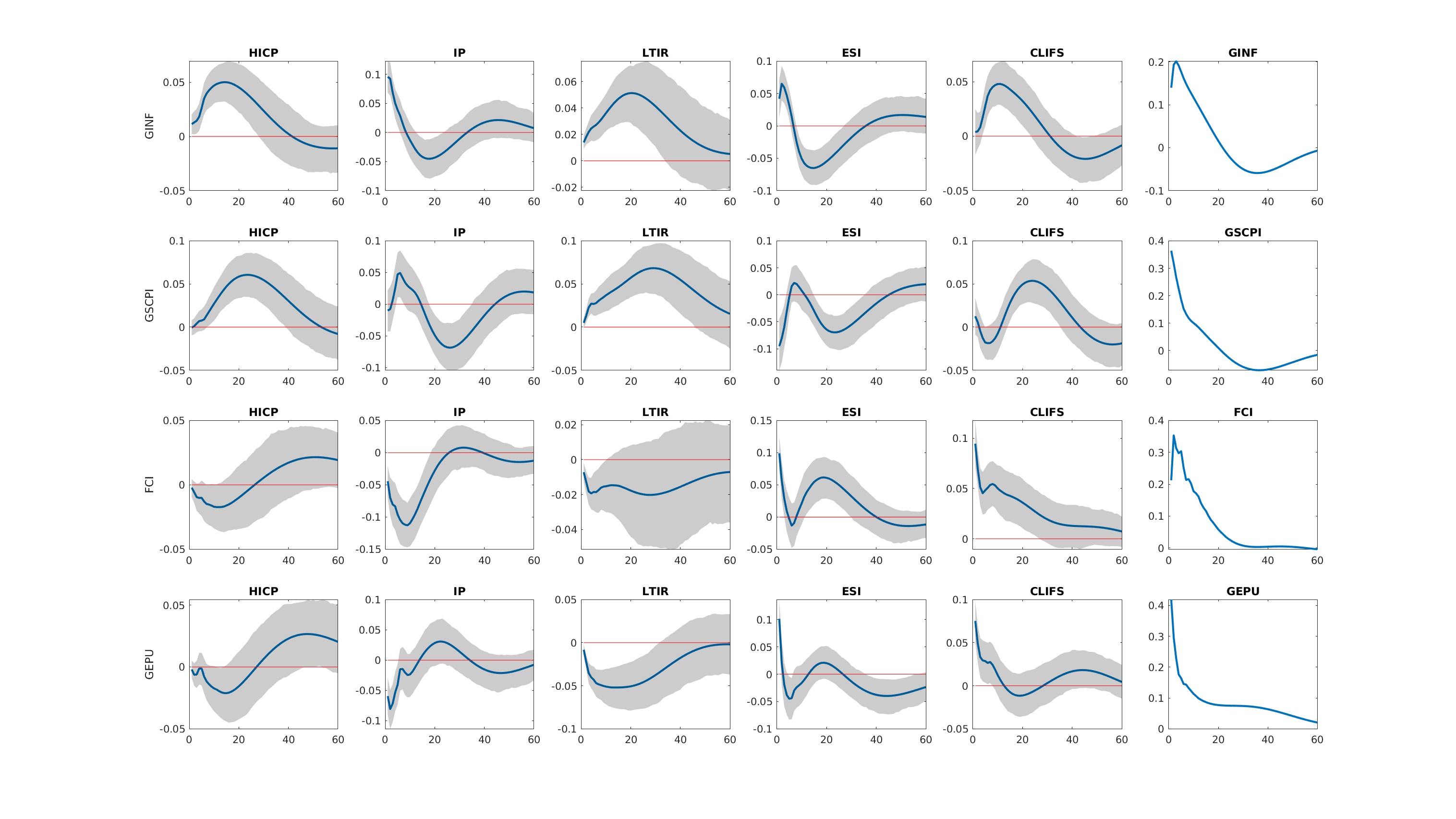}
\caption{Impulse response functions (posterior median and 68\% bands) of mean factors to shocks in the global variables, based on the benchmark FAVAR specification. Each row represents a global shock (GINF, GSCPI, FCI, GEPU); the own-response is shown in the last column. The first five columns show the responses of the five factors extracted from country-level euro-area data.} \label{fig:FAVAR_IRF}
\end{figure}

\begin{figure}[H]
\centering
\captionsetup{width=\linewidth}
\includegraphics[width=0.9\linewidth, trim={9cm 3cm 4cm 3cm}]{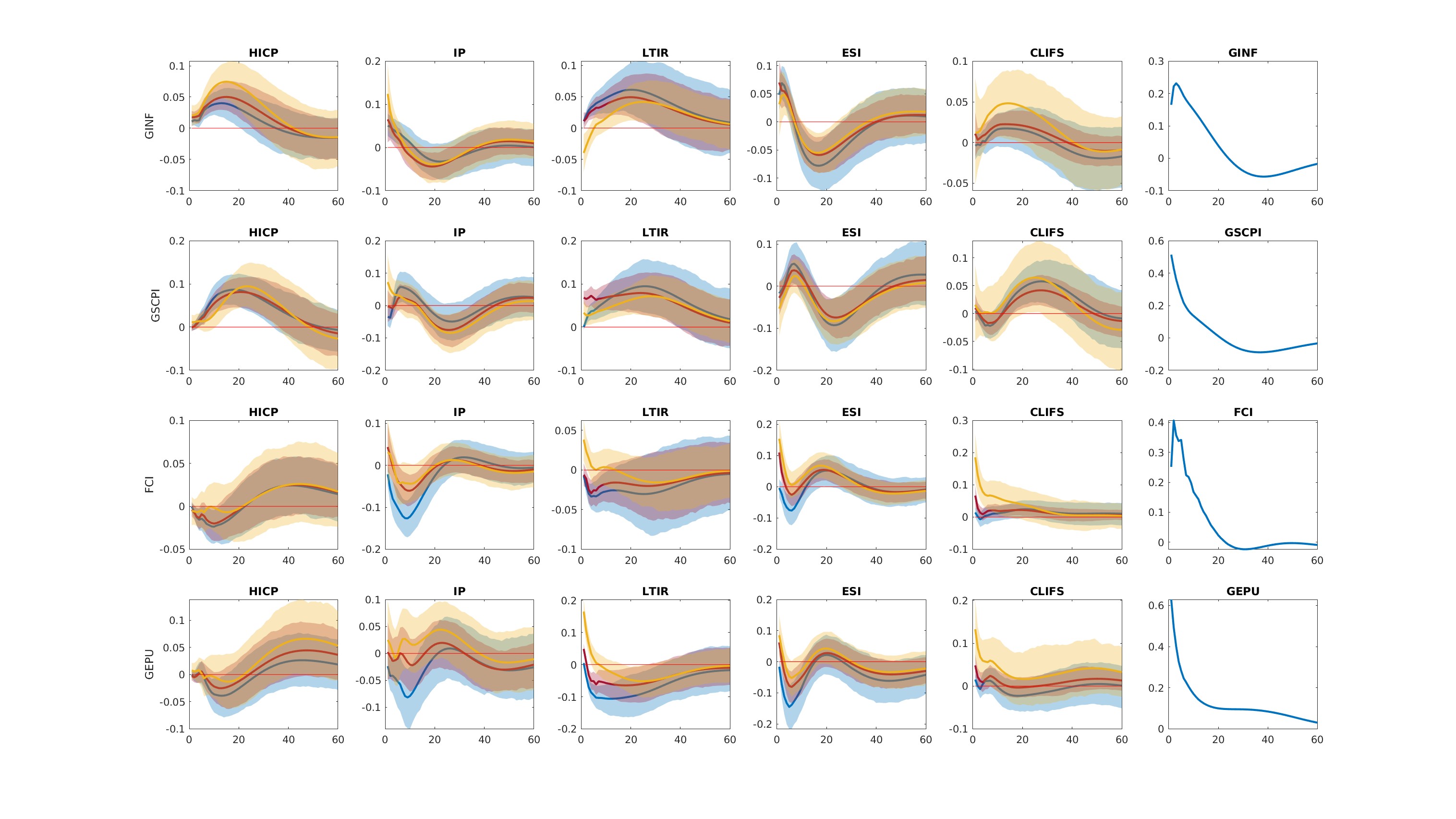}
\caption{Impulse response functions (posterior median and 68\% bands) of 10th, 50th, and 90th percentile factors to global shocks from the benchmark QFAVAR specification. Each row represents a global shock (GINF, GSCPI, FCI, GEPU); the own-response is shown in the last column. The first five columns show the responses of each of the three percentiles of the five factors extracted from country-level euro-area data. Responses of the 10th percentile factors are in blue, reponses of the 50th percentile factors are in red, and responses of the 90th percentile factors are in yellow.} \label{fig:QFAVAR_IRF}
\end{figure}

We compute generalized impulse response functions\footnote{There are numerous ways of turning a reduced-form VAR model into a structural econometric model by means of imposing sensible and plausible identification restrictions. These identification schemes range from recursive and long-run restrictions, to the currently popular methodologies of sign restrictions and identification via instrumental variables. However, the QFAVAR has the particularity that it models different quantiles of the same variables jointly. A global shock will not necessarily have the same effect on all quantiles of a given variable. Even if economic theory, intuition or common sense are available to help choose identification restrictions in VARs, these restrictions will hold on average (that is, in the median) and there are no guarantees that structural relationships remain the same during ``exceptional times'' (that is, at the tails of a distribution). Therefore, the QFAVAR requires careful consideration of plausible restrictions that would help identify structural shocks. We leave this exercise for future research, and in this paper we focus exclusively on generalized impulse response functions that require minimal assumptions about signs, magnitudes or other important features of shocks.} of the FAVAR and QFAVAR factors to changes in global inflation, the global supply chain pressure index, financial conditions, and global economic policy uncertainty. While factors from these models are unobserved, in the case of multi-country analysis the indicator-specific factors we estimate by averaging the same series for all nine countries, can be thought of as proxying aggregate euroarea indicators. Therefore, responses of mean (in the FAVAR) and quantile (in the QFAVAR) factors of HICP, IP, LTIR, ESI and CLIFS to global shocks are particularly interesting to policy-makers interested in characterizing aggregate-level dynamics instead of country-level heterogeneities. The FAVAR and QFAVAR impulse response functions (IRFs) are presented in figures \ref{fig:FAVAR_IRF} and \ref{fig:QFAVAR_IRF}, respectively. Note that quantile IRFs can be interpreted in two ways. On one hand, the factor responses reflect features of the same distribution. Consequently, responses of quantile factors suggest changes in skewness and kurtosis, as well as upside and downside risks. If the quantile responses overlap completely, this is indicative of a level shift in the entire distribution. On the other hand, one can interpret the responses as distinct quantities, through the lens of scenario analysis. In this case, depending on the interpretation of the underlying variables, the 10th (90th) percentile factor can be interpreted as an adverse (favorable) benchmark scenario and, hence, contribute important additional information beyond the median IRFs. As a result, both interpretations provide useful supplements to existing models in the policy-making process. 

In the FAVAR responses, shown in \autoref{fig:FAVAR_IRF}, an increase in global inflation coincides with a positive impact response for all five factors. Given that we compute generalized IRFs, global inflation might largely be driven by demand-side forces. The profile of the corresponding quantile factor responses is empirically interesting. During the first 20 months following a global inflation shock, the 90th percentile factor of inflation reacts more strongly than the median and the 10th percentile factor. During the first five months, this pattern repeats for industrial production. This fact suggests that the distributions of inflation and industrial production become more leptokurtic as well as positively skewed following the shock. Taking the scenario perspective instead, the 90th percentile factor responses represent a benchmark scenario with larger global spillovers and a stronger response of inflation compared to what is the case in the FAVAR. For the global supply chain pressure indicator, inflation does not respond on impact. During subsequent months, inflation then responds positively, peaking after about two years. In the QFAVAR, however, we see that the 90th percentile inflation factor does respond positively on impact, indicating upside risk to the inflation outlook. In the following months, the median and 10th percentile factor follow up, shifting the entire inflation distribution toward higher realizations. For industrial production, the FAVAR first indicates a contraction on impact. The median factor mirrors this behaviour. The left-tail factor responds negatively and the right-tail factor responds positively on impact, translating into an increase in uncertainty about the industrial production outlook and a more leptokurtic distribution. In subsequent periods, the 90th percentile factor contracts more strongly than the 10th percentile factor, inducing a negative skew in the distribution, accompanied by marked output contraction over the medium turn. Alternatively, one could construct a scenario combining the response of the 90th percentile of inflation and median industrial production factor instead. In this scenario, the shock would lead to an output response similar to what is the case in the FAVAR, but with higher inflation, worsening the policy trade-off. The responses of economic sentiment, financial stress, and especially the long-term interest rate all indicate higher uncertainty about the outlook than under the FAVAR.

In case of tightening global financial conditions, the left tail and median inflation factors respond similarly to the mean factors during the first months after the tightening. The right tail factor has a slightly more muted response, inducing positive skew. Overall, the distribution shifts toward lower realizations. What is more striking is the response of industrial production. Compared to the mean model, the median and 90th percentile factor respond less, but the 10th percentile factor responds with a more extreme output contraction. The distribution hence becomes strongly negatively skewed and suggests non-negligible downside risks to the production outlook. Policy-makers might hence pay particular attention to this scenario when monitoring global financial conditions. Combining the 10th percentile inflation and industrial production factors, allows for the construction of a severe scenario in which both prices and output react more negatively than in the FAVAR. The responses of the other variables also show richer dynamics than under the FAVAR. Although the sentiment outlook becomes overall more uncertain, the interest rate and financial stress exhibit considerable positive skew. 

Finally, an increase in global economic policy uncertainty is another interesting case. In the FAVAR, inflation and especially industrial production respond negatively during the first 30 and 10 months, respectively. For inflation, this pattern is broadly mirrored by the quantile factors. The 90th and 10th percentile factors respond less and more strongly than the mean factors, respectively. Overall, the inflation distribution thus becomes more leptokurtic. In the case of industrial production, however, the median responds little, the 90th percentile factor responds positively, and the 10th percentile factor responds strongly negatively, suggesting negative skew and pervasive uncertainty about the outlook. Taking the scenario view, one could construct a severe (benign) scenario by combining the 10th (90th) percentile responses of the inflation and industrial production factor. The severe scenario would then correspond to a case with a more pronounced decline in inflation as well as output. Although the ESI shows signs of downside risks to the sentiment outlook, an increase in global economic policy uncertainty correlates with pronounced upside risks for the long-term interest rate and financial stress. 

\begin{figure}[H]
\centering
\includegraphics[width=\textwidth, trim={7cm 3cm 7cm 3cm}]{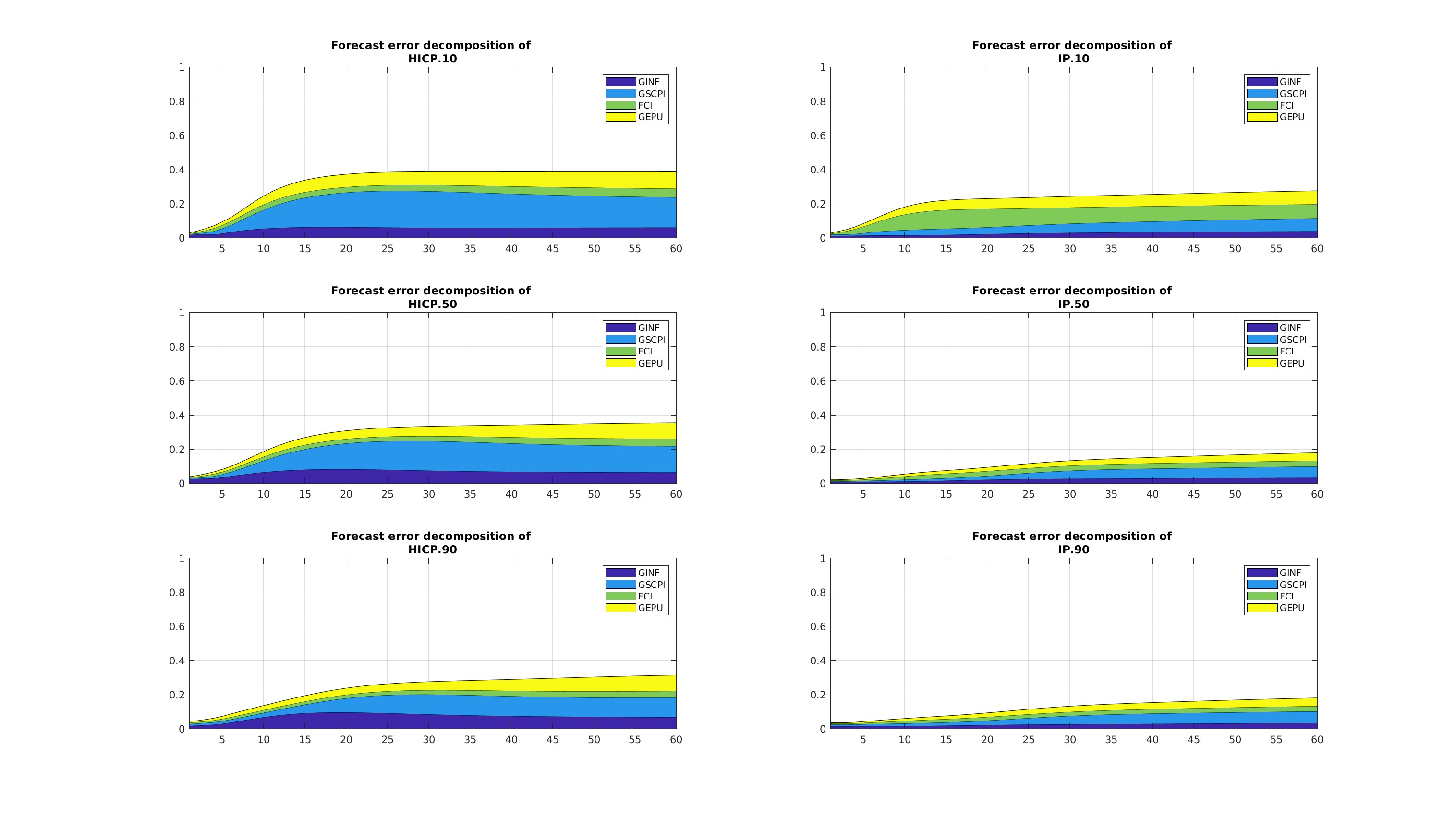}
\caption{Forecast error variance decompositions of the 10th, 50th, and 90th percentile factors for inflation and industrial production.} \label{fig:QFAVAR_FEVD}
\end{figure}

To provide deeper insights into the importance of the four global variables as drivers of the inflation and output responses, figure \ref{fig:QFAVAR_FEVD} displays generalized forecast error variance decompositions of the different quantile levels of the inflation and output factors. For the sake of clarity we only show the contribution of the global variables, which is why the contribution shares do not add up to 100\%. A few interesting patterns stand out. Overall, changes in global variables explain the most variation in the 10th percentile factors of inflation and IP, and they contribute less to the forecast error of the median and 90th percentile factors. Generally, they are more important for inflation than industrial production, with the global variables accounting for roughly 40\% of the forecast error variance of the 10th percentile inflation factor and close to 30\% of the 10th percentile industrial production factor. For inflation, global inflation is the most important global variable on impact. This pattern also holds true for the other quantile levels. In the median and longer run, the global supply chain pressure index becomes dominant and contributes by far the largest share. At the 10th percentile, it accounts for almost 20\%.

Interestingly, the timing of when the supply chain pressure index becomes dominant differs across quantile levels. At the 10th, 50th, and 90th percentile levels it contributes the largest share after roughly 5, 10, and 15 months, respectively. Global economic policy uncertainty is the third most important global variable, and global financial conditions are the least important global variable driving the inflation quantile factors. This picture is different for industrial production. At the 10\% level, the FCI is the most important global variable for most of the forecast horizon, contributing about 10\% to the total variation. The other global factors have roughly equal share, with global inflation contributing the least. These findings again show that financial conditions emerge as an important driver of output at risk - once modelled in a multivariate system - adding to the open debate. At the median and 90\% level the global supply chain pressure indicator is dominant, followed by global economic policy uncertainty. Together they account for more than half of the total contribution by global variables. The FCI and global inflation contribute relatively equal shares. Overall, these findings point towards non-negligible asymmetries in the inflation and industrial production process, with downside risks being particularly responsive to global developments.

\subsubsection{Country-level responses to global shocks}

Finally, our proposed QFAVAR allows mapping the quantile factor responses back to the panel of individual country-level variables. To do so, we simply project the quantile factor responses back to the measurement equation using the loading matrices $\bm \Lambda$ and $\bm \Gamma$. Note that $\bm \Gamma$ plays a particular role for the quantile dynamics of the country-level variables. With $\bm \Gamma=\bm 0$, the global factors have no direct impact on the country-level variables and only affect their quantile responses indirectly through the factors. The entire dynamics in the measurement equation are then driven by the common component. Because we allow the individual variables to load on their group-specific factors only, this also implies that the country-level IRFs are just the factor-level IRFs rescaled by the individual loadings. In this case, heterogeneity between the individual EA countries and across quantile levels is limited to the magnitude of the responses. To allow for more flexible degrees of heterogeneity, in this section we focus on the case where $\bm \Gamma$ is unrestricted, and we report the full set of responses for the $\bm \Gamma=\bm 0$ case in the appendix as a robustness exercise. In addition, given the large set of countries and variables under study, we focus on interesting subsets of countries and variables in the main body of this paper and refer the reader to the appendix for the full set of responses.
 
\begin{figure}[H]
\centering
\includegraphics[width=0.85\textwidth, trim={15cm 3cm 12cm 3cm}]{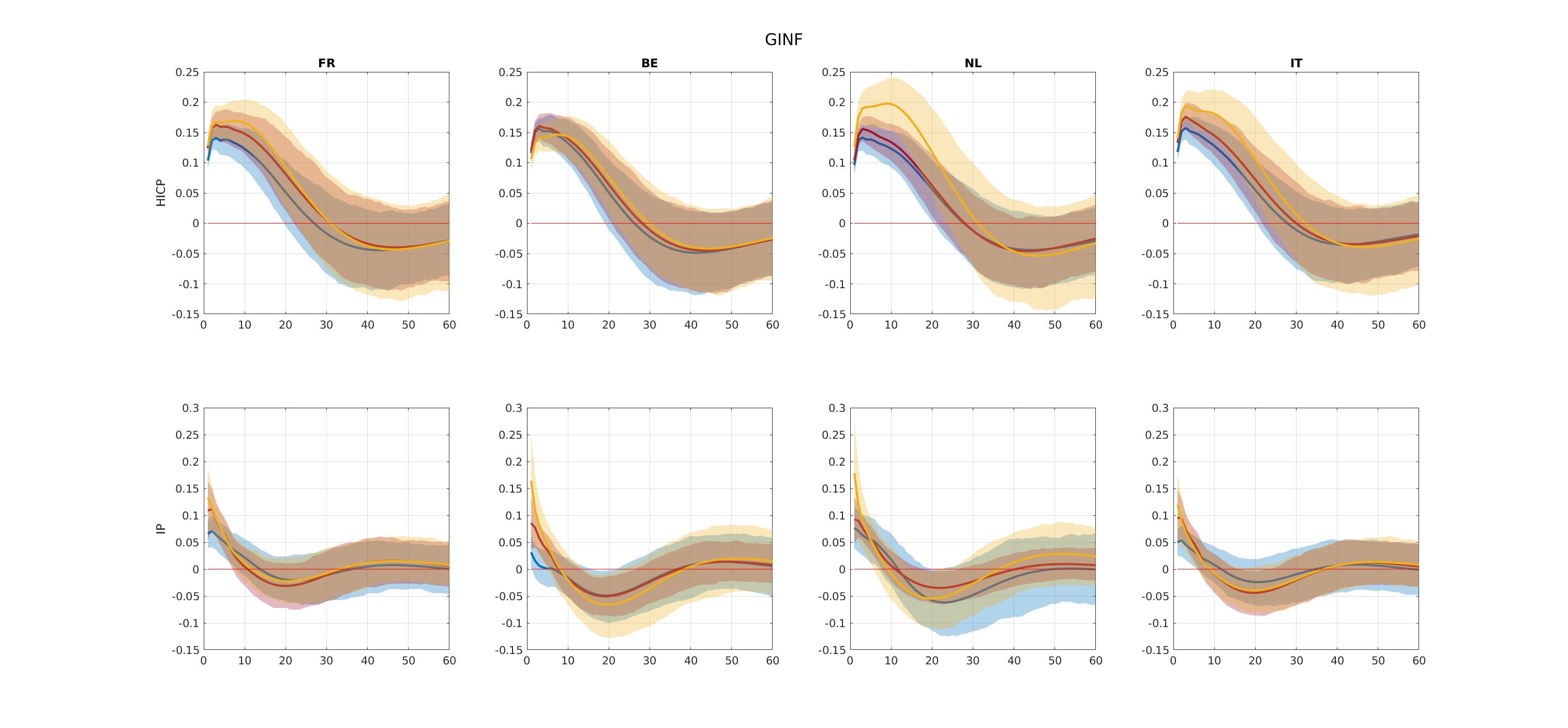}
\caption{Impulse response functions (median and 68\% bands) at the 10th, 50th, and 90th percentiles of country-level inflation (first row) and industrial production (second row). Shock is to the global inflation (GINF). Responses at the 10th percentiles are in blue, responses at the 50th percentiles are in red, and responses at the 90th percentiles are in yellow.} \label{fig:QFAVAR_IRF_meas_INF}
\end{figure}

\begin{figure}[H]
\centering
\includegraphics[width=0.85\textwidth, trim={15cm 3cm 12cm 3cm}]{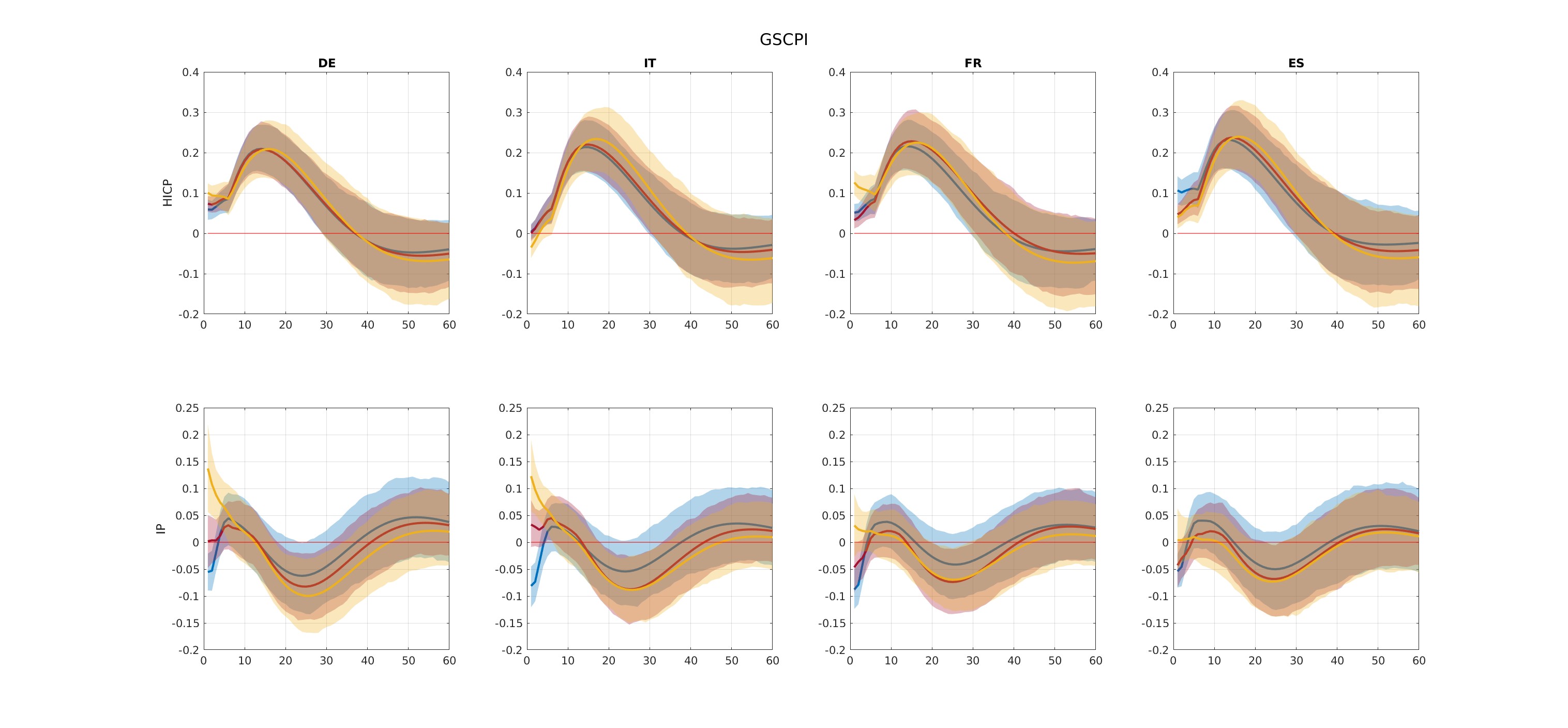}
\caption{Impulse response functions (median and 68\% bands) at 10th, 50th, and 90th percentiles of country-level inflation (first row) and industrial production (second row). Shock is to the global supply chain pressure index (GSCPI). Responses at the 10th percentiles are in blue, reponses at the 50th percentiles are in red, and responses at the 90th percentiles are in yellow.} \label{fig:QFAVAR_IRF_meas_GSCPI}
\end{figure}

\begin{figure}[H]
\centering
\includegraphics[width=0.8\textwidth, trim={7cm 3cm 7cm 3cm}]{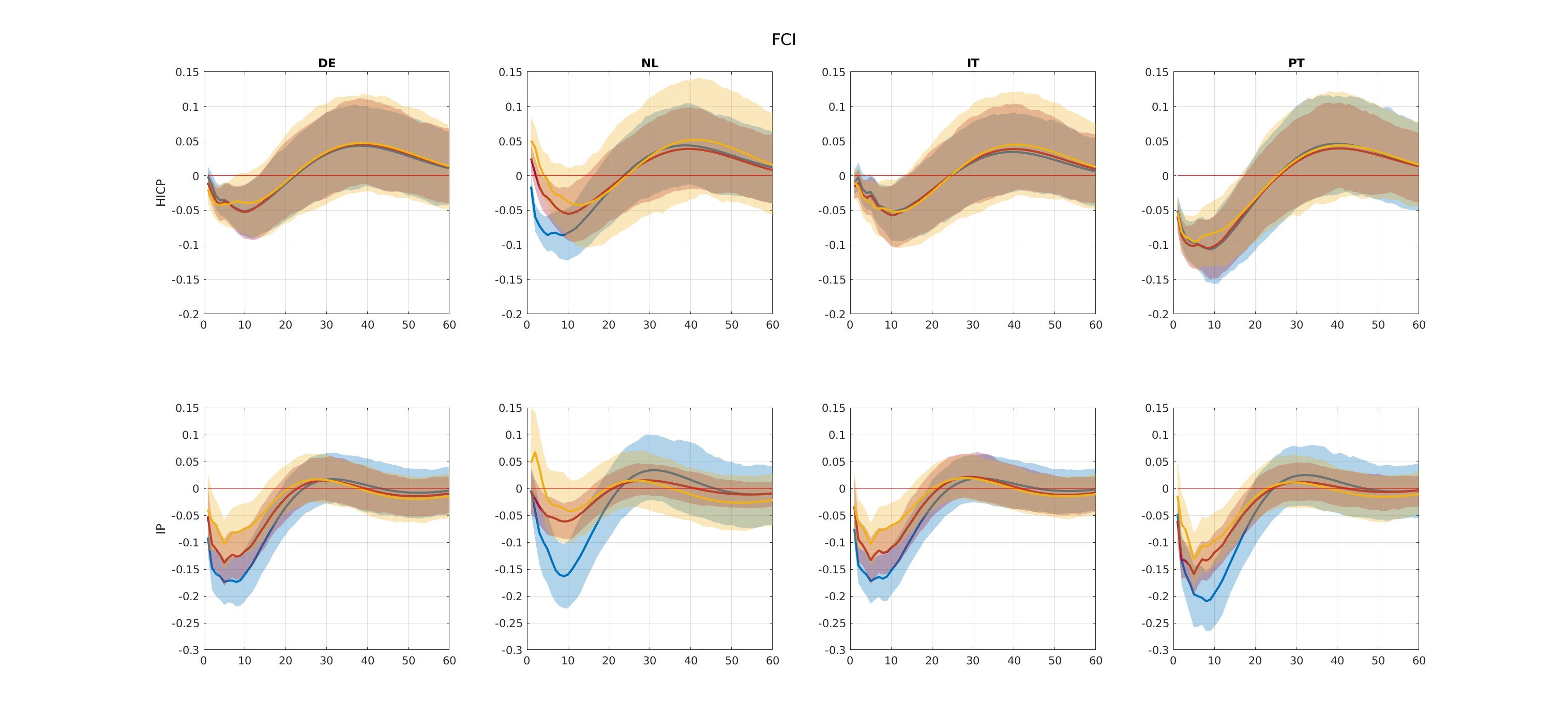}
\caption{Impulse response functions (median and 68\% bands) at 10th, 50th, and 90th percentiles of country-level inflation (first row) and industrial production (second row). Shock is to the financial conditions index (FCI). Responses at the 10th percentiles are in blue, reponses at the 50th percentiles are in red, and responses at the 90th percentiles are in yellow.} \label{fig:QFAVAR_IRF_meas_FCI}
\end{figure}

\begin{figure}[H]
\centering
\includegraphics[width=0.8\textwidth, trim={7cm 3cm 7cm 3cm}]{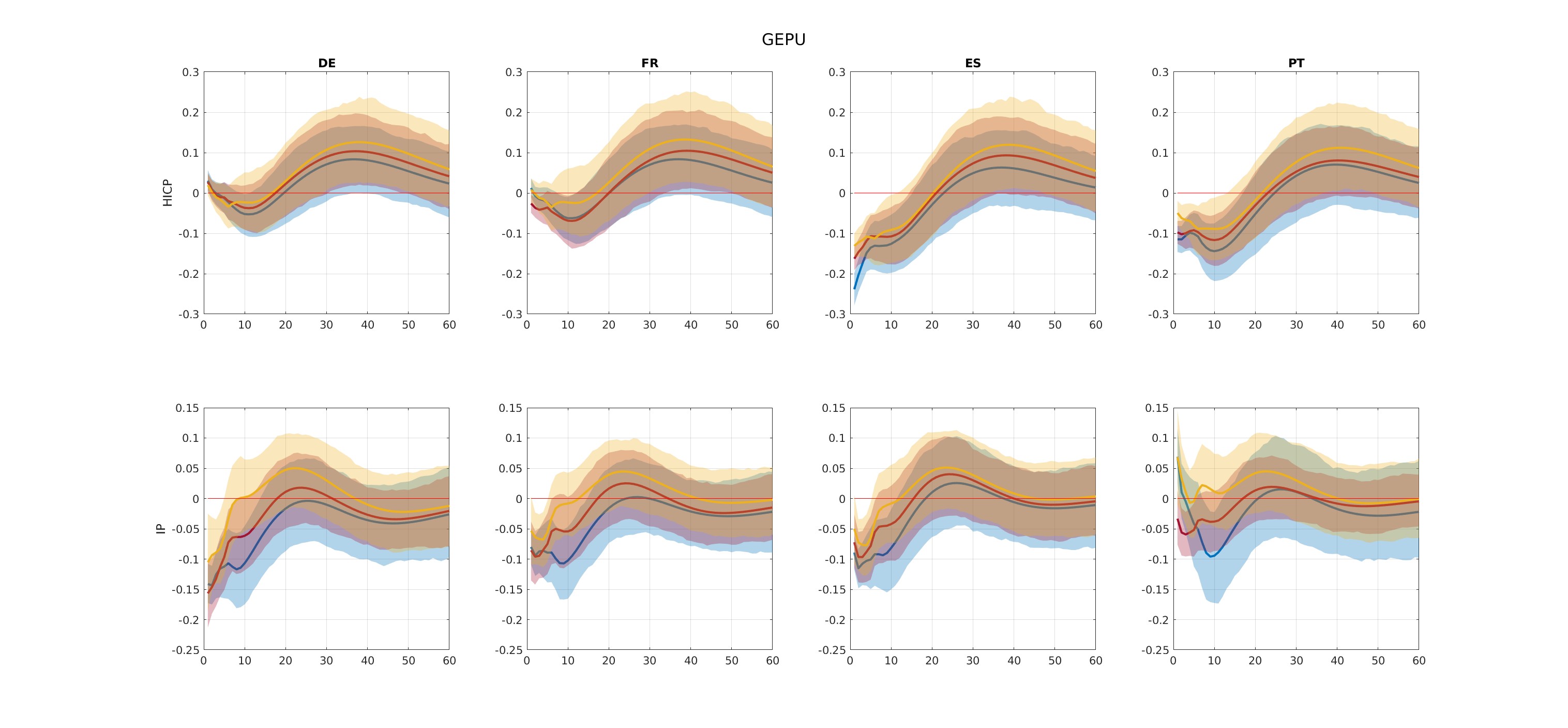}
\caption{Impulse response functions (median and 68\% bands) at 10th, 50th, and 90th percentiles of country-level inflation (first row) and industrial production (second row). Shock is to the global economic policy uncertainty (GEPU). Responses at the 10th percentiles are in blue, reponses at the 50th percentiles are in red, and responses at the 90th percentiles are in yellow.} \label{fig:QFAVAR_IRF_meas_GEPU}
\end{figure}

Figures \ref{fig:QFAVAR_IRF_meas_INF}, \ref{fig:QFAVAR_IRF_meas_GSCPI}, \ref{fig:QFAVAR_IRF_meas_FCI}, and \ref{fig:QFAVAR_IRF_meas_GEPU} present the generalized IRFs for HICP and IP following an increase in the four global variables. Starting with global inflation, we show the IRFs for France, which experienced relatively low inflation compared to the European average, Belgium, which had inflation rates close to the European average, and the Netherlands and Italy, which were among the countries with the highest inflation rates in the EA during the recent high-inflation period. A few interesting features stand out. In terms of dynamics, the inflation responses in France and Belgium look roughly similar; however, the inflation response in France is more leptokurtic and negatively skewed. Overall, the inflation outlook in France is more uncertain, but lower inflation realizations are relatively more likely compared to Belgium. The opposite is true for both, the Netherlands and Italy. Here the inflation distribution is characterized by a marked response of the upper tail and hence upside risk to the inflation outlook following an increase in global inflation. Although the median responses are comparable across both the low- and high-inflation countries, what sets them apart is pronounced heterogeneity in the tails of the distributions. Policy makers in the euro-area might hence want to pay particular attention to cross-country heterogeneity following global inflation shocks. For industrial production, the dynamics are rather similar across countries. In Belgium, the production outlook is more uncertain overall on impact, and in the Netherlands upsides risks are slightly more pronounced.

 
To showcase the effects of an increase in the global supply chain pressure indicator, we select Germany and Italy as examples of a core and periphery country with a large manufacturing share of overall industrial output and we select France and Spain as large core and periphery countries with a comparatively low manufacturing share. On impact, the inflation outlook becomes more uncertain in all countries but Italy and is positively skewed in France and Spain. In addition, although inflation picks up in Germany, France, and Spain on impact, in Italy it only does with delay. A different pictures emerges for industrial production. In all countries, the production outlook becomes uncertain on impact; however, this uncertainty is pervasive in the countries with large manufacturing share (Germany and Italy). 

Next, we turn to the IRFs in response to tightening global financial conditions. Here we focus on Germany and the Netherlands as examples of large and small core countries, and we focus on Italy and Portugal as examples of large and small periphery countries. On impact, inflation responds negatively in all countries. In Portugal, inflation responds more negatively overall, but the response in the Netherlands is characterized by greater uncertainty and a particularly marked response at the 10th percentile. Tail risks are hence more pronounced than in the other countries. Equally interesting are the responses of industrial production. On average the 90th percentile responds the least and the 10th percentile the most, indicating an overall increase in uncertainty in all countries. In Germany, Italy, and Portugal, the response of the median and 90th percentile are rather similar. In the Netherlands, the response is more muted and the 90th percentile responds positively on impact. The largest differences emerge for the response at the 10th percentile. The response is similar in Germany and Italy but more pronounced in Portugal and especially the Netherlands. In this stylized study, smaller countries are hence more susceptible to downside risks to the production outlook, following an increase in global financial conditions. More generally, these results are well in line with our findings from above as well as the literature that finds that financial conditions are an important driver of downside risks to output \citep{Adrianetal2019}.

Finally, we turn attention to global economic policy uncertainty. Here we show the responses of Germany, France, Spain, and Portugal. The inflation response of Germany and France are roughly similar and symmetric. Both countries only respond little on impact and the dynamics are generally more muted compared to the other two countries. What stands out is that the periphery countries (Portugal and Spain) respond negatively on impact. The responses of industrial production show greater heterogeneity across quantile levels. With the exception of Portugal, the on-impact responses are stronger from the 90th to the 10th percentile. As before, this is indicative of increasing uncertainty about the outlook, a feature hidden in a classical FAVAR. 

Our stylized analysis focuses on generalized impulse response functions and hence does not allow statements linked to specific structural shocks. Nonetheless, the country-level IRFs are evidence of strong heterogeneity, not only across quantile levels, but also across individual EA countries. Given the structure of the monetary union and the conduct of a common monetary policy, the QFAVAR offers policy-makers a tool to monitor not only these heterogeneities, but also risks to fragmentation of the broader transmission of economic shocks. Taking the scenario-analysis view based on the proposed QFAVAR might hence be a particularly useful addition to the policy analysis toolbox.


 

\subsubsection{Quantile connectedness}

To conclude the empirical section, we extend our FEVD analysis and derive quantile connectedness measures similar to \cite{DieboldYilmaz2014}. In a first step, we use the state space form of our model to map the FEVDs from the state equation to the panel of individual country-level variables. In a second step, we construct a variance decomposition matrix for each forecast horizon, 

\begin{equation}\label{fevdtab1}
D^h = \begin{bmatrix}
d_{1,1}^h & \dots & d_{1, l+k}^h \\
\vdots  & \ddots      & \vdots \\
d_{mnr,1}^h & \dots & d_{mnr, l+k}^h \\
\end{bmatrix},
\end{equation}

\noindent where element $d_{i,j}^h$ denotes the fraction of variable $i^{th}$ forecast error attributed to state variable $j$ at forecast horizon $h$. Alternatively, we relate the variance decomposition matrix, $D^h$, to a weighted network adjacency matrix, $A$. Compared to \cite{DieboldYilmaz2014}, however, the matrices $D^h$ and $A$ are not directly equivalent, because our state-space system has $l+k$ shocks and $n\times m \times r$ country-level variables. To represent the variance decomposition matrix as directed network, we stack the observed variables, quantile factors, and global variables, and we construct the augmented variance decomposition matrix,

\begin{equation}
\tilde{D}^h = \begin{bmatrix}
\bm 0 & D^h \\
\bm 0 & D^h_s
\end{bmatrix}= A,
\end{equation}

\noindent where $D_s^h$ denotes the variance decomposition matrix for the state equation that is constructed analogously to $D^h$. In line with \cite{DieboldYilmaz2014} we now define 

\begin{equation}
C_{i\leftarrow j}^h = \tilde{d}_{i,j}^h
\end{equation}

\noindent as the pairwise directional connectedness from $j$ to $i$. $D^h$ hence stores the directional connectedness from the global variables and quantile factors to the country-level variables, and $D_s^h$ contains the pairwise connectedness among the global variables and quantile factors. The $\bm 0$ entries are an artefact of the QFAVAR's structure and imply that there is no directional connectedness among the country-level variables or from the country-level variables to the global variables and quantile factors.\\

\begin{figure}[H]
\centering
\includegraphics[width=0.95\linewidth, trim={15cm 3cm 12cm 2.5cm}]{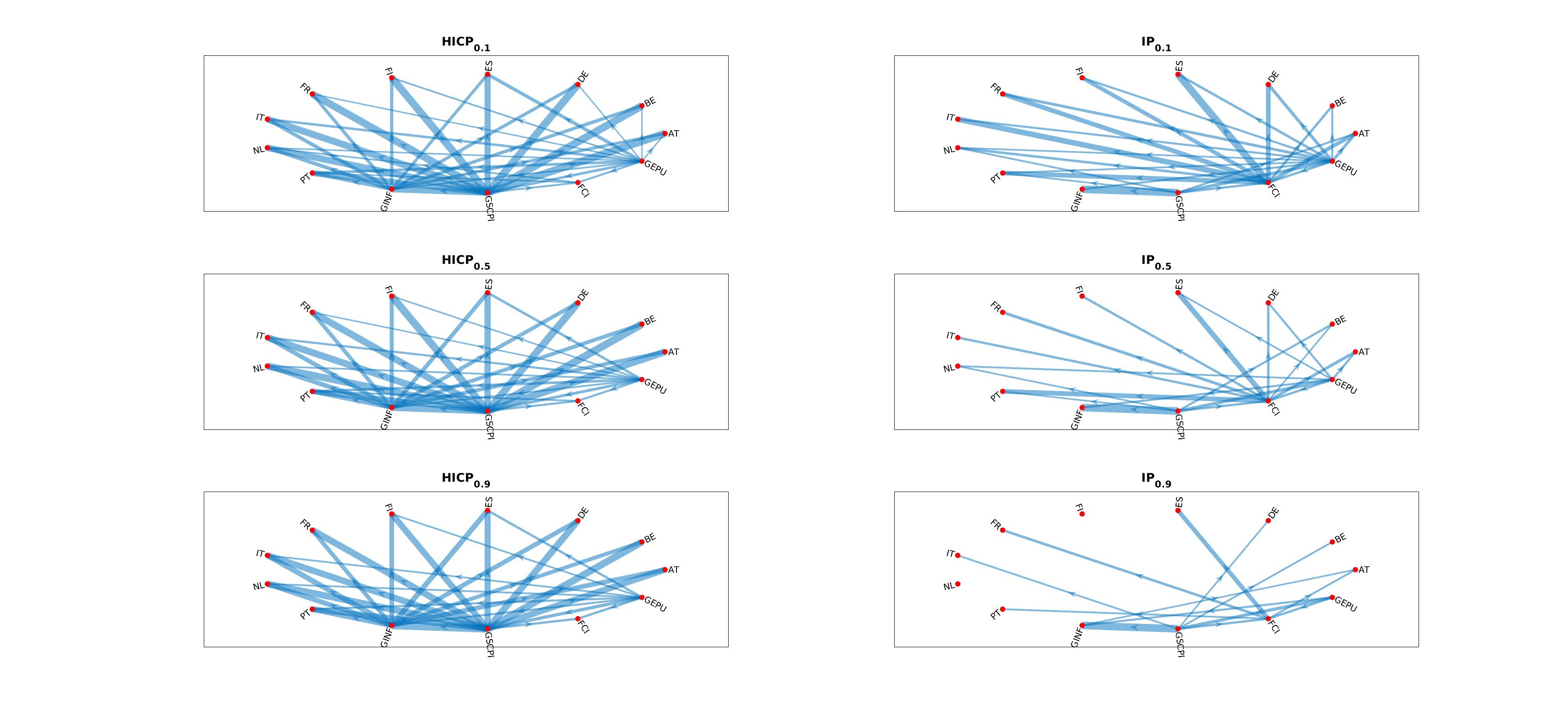}
\caption{Pairwise Directional Connectedness Networks.} \label{fig:QFAVAR_FEVD_network}
\end{figure}

To illustrate the QFAVAR connectedness, figure \ref{fig:QFAVAR_FEVD_network} displays the directed network graphs implied by $\tilde{D}^h$. Given the large dimensions of our model, in line with the previous sections we focus on interesting sub graphs. The left column of figure \ref{fig:QFAVAR_FEVD_network} contains the connectedness of country-level inflation with the global variables for the three quantile levels of interest, $q_r\in[0.1,0.5,0.9]$, for $h=24$. The right column contains the analogous directed graphs for country-level industrial production. In addition, we exclude edges for which the directional connectedness is smaller than 5\%. In general, the thicker the edge, the stronger the directional connectedness. A few features stand out immediately. For inflation, country-level inflation overall is strongly connected to the global variables. The global supply chain pressure index, and to a lesser extent global inflation, have the largest directional connectedness to country-level inflation for all quantile levels. Across quantile levels, the patterns are relatively similar; however, although global economic policy uncertainty is connected to inflation in all countries at the 10\% level, it only connects to a subset of countries at the 50\% and 90\% level. In contrast, for industrial production the pattern is very heterogeneous across quantile levels. At the 10\% level the FCI has the strongest connectedness, especially to Spain. Global economic policy uncertainty emerges as the second most connected global variable. At the 50\% level, the FCI connects to industrial production in all countries but the Netherlands. Global economic policy uncertainty only connects to industrial production in the Netherlands, Austria, and Germany. At the 90\% level the global FCI only connects to Spain, France, Austria, and Portugal, and the global supply chain pressure index now connects to industrial production in Italy, Germany, and Belgium. Overall, though the connectedness networks reflect the observations for the FEVD of the quantile factors in section \ref{sec:FEVD}, they allow us to observe previously hidden heterogeneity across EA countries.

\section{Conclusions}

We develop a new quantile factor augmented vector autoregressive (QFAVAR) model that is a natural extension of the popular FAVAR for targeting specific quantiles of the distribution of macroeconomic data. The advantage of the factor-based approach over quantile VAR modeling is that it is both flexible and parsimonious. The factors not only summarize cross-sectional correlations across different countries, but also across different quantiles of the data distribution. Using a Bayesian perspective, estimation of the QFAVAR adds only a minor level of complexity relative to a Markov chain Monte Carlo algorithm for the classical FAVAR model \citep{Bernankeetal2005}. As MCMC can be computationally demanding in high dimensions, we also develop a simple two-step variational Bayes estimator that is appropriate for the computational demands associated with recursive out-of-sample forecasting exercises.

The proposed QFAVAR is fully parametric (likelihood-based), meaning that our proposed specification can easily extend to incorporate other formulations without inducing huge estimation and setup costs. First, we can obtain quantile dynamic factor models and univariate quantile autoregressions as special cases of the QFAVAR simply by restricting certain parameters of the state-space form of the model. Second, we can incorporate interesting features such as time-varying parameters and stochastic volatility in the measurement and/or state equations of the model in order to allow for more flexible inference. Finally, although our approach to impulse response analysis follows a more neutral approach, by focusing on generalized impulse response functions, the QFAVAR can trivially be treated as a formal structural VAR model. We leave these extensions for future research.

\newpage
\begin{appendix}
\renewcommand{\theequation}{A.\arabic{equation}} \setcounter{equation}{0} %
\renewcommand{\thetable}{A\arabic{table}} \setcounter{table}{0}
\renewcommand{\thefigure}{A\arabic{figure}} \setcounter{figure}{0}
\setcounter{footnote}{0}
\section{Technical Appendix: Derivation of the linear state-space form}
Throughout the following, lower case letters indicate scalars, bold lower case letters indicate vectors, and bold upper case letters denote matrices. Let $y_{ij,t}$ denote a macroeconomic or financial indicator $i = 1,\dots,m$ for country $j=1,\dots,n$ that is observed for time $t=1,\dots,T$. Additionally, let $q\in(0,1)$ denote a given quantile level.  The quantile factor regression for variable $y_{ij,t}$ for quantile level $q$ is of the form
\begin{equation}\label{measij}
y_{ijt} = c_{ij(q)} + \lambda_{ij(q)} f^{i}_{t(q)} + \bm \gamma_{ij(q)} \bm g_t + u_{ij,t(q)}, \:\:\:\: u_{ij,t(q)} \sim AL(0,\sigma_{ij(q)},q)
\end{equation} 
Stacking all variables over $i,j$, $i=1,...,m$, $j=1,...,n$, into the vector $\bm y_{t} = \left[ y_{11t},...,y_{1nt},...,y_{m1t},...,y_{mnt} \right]^{\prime}$ we obtain the model
\begin{equation}
\bm y_{t} = \bm c_{q} + \bm \lambda_{(q)} \bm f_{t(q)} + \bm \gamma_{(q)} \bm g_{t} + \bm u_{t(q)}, \label{measperq}
\end{equation}
where $\bm c_{(q)} = \left[ c_{11(q)},...,c_{1n(q)},...,c_{m1(q)},...,c_{mn(q)} \right]^{\prime}$ is an $nm \times 1$ vector, $\bm \gamma_{(q)} = \left[ \bm \gamma_{11(q)}^{\prime},...,\bm \gamma_{1n(q)}^{\prime},...,\bm \gamma_{m1(q)}^{\prime},...,\bm \gamma_{mn(q)}^{\prime} \right]^{\prime}$ is an $nm \times k$ matrix, $\bm u_{t(q)} = \left[ u_{11t(q)},...,u_{1nt(q)},...,u_{m1t(q)},...,u_{mnt(q)} \right]^{\prime}$ is an $nm \times 1$ vector, 
\begin{equation*}
\bm \lambda_{(q)} = diag\left(\bm \lambda_{1(q)},\bm \lambda_{2(q)},...,\bm \lambda_{m(q)} \right) =  \left[ 
\begin{array}{cccc}
\left[ \begin{array}{c}
\lambda_{11(q)} \\
\vdots \\
\lambda_{1n(q)}
\end{array} \right] & \bm 0 & ... & \bm 0 \\
\bm 0   &  \left[ \begin{array}{c}
\lambda_{21(q)} \\
\vdots \\
\lambda_{2n(q)}
\end{array} \right] & \ddots & \vdots \\
   & & & \\
\vdots &  \ddots & \ddots & \bm 0 \\
   & & & \\
\bm 0  &  ... & \bm 0 & \left[ \begin{array}{c}
\lambda_{m1(q)} \\
\vdots \\
\lambda_{mn(q)}
\end{array} \right]
\end{array} \right]
\end{equation*}
and $\bm f_{t(q)}=   \left[
\begin{array}{c}
f_{t(q)}^{1} \\
f_{t(q)}^{2} \\
\vdots \\
f_{t(q)}^{m}
\end{array} \right]$.
Equation \eqref{measperq} shows the FAVAR measurement equation for each quantile level $q$. Stacking across $r$ quantiles $q=q_{1},...,q_{r}$ we obtain
\begin{equation}
\bm Y_{t} = \bm c +  \bm \Lambda \bm F_{t} + \bm \Gamma \bm g_{t} + \bm u_{t}, \label{measfinal}
\end{equation}
where $\bm c = \left[ \bm c_{(q_{1})}^{\prime},...,\bm c_{(q_{r})}^{\prime} \right]^{\prime}$ is an $(nmr \times 1)$ vector, $\bm \Gamma = \left[\bm \gamma_{(q_{1})}^{\prime},...,\bm \gamma_{(q_{r})}^{\prime} \right]^{\prime}$ is an $(nmr \times k)$ matrix, $\bm F_{t} = \left[\bm f_{t(q_{1})}^{\prime},...,\bm f_{t(q_{r})}^{\prime} \right]^{\prime}$ is an $(mr \times 1)$ vector and $\bm \Lambda = \left[\bm \lambda_{(q_{1})}^{\prime},...,\bm \lambda_{(q_{r})^{\prime}} \right]^{\prime}$ is an $(nmr \times mr)$ matrix.

Augmenting the measurement equation \eqref{measfinal} with an identity for $\bm g_{t}$ and combining it with the state equation \eqref{state} we obtain the state-space form of the QFAVAR
\begin{eqnarray}
\left[
\begin{array}{c}
\bm Y_{t} \\
\bm g_{t}
\end{array}
\right]   & = & \bm c + \left[
\begin{array}{cc}
\bm \Lambda & \bm \Gamma \\
\bm 0  &\bm I
\end{array}
\right]   
\left[
\begin{array}{c}
\bm F_{t} \\
\bm g_{t}
\end{array}
\right] + 
\left[
\begin{array}{c}
\bm u_{t} \\
\bm 0
\end{array}  
\right], \label{measurementApp} \\
\left[
\begin{array}{c}
\bm F_{t} \\
\bm g_{t}
\end{array}
\right] & = & \bm v +  \bm \Phi_{1} \left[
\begin{array}{c}
\bm F_{t-1} \\
\bm g_{t-1}
\end{array}
\right] + ... + \bm \Phi_{p} \left[
\begin{array}{c}
\bm F_{t-p} \\
\bm g_{t-p}
\end{array}
\right] + \bm \varepsilon_{t}. \label{stateApp}
\end{eqnarray}
\noindent Each element of the vector $\bm u_{t}$ is distributed as independent univariate asymmetric Laplace. As explained in the main text we can write the asymmetric Laplace as a Gaussian-Exponential location-scale mixture, in which case we the state-space model above is in conditionally normal form and sampling of the state-vector $\left[\bm F_{t}^{\prime}, \bm g_{t}^{\prime} \right]^{\prime}$ is feasible using the simulation smoother of \cite{CarterKohn1994}. Finally, note that sampling of the state-form requires first-order Markov dependence of the state variable. In equation \eqref{stateApp} above the state vector follows a VAR(p) but we can use standard tools for writing it in VAR(1) companion form, \cite[see][Chapter 2]{Helmut2005}. Detailed derivations of MCMC and variational Bayes  algorithms for inference, are provided in the online supplement.

\end{appendix}

\newpage

\clearpage
\bibliographystyle{apa}
\addcontentsline{toc}{section}{\refname}
\bibliography{QFAVAR}


\newpage

\setcounter{page}{1}

\begin{center}
{\Huge Online supplement to ``Monitoring multicountry macroeconomic risk''}\\
\bigskip
Dimitris Korobilis and Maximilian Schr\"{o}der
\end{center}

\setcounter{section}{0}
\setcounter{equation}{0}
\setcounter{table}{0}
\setcounter{figure}{0}
\setcounter{footnote}{0}
\section{Bayesian estimation of the Quantile FAVAR model}
\subsection{Derivation of the linear, Gaussian state-space form}

Throughout the following, lower case letters indicate scalars, bold lower case letters indicate vectors, and bold upper case letters denote matrices. Let $y_{ij,t}$ denote a macroeconomic or financial indicator $i = 1,\dots,m$ for country $j=1,\dots,n$ that is observed for time $t=1,\dots,T$. Additionally, let $q\in(0,1)$ denote a given quantile level.  The measurement equation of the QFAVAR for variable $y_{ij,t}$ for quantile level $q$ is then of the form

\begin{equation}\label{meas1app}
y_{ij,t} = c_{ij(q)} + \beta_{ij(q)}y_{ij,t-1}  + \bm\lambda_{ij(q)}' \bm f_{t(q)} + \bm \gamma_{ij(q)}' \bm g_t + u_{ij,t(q)}, \:\:\:\: u_{ij,t(q)} \sim AL(0,\sigma_{ij(q)},q)
\end{equation}
 
\noindent where $\bm g_t$ denotes the set of observed global variables, $\bm f_t(q)$ denotes the quantile factors, and $\bm\gamma_{ij(q)}$ and $\bm \lambda_{ij(q)}$ are conformable loading vectors. $c_ij(q)$ and $\beta_{ij(q)}$ denote a constant and an autoregressive coefficient, respectively. To ease notation, we will suppress $c_{ij(q)}$ and $\beta_{ij(q)}$ in the following expressions without loss of generality. Finally, $AL(0,\sigma_{ij(q)},q)$ denotes the univariate asymmetric Laplace density with the location parameter set to $0$, scale parameter $\sigma_{ij(q)}$, and asymmetry parameter $q$. The use of the univariate asymmetric Laplace later implies a diagonal covariance matrix in the measurement equation and thus mirrors the standard identifying assumption used in linear Gaussian factor models. This distribution is parametrized as

\begin{equation}
u_{ij,t(q)} \sim \frac{q(1-q)}{\sigma_{ij,t}} \left[e^{(1-q)\frac{u_{ij,t(q)}}{\sigma_{ij(q)}}} \mathbb{I}(u_{ij,t(q)} \leq 0) + e^{-q\frac{u_{ij,t(q)}}{\sigma_{ij(q)}}} \mathbb{I}(u_{ij,t(q)} > 0)\right].
\end{equation}

The first step towards tractable Bayesian inference is to rewrite the asymmetric Laplace likelihood as a conditionally Gaussian likelihood, which greatly simplifies inference. Following \cite{YuMoyeed2001} the AL distribution can equivalently be expressed as a normal-exponential mixture of the form 

\small\begin{equation}
u_{ij,t(q)}|z_{ij,t(q)} \sim \frac{1}{\sqrt{2 \pi z_{ij,t(q)} \sigma_{ij(q)}\kappa_{2(q)}^2}}exp \left \{  -\frac{ \left(y_{ij,t} -\bm \lambda_{ij(q)}'\bm f_{t(q)} -\bm \gamma_{ij(q)}' \bm g_t - \kappa_{1(q)}z_{ij,t(q)} \right)^2 }{2z_{ij,t(q)}\sigma_{ij(q)}\kappa_{2(q)}^2}  \right \} exp \left \{  - \frac{z_{ij,t(q)}}{\sigma_{ij(q)}} \right  \},
\end{equation}

\noindent or compactly $u_{ij,t(q)}|z_{ij,t(q)} \sim N(\kappa_1(q)z_{ij,t(q)},\kappa_{2(q)}^2 \sigma_{ij(q)} z_{ij,t(q)})$, with $z_{ijt,(q)} \sim Exp(\sigma_{ij(q)})$, where $Exp(\bullet)$ denotes the exponential distribution, $\kappa_{1(q)} = \frac{1-2q}{q(1-q)}$, and $\kappa_{2(q)}^2 = \frac{2}{q(1-q)}$. Using this scale mixture of normals representation, we can rewrite \ref{meas1app} as

\begin{equation}\label{meas2app}
y_{ij,t} =   \bm\lambda_{ij(q)}' \bm f_{t(q)} + \bm \gamma_{ij(q)}' \bm g_t + \kappa_{1(q)} z_{ij,t(q)} + \kappa_{2(q)}  \sqrt{\sigma_{ij(q)} z_{ij,t(q)}}\nu_{ij,t}, \: \: \: \: \nu_{ij,t} \sim N(0,1).
\end{equation}

\noindent For a given quantile level $q$, we can now collect the set of variables and stack their respective measurement equations across countries, $j$, and indicators, $i$

\begin{equation}\label{block1sApp}
\left[\begin{array}{c}
 y_{11,t} \\
 \vdots \\
 y_{1n,t} \\ \hdashline 
 y_{21,t} \\
 \vdots		\\
  y_{2n,t} \\ \hdashline
  \vdots \\
 y_{mn,t}  
\end{array} \right] =
\left[\begin{array}{cc}
\bm\lambda_{11(q)}' & \bm\gamma_{11(q)}' \\
\vdots & \vdots \\
\bm\lambda_{1n(q)}' & \bm\gamma_{1n(q)}' \\\hdashline
\bm\lambda_{21(q)}' & \bm\gamma_{21(q)}' \\
\vdots & \vdots\\
\bm\lambda_{2n(q)}' & \bm\gamma_{2n(q)}' \\ \hdashline
\vdots & \vdots\\
\bm\lambda_{mn(q)}' & \bm\gamma_{mn(q)}' 
\end{array} \right]
\left[\begin{array}{c}
\bm f_{t(q)} \\ 
  \bm g_{t}  
\end{array} \right]
\left[\begin{array}{c}
\tilde{z}_{11,t(q)} \\
\vdots \\
\tilde{z}_{1n,t(q)} \\\hdashline
\tilde{z}_{21,t(q)} \\
\vdots \\
\tilde{z}_{2n,t(q)} \\ \hdashline
\vdots \\
\tilde{z}_{mn,t(q)}
\end{array} \right] +
\left[\begin{array}{c}
\tilde{\nu}_{11,t(q)} \\
\vdots \\
\tilde{\nu}_{1n,t(q)} \\\hdashline
\tilde{\nu}_{21,t(q)}\\
\vdots \\
\tilde{\nu}_{2n,t(q)}\\ \hdashline
\vdots \\
\tilde{\nu}_{mn,t(q)}\\
\end{array} \right]
\end{equation}

\noindent where $\tilde{\nu}_{ij,t(q)}$ collects $\kappa_{2(q)}\sqrt{\sigma_{ij(q)} z_{ij,t(q)}}\nu_{ij,t}$ and $\tilde{z}_{ij,t(q)}$ denotes $\kappa_{1(q)}z_{ij,t(q)}$ to simplify notation. Importantly, different quantile levels maintain the same structure. To model multiple quantile levels simultaneously, we now stack the block of equations in \ref{block1sApp} for different quantile levels. For example, for three arbitrary quantile levels $q = \{q_1,q_2,q_3\}$, this yields

\small\begin{equation}\label{block2sApp}
\left[\begin{array}{c}
\bm y_{1\bullet,t} \\  
 \vdots		\\  
\bm  y_{m\bullet,t} \\ \hline
\bm y_{1\bullet,t} \\  
 \vdots		\\  
 \bm y_{m\bullet,t} \\ \hline
\bm y_{1\bullet,t} \\  
 \vdots		\\  
 \bm y_{m\bullet,t} \\ 
\end{array} \right] =
\left[\begin{array}{cccc}
 \bm\Lambda_{1\bullet,t(q_1)}' & \bm 0 &\bm 0 &  \gamma_{1\bullet,t(q_1)}'\\  
 \vdots	 & \vdots & \vdots	\\  
  \bm\Lambda_{m\bullet(q_1)}' &\bm 0 &\bm 0 &  \gamma_{m\bullet,t(q_1)}'\\ \hline
 \bm 0 & \bm\Lambda_{1\bullet(q_2)}'  &\bm 0 &  \gamma_{1\bullet,t(q_2)}'\\  
 \vdots	 & \vdots & \vdots	\\  
\bm 0 & \bm\Lambda_{m\bullet(q_2)}' &\bm 0 & \gamma_{m\bullet,t(q_2)}'\\ \hline
\bm 0 &\bm 0 &\bm \Lambda_{1\bullet(q_3)}' &  \gamma_{1\bullet,t(q_3)}'\\  
 \vdots	 & \vdots & \vdots		\\  
 \bm  0 &\bm 0 &\bm \Lambda_{m\bullet(q_3)}' &  \gamma_{m\bullet,t(q_3)}'\\ 
\end{array} \right]
 \left[\begin{array}{c}
 \bm f_{t(q_1)} \\ 
  \bm f_{t(q_2)} \\ 
   \bm f_{t(q_3)} \\
   \bm g_t
 \end{array} \right]
  + 
\left[\begin{array}{c}
 \bm\tilde{z}_{1\bullet,t(q_1)} \\  
 \vdots		\\  
 \bm \tilde{z}_{m\bullet,t(q_1)} \\ \hline
\bm \tilde{z}_{1\bullet,t(q_2)} \\  
 \vdots		\\  
\bm \tilde{z}_{m\bullet,t(q_2)} \\ \hline
\bm\tilde{z}_{1\bullet,t(q_3)} \\  
 \vdots		\\  
  \bm\tilde{z}_{m\bullet,t(q_3)} \\ 
\end{array} \right]  
   + 
\left[\begin{array}{c}
 \bm\tilde{\nu}_{1\bullet,t(q_1)} \\  
 \vdots		\\  
 \bm \tilde{\nu}_{m\bullet,t(q_1)} \\ \hline
\bm\tilde{\nu}_{1\bullet,t(q_2)} \\  
 \vdots		\\  
\bm \tilde{\nu}_{m\bullet,t(q_2)} \\ \hline
 \bm\tilde{\nu}_{1\bullet,t(q_3)} \\  
 \vdots		\\  
 \bm\tilde{\nu}_{m\bullet,t(v3)} \\ 
\end{array} \right]   
\end{equation}

\noindent Note that the LHS of equation \ref{block2sApp} repeatedly stacks the observed variables $y_{ij,t}$. A key feature of this formulation is that the conditional quantiles of each variable only load onto the quantile factors that are defined at the same quantile level. This induces sparsity in the loadings matrix and hence keeps the model parsimonious which in turn facilitates inference. 
\newline

With the measurement equations in place, we now turn to defining the state equation. The state equation generally follows a VAR(p) and describes the joint evolution of the quantile factors $\bm f_{t(q)}$ and the observed global variables $\bm g_t$. Without loss of generality, we restrict attention to a VAR(1) in order to ease notation.

\begin{equation}\label{statesApp1}
\left[\begin{array}{c}
 \bm f_{t(q_1)} \\  
\bm  f_{t(q_2)} \\
\bm  f_{t(q_3)} \\
\bm  g_{t} \\ 
\end{array} \right] = \bm v + \bm\Phi
\left[\begin{array}{c}
 \bm f_{t-1(q_1)} \\  
\bm  f_{t-1(q_2)} \\
\bm  f_{t-1(q_3)} \\
\bm  g_{t-1} \\ 
\end{array} \right] + \bm \varepsilon_t, \:\:\:\: \bm\varepsilon_t \sim N(\bm 0, \Omega)
\end{equation}

\noindent where $\bm \Phi$ denotes a conformable coefficient matrix, $\bm v$ is a vector of constants, and $\bm \varepsilon_t$ denotes the vector of reduced form residuals. While the variables in the measurement equation evolve independently conditional on the factors, the state equation allows for the factors at different quantile levels to co-move with one another and with the global variables. Overall, we hence model the co-movements of a potentially high dimensional quantile surface, by compressing them down to a lower dimensional space. This ensures parsimony and computational tractability. Fully cast into state space form, the system is given by

\begin{align}
\left[\begin{array}{c}
\bm y_t \\
\bm g_{t}
\end{array} \right]
&= \left[\begin{array}{cc}
 \bm\Lambda &  \bm\Gamma \\
 0          & \bm I
\end{array} \right] 
\left[\begin{array}{c}
\bm F_{t} \\
\bm g_{t}
\end{array} \right]
 + \bm\tilde{\nu}_t \\
\left[\begin{array}{c}
\bm F_{t} \\
\bm g_{t}
\end{array} \right] & = \bm \Phi
\left[\begin{array}{c}
\bm F_{t-1} \\
\bm g_{t-1}
\end{array} \right] + \bm\varepsilon_t, \:\:\:\: \bm\varepsilon_t \sim N(\bm 0, \bm \Omega)
\end{align}

\noindent where $\bm y_t = [y_{11,t(q_1)}',\dots,y_{mn,t(q_3)}']'$, $\bm F_t = [\bm f_{t(q_1)}',\dots, \bm f_{t(q_3)}']'$, $\bm \tilde{\nu}_t = [\tilde{\nu}_{11,t(q_1)}', \dots,\tilde{\nu}_{mn,t(q_r)}']' $, and $\bm\Lambda$ and $\bm\Gamma$ are conformable coefficient matrices collecting $\bm \lambda_{ij(q)}$ and $\bm\gamma_{ij(q)}$, respectively, as shown above. Further, $\bm \Omega$ denotes the covariance matrix of the reduced form residuals. The resulting QFAVAR is hence a linear Gaussian state-space model and standard filtering and smoothing algorithms apply. 
\newline

\noindent As one additional extension, in our application we introduce stochastic volatility to the state equation and hence allow the diagonal elements of $\Omega_t$ to be time-varying. Again, standard methods for estimating models with stochastic volatility apply so we leave the details to the appendix on estimation.

\subsection{Identification}
As is common in standard linear Gaussian factor models, while the common component, $\bm \Lambda \bm F_t$, is identified the loadings, $\bm \Lambda$, and factors, $\bm F_t$, are not identified individually. While forecasting exercises can still proceed even without identification, identification is necessary for structural analysis and the computation of IRFs. In our empirical exercise, we facilitate factor identification with three strategies.
\begin{enumerate}
\item First, we group the variables $y_{ij,t}$ by their economic interpretation and extract only one factor per block. The loadings matrix $\bm \Lambda_{(q)}$ collecting all loadings at a given quantile level is hence block-diagonal. Specifically, for $\bm f_{t(q)} = [f_{t(q)}^1, f_{t(q)}^2, \cdots, f_{t(q)}^i,\cdots, f_{t(q)}^m]'$ the corresponding loadings are given by

\begin{equation}
\bm\lambda_{ij(q)}' = [0, 0, \cdots, \lambda_{ij(q)}, \cdots, 0],
\end{equation}

and

\begin{equation}
\bm\Lambda_{i\bullet(q)}' = 
\left[\begin{array}{cccccc}
 0 & 0 & \cdots & \lambda_{i1(q)} & \cdots & 0 \\  
0 & 0 & \cdots & \lambda_{i2(q)}& \cdots & 0  \\
\vdots & \vdots &  & \vdots & & \vdots   \\
0 & 0 & \cdots & \lambda_{in(q)} & \cdots& 0  \\
\end{array} \right] 
\end{equation}

\noindent For two reasons, this has the additional benefit of giving the factors a specific interpretation. First, for e.g. all industrial production series, we will identify one industrial factor per quantile level. A similar strategy is already suggested in the original FAVAR framework proposed by \cite{Bernankeetal2005}. Second, grouping series that move similarly and are conceptually related, such as industrial production from different EA countries, ensures that the quantile factors can be assigned to the corresponding quantile levels and maintain their economic interpretation. Grouping e.g. industrial production and unemployment instead would mix across different economic concepts as the resulting e.g. $10^{th}$ percentile factor would be extracted from low unemployment, but also low industrial production. This greatly complicates interpretation. Following our strategy instead, we interpret the factors as e.g. the quantile industrial production factors.  \\
Note, however, that this choice also has clear implications for the dynamics of the model. In the case of $\bm\Gamma =\bm 0$, $\bm \beta =\bm 0$, and $\bm c =\bm 0$, i.e. the absence of global variables and constants in the measurement equation, the conditional quantiles of $\bm y_t$, $\bm y_t(q)$, are governed by the common component, $\bm\Lambda \bm f_t$, only. Given that we allow for only one factor per group of variables, this implies that the conditional quantiles $y_{ijt(q)}$ are just rescaled versions of the factors, $\bm f_t$. The conditional quantiles of a given variables for different countries, $\bm y_{i\bullet t(q)}$ then also feature identical dynamics and only differ in their scale. In order to highlight more meaningful heterogeneities, it is hence important to allow for additional RHS variables in the measurement equation, such as global variables.  
\item To fix the scale of the quantile factors within their groups, we further restrict one loading per quantile to unity. This yields e.g.

\begin{equation}
\bm\Lambda_{i\bullet(q)}' = 
\left[\begin{array}{cccccc}
 0 & 0 & \cdots & 1 & \cdots & 0 \\  
0 & 0 & \cdots & \lambda_{i2(q)}& \cdots & 0  \\
\vdots & \vdots &  & \vdots & & \vdots   \\
0 & 0 & \cdots & \lambda_{in(q)} & \cdots& 0  \\
\end{array} \right] 
\end{equation}

\item To identify the sign of the quantile factors, we extract quantile factors using the probabilistic quantile factor algorithm proposed by \cite{KorobilisSchroeder2022} from the same set of variables prior to estimation. Each iteration of the algorithm, we then check whether the QFAVAR factors are positively correlated with the corresponding VBQFA factors and invert them if this is not the case.

\end{enumerate}

\subsection{Markov Chain Monte Carlo estimation}
The Gibbs sampler requires deriving conditional posteriors. As outlined above, we formulated the QFAVAR as a linear Gaussian state space model and hence standard practices apply. 

\subsubsection{Prior Distributions}
In our proposed approach, we specify the following prior distributions

\begin{align}
\sigma_{ij(q)} & \sim G^{-1}(r_0,s_0) \\
z_{ij,t(q)} & \sim Exp(\sigma_{ij(q)})
\end{align} 

\noindent To simplify notation, let $\bm\Phi$ collect the coefficient vectors and matrices in the measurement equations, i.e. $\bm \Lambda$ and $\bm \Gamma$. Correspondingly, let $\hat{\bm F_t} = [\bm F_t', \bm g_t']$. As outlined in the main body of the text, we impose shrinkage on $\bm\Phi$ by means of the horseshoe prior following \cite{MakalicSchmidt2016}, given by 

\begin{align}
\bm\phi_{ij(q)}| \left\{\bar{\lambda}_{ij(q),k,\phi}, \upsilon_{ij(q),k,\phi}\right \}_{k=1}^{l_\phi}, \tau_{ij(q),\phi}, \xi_{ij(q),\phi}  & \sim  N(\bm 0,\sigma_{ij(q)} \tau^2_{ij(q),\phi} \bar{\bm\Lambda}_{ij(q),\phi}), \\
\bar{\lambda}^{2}_{ij(q),k,\phi} | \upsilon_{ij(q),k,\phi} & \sim G^{-1}\left(\frac{1}{2},\frac{1}{\upsilon_{ij(q),k,\phi}} \right), \:\:\: \text{for } k = 1,\cdots,l_\phi,\\
\upsilon_{ij(q),k,\phi} & \sim G^{-1}\left(\frac{1}{2},1\right), \:\:\: \text{for } k = 1,\cdots,l_\phi,\\
\tau^2_{ij(q),\phi} | \xi_{ij(q),\phi} & \sim G^{-1}\left(\frac{1}{2},\frac{1}{\xi_{ij(q),\phi}}\right),\\
\xi_{ij(q),\phi} & \sim G^{-1}\left(\frac{1}{2},1\right),
\end{align} 
  
\noindent where $\bar{\Lambda}_{ij(q),\phi} = diag(\bar{\lambda}_{ij(q),1,\phi}^2,\cdots,\bar{\lambda}_{ij(q),l,\phi}^2)$, subscript $k$ denotes the element of $\bm \phi_{ij(q)}$ for $k = 1, \cdots, l_\phi$, where $l_\phi$ is equal to the number of factors, $n_f$, number of global variables, $n_g$, plus the constant. Further, subscript $\phi$ indicates that the parameters of the shrinkage prior apply to the elements of the measurement equation. Similarly, because the VAR in the state equation \ref{statesApp1} is heavily parametrized, we define another set of horseshoe priors to the elements of $\Psi$.

\begin{align}
\bm\psi_{r}| \left\{\bar{\lambda}_{s,\psi}, \upsilon_{s,\psi}\right \}_{s=1}^{l_\psi}, \tau_{r,\psi}, \xi_{r,\psi}  & \sim  N(\bm 0,\omega_{r} \tau^2_{r,\psi} \bar{\bm\Lambda}_{r,\psi}), \\
\bar{\lambda}^{2}_{r,s,\psi} | \upsilon_{r,s,\psi} & \sim G^{-1}\left(\frac{1}{2},\frac{1}{\upsilon_{r,s,\psi}} \right), \:\:\: \text{for } s = 1,\cdots,l_\psi,\\
\upsilon_{r,s,\psi} & \sim G^{-1}\left(\frac{1}{2},1\right), \:\:\: \text{for } s = 1,\cdots,l_\psi,\\
\tau^2_{r,\psi} | \xi_{r,\psi} & \sim G^{-1}\left(\frac{1}{2},\frac{1}{\xi_{r,\psi}}\right),\\
\xi_{r,\psi} & \sim G^{-1}\left(\frac{1}{2},1\right),
\end{align} 

\noindent where $r = 1,\cdots,R$, with $R=n_f+n_g$ indexes the state equation, $s = 1,\cdots,l_\psi$ indexes the individual parameter in the respective state equation and $l_\psi = (n_f+n_g)\times p$, where $p$ denotes the number of lags, and subscript $\psi$ indicates that the prior hyperparameters belong to the state equation.  
\newline

To be able to estimate the QFAVAR with and without stochastic volatility in the state equation, we take the following approach to sample the elements of the covariance matrix. Let

\begin{equation}
\Omega_t = A H_tA'
\end{equation}

\noindent with $H_t = diag(e^{h_{1,t}},\cdots,e^{h_{R,t}})$. Further, $A$ is a lower unitriangular coefficient matrix. Sampling the elements of $A$ this way has a drawback. In principle, the triangular structure implies that the order of the states in the state equation matters. In a standard VAR setting, the likelihood quickly dominates the prior for the covariance matrix, such that inference likely remains unaffected. In our setting, it is unclear how much information the data contains about the quantile factors. Depending on the exercise inference might hence suffer. To sample the coefficients in $A$, we impose a normal prior. 

\begin{align}
\bm a_{r} \sim N(\bm \mu_{r,a}, \bm\Sigma_{r,a})
\end{align} 

\noindent with $r$ denoting the columns of $A$. 
What remains is to define the prior for $h_{r,t}$. In the case without stochastic volatility, we set

\begin{equation}
\log(h_{r,t}) = \log(h_{r}) = G^{-1}(r_h,s_h)
\end{equation}

\noindent with stochastic volatility, we introduce the state equation

\begin{equation}
\log(h_{r,t}) = \log(h_{r,t-1}) + v_t,\:\:\:\: v_t\sim N(0,\sigma^2_\omega)
\end{equation}
\noindent with $h_{r,1}\sim N(0,V_0)$ and prior $\sigma_\omega^2 \sim IG(r_\omega,s_\omega)$. 
\subsubsection{Estimation Algorithm}

Conditional on all parameter matrices being known (e.g. $\bm \Lambda$, $\bm S$, $\bm \Phi$), sampling of the state vector $\left[\bm F_{t}^{\prime}, \bm g_{t}^{\prime} \right]^{\prime}$ can be obtained using a simulation smoother such as the one proposed by \cite{CarterKohn1994}. Conditional on these states, parameters can be obtained by using standard methodologies for linear and quantile regression models. First, we need to obtain the parameters in the measurement equation. As mentioned above, the univariate asymmetric Laplace distributions of the individual measurement equations imply a diagonal covariance matrix. Conditioning on the factors $f^i_{t(q)}$ and treating these as observed, the individual measurement equations for $y_{ijt}$ are independent. They can hence be treated as $m \times n \times q$ individual univariate quantile regressions. We can thus obtain samples from all parameters of equation \eqref{meas1app} using standard formulas provided in \cite{Khare2012}. Next, we obtain samples from the parameters of the state equation using standard formulas for Bayesian vector autoregressions. Indeed, once we condition on $\left[\bm F_{t}^{\prime}, \bm g_{t}^{\prime} \right]^{\prime}$, equation \eqref{statesApp1} becomes a VAR and the conditional posteriors of $\bm v$, $\bm \Phi$, $\bm \Omega$ can be derived in a straightforward way \citep[see][for a thorough examination of Bayesian inference in VARs]{KoopKorobilis2010}.


All sampling steps of the MCMC algorithm are provided in \autoref{alg:cap}.

\RestyleAlgo{ruled} 
\begin{algorithm}
\scriptsize
\caption{MCMC algorithm for the estimation of the QFAVAR}\label{alg:cap}
\Begin
{
Define $ \kappa_{1(q)} = \frac{1-2q}{q(1-q)}$ and $ \kappa_{2(q)}^2 = \frac{2}{q(1-q)}$ for each quantile level $q$.\\
$[\bm 1]$  Sample the parameters in the measurement equation leveraging results from Bayesian quantile regression:\\
\For {$q=1:nq$}
{
\For {$i=1:m$}
{
\For {$j=1:n$}
{
$[\text{i}]$ Sample $\bm \phi_{ij(q)}$ from the full conditional $(\bm \phi_{ij(q)}|-) \sim N(\bm \mu^\phi_{ij(q)},\bm \Sigma_{ij(q)}^\phi)$, with $\bm\mu^\phi_{ij(q)} = \bm \Sigma_{ij(q)}^\phi \left \{ \hat{\bm F}'diag(\tilde{\bm\nu}_{ij(q)}^{-1})\tilde{\bm y}_{ij(q)} \right \}$, $\bm \Sigma_{ij(q)}^\phi = \left \{   \hat{\bm F}' diag(\tilde{\bm\nu}_{ij(q)})^{-1}\hat{\bm F}  +   \tilde{\bm \Lambda}_{ij(q)}^{-1}    \right \}^{-1}$, \\
where $\tilde{\bm\nu}_{ij(q)} = \sigma_{ij(q)}\kappa_{2(q)}^2 \bm z_{ij,\bullet(q)}$, $\tilde{\bm y}_{ij(q)}= \bm y_{ij,\bullet(q)}-\kappa_{1(q)}\bm z_{ij,\bullet(q)}$, and $\tilde{\bm \Lambda}_{ij(q)}^{-1} = diag(\bar{\bm\lambda}_{ij(q),\bullet,\phi}^2 \tau_{ij(q),\phi}^2)^{-1}$. \\

$[\text{ii}]$ Sample the parameters corresponding to the horseshoe prior from $\left(\bm \bar{\lambda}_{ij(q),k,\phi}^2|-\right) \sim IG\left(1,\frac{\bm\phi_{ij,k(q)}^2}{2\tau_{ij(q),\phi}^2} + \frac{1}{\upsilon_{ij(q),k,\phi}}\right)$, 
$\left({\upsilon}_{ij(q),k,\phi}|-\right) \sim IG\left(1,1 + 1/(  \bar{\bm\lambda}_{ij(q),k,\phi}^2) \right)$,
$\left(\tau_{ij(q),\phi}^2|-\right) \sim IG\left(\frac{l_\phi+1}{2}, \frac{1}{\xi_{ij(q),\phi}} + \sum_k^{l_\phi}\frac{\bm\phi_{ij(q)}^2}{2\bar{\bm\lambda}_{ij(q),k,\phi}^2} \right)$,
$\left({\xi}_{ij(q),\phi}|-\right) \sim IG\left(1 , 1 + \frac{1}{\tau_{ij(q),\phi}^2} \right)$, \\
for $k = 1,\cdots,l_\phi$. 

$[\text{iii}]$ Sample the latent quantile variables from  $\left(z_{ij,t(q)}|-\right) \sim GIG\left(    \chi_z, \psi_z\right)$, with $\chi_z = \frac{ \left \{ y_{ij,t(q)} - \hat{ \bm\phi}_{ij(q)}'\bm F_t \right\}^2 }{\kappa_{1(q)}^2 + 2\kappa_{2(q)}^2}$, $\psi_z = \frac{\kappa_{1(q)}^2 +  2 \kappa_{2(q)}^2}{\sigma_{ij(q)}\kappa_{2(q)}^2}$\\
for all $t = 1,\cdots,T$.

$[\text{iv}]$ Sample the factor regression variances $\left(\sigma_{ij(q)}|-\right) \sim IG\left( r_\sigma, s_\sigma \right)$, where $r_\sigma = r_0 + 3T/2$ and $s_\sigma = s_0 + \sum_{t=1}^T\frac{ \left \{y_{ij,t(q)} - \hat{ \bm\phi}_{ij(q)}'\bm F_t - \kappa_{1(q)}^2 z_{ij,t(q)} \right \}^2 }{2 z_{ij,t(q)} \kappa_{2(q)}^2}$

}
}
}
$[\text{2}]$ Sample the quantile factors $F_t$. Define $\tilde{\bm y}_{ij(q)}= \bm y_{ij,\bullet(q)}-\kappa_{1(q)}\bm z_{ij,\bullet(q)}$ and stack the loadings according to \ref{block2sApp}. Sample $F_t$ using the Kalman filter \& smoother. \\
$[\text{3}]$ Sample the diagonal elements, $h_{r,t}$ of covariance matrix $\Omega_t$ in state equation \ref{statesApp1}.\\
\eIf{$\text{stochastic volatility}$}
{
    set $y_{r,t}^* = log\left \{ \left( \bm F_{r,t} - \bm F_{t-1}\bm\Psi_{r}' \right)^2 \right \}$ and sample $log (h_{r,t})$ using the \cite{chan2013moving} filter. And update $\sigma_{t,\omega}^2$ from $(\sigma_{t,\omega}^2|-) \sim IG \left( r_\omega + 0.5(T-p-1), s_\omega + 0.5\cdot \sum_{t=1}^T (h_{r,t}-h_{r,t-1})^2    \right)$ \\
for every $r=1,\cdots,R$.
}{
Sample from $(\sigma_{t,\omega}^2 = \sigma_\omega^2|-) \sim IG \left( r_h + 0.5\cdot(T-p), s_h + 0.5\cdot \sum_{t=1}^T (\bm F_{r,t} - \bm F_{t-1}\bm\Psi_{r}')^2 \right)$ \\
for all $r=1,\cdots,R$.    
}
$[\text{4}]$ Sample the off-diagonal elements of $\bm\Omega_t$, $\bm A$, and the VAR coefficients $\bm\Psi$\\
\For {$r=1:R$}
{
$[\text{i}]$ Sample $\bm\psi_r$ and $\bm a_r$ jointly from the full conditional $(\bm \psi_{r}, \bm a_r|-) \sim N(\bm \mu^\psi_{r},\bm \Sigma_{r}^\psi)$, with $\bm\mu^\psi_{r} = \bm \Sigma_{r}^\psi \left \{ \bm X'diag(\sigma_{t,\omega}^{2})^{-1}\hat{\bm f}_{r} \right \}$, $\bm \Sigma_{r}^\psi = \left \{   \bm X' diag(\sigma_{t,\omega}^{2})^{-1}\bm X  +   \tilde{\bm \Lambda}_{r}^{-1}    \right \}^{-1}$,  where\\
$\bm X = [ L \bm F', L \bm G', \bm  E_{1:r-1}']'$, $\bm e_{r} = \hat{\bm f}_{r} - \bm \hat{\bm F} \bm\psi_r' - \bm  E_{1:r-1}' \bm a_r'$, and $\tilde{\bm \Lambda}_{r}^{-1} = diag\left(  [(\bar{\bm\lambda}_{r,\bullet,\psi}^2,' \tau_{r,\psi}^2)^{-1}, \bm \mu_{r,a}/\bm \Sigma_{r,a} ]  
\right)$. \\

$[\text{ii}]$ Sample the parameters of the horseshoe prior for $\psi_r$ from $\left(\bm \bar{\lambda}_{r,s,\psi}^2|-\right) \sim IG\left(1,\frac{\bm\psi_{r,s}^2}{2\tau_{r,\psi}^2} + \frac{1}{\upsilon_{r,s,\psi}}\right)$, 
$\left({\upsilon}_{r,s,\psi}|-\right) \sim IG\left(1,1 + 1/(  \bar{\bm\lambda}_{r,s,\psi}^2) \right)$,
$\left(\tau_{r,\psi}^2|-\right) \sim IG\left(\frac{l_\psi+1}{2}, \frac{1}{\xi_{r,\psi}} + \sum_k^{l_\psi}\frac{\bm\psi_{r}^2}{2\bar{\bm\lambda}_{r,s,\psi}^2} \right)$,
$\left({\xi}_{r,\psi}|-\right) \sim IG\left(1 , 1 + \frac{1}{\tau_{r,\psi}^2} \right)$, \\
for $s = 1,\cdots,l_\psi$. 
}
}
Note: We use the rate parametrization of the $IG$ distribution throughout. We use $L$ to denote the lag operator.
\end{algorithm}


\subsection{Variational Bayes estimation}

We also propose a 2-step variational QFAVAR algorithm which is appropriate for forecasting or other computationally cumbersome applications of the model. For a general introduction to variational Bayes (VB) see \cite{Bleietal2017}. Generally, variational inference and MCMC both provide approximations to a given posterior distribution. While MCMC provides approximations through sampling, VB approximates the objective by solving an optimization problem.While in practice both algorithms end up being iterative, VB is computationally less intensive than MCMC due to the fact that it requires significantly less iterations. As a caveat, while MCMC approximates the full posterior distribution and its uncertainty, VB only provides the posterior mean and (a possibly biased estimate of the) posterior variance. Therefore, VB is particularly useful for tasks where precise inference is less of a concern, such as big data applications, large scale forecasting exercises, or real-time monitoring tasks. \newline

Our algorithm evolves as follows. We first extract the quantile factors, $F_t$, using the VBQFA algorithm proposed in \cite{KorobilisSchroeder2022}. These are the same factors that we use to identify the sign of the factors in our proposed MCMC algorithm. Conditional on these factors, we then update the loadings and the remaining parameters in measurement equation \ref{meas1app}. Finally, we update the parameters of the VAR in state equation \ref{statesApp1} and generate forecasts. Treating the quantile factors estimated in the first stage as observed, the model parameters and latent variables are given by $\bm \theta = (\bm \Phi, \bm \sigma, \bm z, \bm \Psi, \bm \Omega , \bm \tau^2_\phi,\bm \xi_\phi, \bar{\bm \lambda}^2_\phi, \bm \upsilon_\phi,\bm \tau^2_\psi,\bm \xi_\psi, \bar{\bm \lambda}^2_\psi, \bm \upsilon_\psi)$. For a family of tractable densities $q(\bm\theta)$, we aim to find a density $q^\star$ that best approximates the posterior $p(\bm \theta| \bm x)$ by minimizing

\begin{equation}
q^\star(\bm \theta | \bm x) = \underset{q\in \mathcal{Q}}{\text{argmin}} \: \mathbb{D}_{KL} \left( q(\bm \theta || p(\bm \theta |\bm x)\right),
\end{equation}

\noindent which is equivalent to maximizing

\begin{equation}
ELBO = \mathbb{E}_{q(\bm \theta | \bm x)}\left[ \text{log} \: p(\bm x|\bm \theta)\right] + \mathbb{E}_{q(\bm \theta | \bm x)}\left[ \text{log} \: p(\bm \theta)\right] - \mathbb{E}_{q(\bm \theta | \bm x)}\left[ \text{log} \: q(\bm \theta| \bm x)\right],
\end{equation}

\noindent where $KL$ denotes the Kullback-Leibler divergence. Note that we need to optimize over a family of distribution functions. Finding the solution to the problem hence requires the application of variational calculus. Usually, optimization can be simplified by factorizing the variational posterior into $L$ groups of independent densities. In our proposed two-step estimator, we apply the following factorization

\begin{align}\label{VBpostAPP}
\hspace{-1.5cm} q(\bm \theta | \bm x)  & \equiv q\left( \bm \Phi, \bm \sigma, \bm z, \bm \Psi, \bm \Omega , \bm \tau^2_\phi,\bm \xi_\phi, \bar{\bm \lambda}^2_\phi, \bm \upsilon_\phi,\bm \tau^2_\psi,\bm \xi_\psi, \bar{\bm \lambda}^2_\psi, \bm \upsilon_\psi \right) \\
               & = \prod_{q=1}^{nq} \prod_{j=1}^{m} \prod_{i=1}^{n} \left[ q(\bm \phi_{ij(q)} | \bm x) q(\sigma_{ij(q)} | \bm x) q(\tau_{ij(q),\phi}^2 | \bm x) q(\xi_{ij(q),\phi} | \bm x) \prod_{k=1}^{l_\phi}   q(\bar{\lambda}_{ij(q),k,\phi}^2 | \bm x)  q(\upsilon_{ij(q),k),\phi} | \bm x) \prod_{t=1}^{T} q(z_{ij,t(q)} | \bm x)  \right]\nonumber \\ 
               & \:\:\:\:\:\: \cdot  \prod_{r=1}^{R} \left[ q(\bm \psi_{r},\bm a_{r} | \bm x) q(\tau_{r,\psi}^2 | \bm x) q(\xi_{r,\psi} | \bm x) q(\omega_{r} | \bm x) \prod_{s=1}^{l_\psi} q(\bar{\lambda}_{r,s,\psi}^2 | \bm x)  q(\upsilon_{r,s,\psi} | \bm x)   \right],   \nonumber
\end{align}

\noindent which implies partial posterior independence between the regression parameters and residual variances in the individual measurement equations, the VAR parameters and covariance matrix elements in the state equations, as well as the hyperparameters of the horseshoe priors. With the additional assumption that the priors are conditionally independent, we can write the joint prior as

\begin{align}\label{VBpriorAPP}
\hspace{-1.5cm} p(\bm \theta) &= p\left( \bm \Phi, \bm \sigma, \bm z, \bm \Psi, \bm \Omega , \bm \tau^2_\phi,\bm \xi_\phi, \bar{\bm \lambda}^2_\phi, \bm \upsilon_\phi,\bm \tau^2_\psi,\bm \xi_\psi, \bar{\bm \lambda}^2_\psi, \bm \upsilon_\psi \right) \\
& = \prod_{q=1}^{nq} \prod_{j=1}^{m} \prod_{i=1}^{n} \Bigg[ p(\bm\phi_{ij(q)}| \bar{\bm\lambda}_{ij(q),\phi}^2, \tau_{ij(q),\phi}^2,\sigma_{ij(q)}) p(\sigma_{ij(q)}) p(\tau_{ij(q),\phi}^2|\xi_{ij(q),\phi})p(\xi_{ij(q),\phi})  \Bigg. \nonumber \\ 
 & \:\:\:\:\:\: \cdot\Bigg. p(\bm z_{ij(q)})\prod_{k=1}^{l_\phi}   p(\bar{\lambda}_{ij(q),k,\phi}^2|\upsilon_{ij(q),k,\phi}) p(\upsilon_{ij(q),k,\phi})  \Bigg]  \nonumber \\
 & \:\:\:\:\:\:  \cdot \prod_{r=1}^{R}  \Bigg[ \Bigg. p(\bm \psi_r|\bar{\bm\lambda}_{r,\psi}^2, \tau_{r,\psi}^2,\omega_{r})    p( \omega_{r})   p(\bm a_r)  p(\tau_{r,\psi}^2|\xi_{r,\psi})p(\xi_{r,\psi})  \prod_{s=1}^{l_\psi}  p(\bar{\lambda}_{r,s,\psi}^2|\upsilon_{r,s,\psi}) p(\upsilon_{r,s,\psi})  \Bigg].  \nonumber
\end{align}

Given this partitioning, it can be shown that the solution to the optimization problem can be obtained by sequentially iterating over the densities

\begin{equation}\label{VBdensAPP}
q(\bm \theta_l  | \bm x) \propto \text{exp} \:\: \mathbb{E}_{q(\bm \theta_{(-l)} | \bm x)} \left( \text{log} \:\: p(\bm \theta_l | \bm \theta_{(-l)}, \bm x) \right)
\end{equation}
where $\bm\theta_{(-l)}$ denotes all elements of $\bm\theta$, excluding those in the $l^{th}$ group, $l=1,\cdots,L$. Consequently, the variational posterior can be obtained by calculating the variational expectation of the conditional posterior. Generally, the accuracy of the variational approximation depends on how well the chosen partitioning matches the independence structure of the parameters in the target posterior. In essence, this is where we trade off computational tractability and speed with accuracy. For a general discussion of this issue see e.g. \cite{Ormerod2017}.
In order to arrive at the final variational densities, we now need to insert \ref{VBpostAPP} and \ref{VBpriorAPP} into \ref{VBdensAPP}. Given that this expression relies on expectations of logarithms of standard densities it is rather easy to derive, however, we suppress it here due to its very lengthy nature. The expressions for $q(\bm\theta_l|\bm x)$ are presented in \autoref{alg2:cap} together with the full variational algorithm. 

\newgeometry{top=1cm,left=1.5cm, right=1.5cm, bottom=2cm}
\RestyleAlgo{ruled} 
\begin{algorithm}
\scriptsize
\caption{2-step Variational Bayes QFAVAR algorithm}\label{alg2:cap}
Define $ \kappa_{1(q)} = \frac{1-2q}{q(1-q)}$ and $ \kappa_{2(q)}^2 = \frac{2}{q(1-q)}$ for each quantile level $q$.\\

$\text{\textbf{Step 1:}}$ Estimate the factors $\hat{F}$ using the variational VBQFA algorithm proposed in \cite{KorobilisSchroeder2022} for each quantile level, $q$, and for the variable blockings in \ref{meas2app}.
\\
\text{\phantom{}}
\\
$\text{\textbf{Step 2:}}$ Conditional on the factors, estimate the remaining parameters of the model.\\
\Begin
{
$[\bm 1]$  Update the parameters in the measurement equation:\\
\For {$q=1:nq$}
{
\For {$i=1:m$}
{
\For {$j=1:n$}
{
$[\text{i}]$ Update $\bm \phi_{ij(q)}$ from $q(\bm\phi_{ij(q)}|\bm x) = N(\bm \mu_{ij(q)}^\phi,\bm\Sigma_{ij(q)}^\phi)$, with $\bm\mu^\phi_{ij(q)} = \bm \Sigma_{ij(q)}^\phi \left \{ \hat{\bm F}'diag(\tilde{\bm\nu}_{ij(q)}^{-1})\tilde{\bm y}_{ij(q)} \right \}$, $\bm \Sigma_{ij(q)}^\phi = \left \{   \hat{\bm F}' diag(\tilde{\bm\nu}_{ij(q)})^{-1}\hat{\bm F}  +   \tilde{\bm \Lambda}_{ij(q)}^{-1}    \right \}^{-1}$, where $\tilde{\bm\nu}_{ij(q)}^{-1} = \kappa_{2(q)}^{-2} \mathbb{E}\left(\sigma_{ij(q)}^{-1}\right) \mathbb{E}\left(\bm z_{ij,\bullet(q)}^{-1}\right)$, 
$\tilde{\bm y}_{ij(q)}= \bm y_{ij,\bullet(q)}-\kappa_{1(q)} \mathbb{E}\left(\bm z_{ij,\bullet(q)}\right)$, and $\tilde{\bm \Lambda}_{ij(q)}^{-1} = diag \left( \mathbb{E} \left( \bar{\bm\lambda}_{ij(q),\bullet,\phi}^{-2} \right) \mathbb{E} \left(\tau_{ij(q),\phi}^{-2} \right)\right)$. \\

\textbf{Assign:} $ \mathbb{E}\left( \bm \phi_{ij(q)} \right) = \bm \mu_{ij(q)}^\phi$ and  $ \mathbb{E}\left( \bm \phi_{ij(q)}^2 \right) = \bm \mu_{ij(q)}^{2,\phi} + diag\left( \Sigma_{ij(q)}^\phi \right) $\\

$[\text{ii}]$ Update the parameters of the horseshoe prior from 

$q(\bm \bar{\lambda}_{ij(q),k,\phi}^2| \bm x) = IG( a_{\bar{\lambda},\phi}, b_{\bar{\lambda},\phi}) = IG\left(1,\frac{\mathbb{E} (\bm\phi_{ij,k(q)}^2)}{2 } + \mathbb{E}\left(\frac{1}{\upsilon_{ij(q),k,\phi}}\right)\right)$, 

$q({\upsilon}_{ij(q),k,\phi}|\bm x) = IG( a_{\upsilon,\phi}, b_{\upsilon,\phi}) = IG\left(1,\mathbb{E}\left(\frac{1}{\bar{\bm\lambda}_{ij(q),k,\phi}^2}\right) + b_\phi^{-2} \mathbb{E}\left( \frac{1}{\tau^2_{ij,\phi}}\right)\right)$,

$q(\tau_{ij(q),k,\phi}^2|\bm x)  = IG( a_{\tau,\phi}, b_{\tau,\phi}) = IG\left(1,  b_\phi^{-2}  \mathbb{E}\left( \frac{1}{\upsilon_{ij(q),k,\phi}}\right) + \mathbb{E}\left( \frac{1}{\xi_{ij(q),\phi}}\right) \right)$,

$q({\xi}_{ij(q),\phi}|\bm x)  = IG( a_{\xi,\phi}, b_{\xi,\phi}) = IG\left(\frac{l_\phi+1}{2}, 1 + \sum_{k=1}^{l_\phi} \mathbb{E}\left(\frac{1}{\tau_{ij(q),k,\phi}^2} \right)\right)$, \\
for $k = 1,\cdots,l_\phi$. 

\textbf{Assign:} $\mathbb{E}\left(  \frac{ a_{\bar{\lambda},\phi}}{ b_{\bar{\lambda},\phi}}  \right)$, $\mathbb{E}\left(  \frac{ a_{\upsilon,\phi}}{ b_{\upsilon,\phi}}  \right)$, $\mathbb{E}\left(  \frac{ a_{\tau,\phi}}{ b_{\tau,\phi}}  \right)$, and $\mathbb{E}\left(  \frac{ a_{\xi,\phi}}{ b_{\xi,\phi}}  \right)$.

$[\text{iii}]$ Update the latent quantile indicators, $\bm z_{ij,t(q)}$, from $q(\bm z_{ij,t(q)}|\bm x)=GIG(1/2,\delta_{z},\rho_{t,z})$ with \\

$\delta_{z}$ = $\mathbb{E}\left(  \frac{1}{\sigma_{ij(q)}} \right)\frac{\kappa_{1(q)}^2}{\kappa_{2(q)}^2}$, 
$\rho_{t,z} =\mathbb{E}\left(  \frac{1}{\sigma_{ij(q)}} \right) \frac{ \left \{  y_{ij,t(q)}- \mathbb{E}(\bm \phi_{ij(q)})' \bm \hat{\bm f_t}    \right \}^2 + \hat{\bm f_t}'\bm\Sigma_{ij(q)}^\phi \hat{\bm f_t} }{\kappa_{2(q)}^2} $, \\
for $t = 1,\cdots,T$\\
\textbf{Assign:} $\mathbb{E}( z_{ij,t(q)})= \frac{\sqrt{\rho_{t,z}} K_{3/2} \left(\sqrt{\delta_{z}\rho_{t,z}} \right)}{\sqrt{\delta_{z}} K_{1/2} \left(\sqrt{\delta_{z}\rho_{t,z}} \right)}$,
$\mathbb{E}\left( \frac{1}{z_{ij,t(q)}}\right)= \frac{\sqrt{\delta_{z}} K_{3/2} \left(\sqrt{\delta_{z}\rho_{t,z}} \right)}{\sqrt{\rho_{t,z}} K_{1/2} \left(\sqrt{\delta_{z}\rho_{t,z}} \right)} - \frac{1}{\rho_{t,z}}$\\
$[\text{iv}]$ Update the factor regression variances, $\sigma_{ij(q)}$, from $q(\sigma_{ij(q)}|\bm x) = IG(r_\sigma,s_\sigma)$ with \\
$r_\sigma=r_0+3T$ and $s_\sigma=s_0 + \sum_{t=1}^T \left[ \mathbb{E}\left( \frac{1}{z_{ij,t(q)}}\right)  \frac{ M}{2\kappa_{2(q)}^2} -\kappa_{1(q)}^2  \frac{y_{ij,t(q)}- \mathbb{E}(\bm \phi_{ij(q)})' \bm \hat{\bm f_t} }{\kappa_{2(q)}^2}  + \left( 1 + \frac{\kappa_{1(q)}^2}{2\kappa_{2(q)}^2} \right) \mathbb{E}( z_{ij,t(q)}) \right]  $ and $M = \left \{ y_{ij,t(q)}- \mathbb{E}(\bm \phi_{ij(q)})' \bm \hat{\bm f_t}    \right \}^2 + \hat{\bm f_t}'\bm\Sigma_{ij(q)}^\phi \hat{\bm f_t}$.\\
\textbf{Assign:} $\mathbb{E}\left( \frac{1}{\sigma_{ij(q)}} \right) = \frac{r_\sigma}{s_\sigma}$.
}
}
}
$[\bm 2]$  Update the diagonal elements of $\Omega$ from $q(\omega_r|\bm x) = IG(r_\omega,s_\omega)$ with\\
$r_\omega = r_h + T/2$ and $s_\omega = s_h + \sum_{t=1}^T (\hat{\bm f}_{r,t} - \hat{\bm F}_{t-1}\bm\psi_{r}')^2 )$, for all $r=1,\cdots,R$. \\
\textbf{Assign:} $\mathbb{E}\left( \frac{1}{\omega_{r}} \right) = \frac{r_\omega}{s_\omega}$.

$[\text{3}]$ Sample the off-diagonal elements of $\bm\Omega$, $\bm A$, and the VAR coefficients $\bm\Psi$\\
\For {$r=1:R$}
{
$[\text{i}]$ Sample $\bm\psi_r$ and $\bm a_r$ jointly from $q(\bm \psi_{r}, \bm a_r|\bm x) = N(\bm \mu^\psi_{r},\bm \Sigma_{r}^\psi)$, with $\bm\mu^\psi_{r} = \bm \Sigma_{r}^\psi \left \{ \bm X'\mathbb{E}(\sigma_{\omega}^{-1})\hat{\bm f}_{r} \right \}$, $\bm \Sigma_{r}^\psi = \left \{   \bm X' \mathbb{E}(\sigma_{\omega}^{-1})\bm X  +   \tilde{\bm \Lambda}_{r}^{-1}    \right \}^{-1}$,  where
$\bm X = [ L \bm F', L \bm G', \bm  E_{1:r-1}']'$, $\bm e_{r} = \hat{\bm f}_{r} - \bm \hat{\bm F} \mathbb{E}(\bm\psi_r)' - \bm  E_{1:r-1}' \mathbb{E}(\bm a_r)'$, and $\tilde{\bm \Lambda}_{r}^{-1} = diag\left(  [\mathbb{E}(\bar{\bm\lambda}_{r,\bullet,\psi}^{-2}), \bm \mu_{r,a}/\bm \Sigma_{r,a} ]  
\right)$. \\

$[\text{ii}]$ Update the parameters of the horseshoe prior from $q(\bar{\lambda}_{r,s,\psi}^2|\bm x)= IG( a_{\bar{\lambda},\psi}, b_{\bar{\lambda},\psi}) = IG\left(1,\frac{\mathbb{E}(\bm\psi_{r,s}^2)}{2} + \frac{1}{\mathbb{E}(\upsilon_{r,s,\psi})}\right)$, 

$q({\upsilon}_{r,s,\psi}|\bm x)= IG( a_{{\upsilon},\psi}, b_{{\upsilon},\psi}) = IG\left(1, \mathbb{E} \left( \frac{1}{\bar{\lambda}_{r,s,\psi}^2} \right)  + b_\psi^{-2} \mathbb{E} \left(  \frac{1}{\tau_{r,\psi}^2} \right)\right)$,

$q(\tau_{r,s,\psi}^2|\bm x) = IG( a_{{\tau},\psi}, b_{{\tau},\psi})= IG\left(1, b_\psi^{-2} \mathbb{E} \left( \frac{1}{{\upsilon}_{r,s,\psi}} \right) + \mathbb{E} \left( \frac{1}{{\xi}_{r,\psi}} \right)\right)$,

$q({\xi}_{r,\psi}|\bm x) = IG( a_{{\xi},\psi}, b_{{\xi},\psi})= IG\left(\frac{l_\psi +1}{2}, 1 + \sum_{s=1}^{l_\psi} \mathbb{E} \left( \frac{1}{\tau_{r,s,\psi}^2} \right) \right)$, \\
for $s = 1,\cdots,l_\psi$.\\
\textbf{Assign:} $\mathbb{E}\left(  \frac{ a_{\bar{\lambda},\psi}}{ b_{\bar{\lambda},\psi}}  \right)$, $\mathbb{E}\left(  \frac{ a_{\upsilon,\psi}}{ b_{\upsilon,\psi}}  \right)$, $\mathbb{E}\left(  \frac{ a_{\tau,\psi}}{ b_{\tau,\psi}}  \right)$, and $\mathbb{E}\left(  \frac{ a_{\xi,\psi}}{ b_{\xi,\psi}}  \right)$.
}
}
Note: We use the rate parametrization of the $IG$ distribution throughout. $K_p(\bullet)$ denotes the Bessel function of order $p$ and $L$ denotes the Lag operator.
\end{algorithm}
\restoregeometry

Finally, note that the horseshoe priors take a slightly different form in our VB algorithm with

\begin{align}
\bm\phi_{ij(q)}| \left\{\bar{\lambda}_{ij(q),k,\phi}, \upsilon_{ij(q),k,\phi}\right \}_{k=1}^{l_\phi}, \tau_{ij(q),\phi}, \xi_{ij(q),\phi}  & \sim  N(\bm 0, \bar{\bm\Lambda}_{ij(q),\phi}), \\
\bar{\lambda}^{2}_{ij(q),k,\phi} | \upsilon_{ij(q),k,\phi} & \sim G^{-1}\left(\frac{1}{2},\frac{1}{\upsilon_{ij(q),k,\phi}} \right), \:\:\: \text{for } k = 1,\cdots,l_\phi,\nonumber\\
\upsilon_{ij(q),k,\phi}|\tau^2_{ij(q),\phi} & \sim G^{-1}\left(\frac{1}{2},\frac{1}{b_\phi^2  \tau^2_{ij(q),\phi}  }\right), \:\:\: \text{for } k = 1,\cdots,l_\phi,\nonumber\\
\tau^2_{ij(q),k,\phi} | \xi_{ij(q),\phi} & \sim G^{-1}\left(\frac{1}{2},\frac{1}{\xi_{ij(q),\phi}}\right),\nonumber\\
\xi_{ij(q),\phi} & \sim G^{-1}\left(\frac{1}{2},1\right),\nonumber
\end{align}

\noindent and

\begin{align}
\bm\psi_{r}| \left\{\bar{\lambda}_{r,s,\psi}, \upsilon_{r,s,\psi}\right \}_{s=1}^{l_\psi}, \tau_{r,\psi}, \xi_{r,\psi}  & \sim  N(\bm 0, \bar{\bm\Lambda}_{r,\psi}), \\
\bar{\lambda}^{2}_{r,s,\psi} | \upsilon_{r,s,\psi} & \sim G^{-1}\left(\frac{1}{2},\frac{1}{\upsilon_{r,s,\psi}} \right), \:\:\: \text{for } s = 1,\cdots,l_\psi, \nonumber\\
\upsilon_{r,s,\psi} | \tau^2_{r,\psi} & \sim G^{-1}\left(\frac{1}{2},\frac{1}{b_\psi^2 \tau^2_{r,\psi}}\right), \:\:\: \text{for } s = 1,\cdots,l_\psi,\nonumber\\
\tau^2_{r,s,\psi} | \xi_{r,\psi} & \sim G^{-1}\left(\frac{1}{2},\frac{1}{\xi_{r,\psi}}\right),\nonumber\\
\xi_{r,\psi} & \sim G^{-1}\left(\frac{1}{2},1\right),\nonumber
\end{align} 

\noindent where $b_\phi$ and $b_\psi$ are hyperparameters, which we set to $0.0001$ in application. 



\newpage

\section{Additional empirical results}
\subsection{Forecast evaluation}
\begin{figure}[H]
\centering
\captionsetup{width=\linewidth}
\includegraphics[width=0.95\linewidth, trim={15cm 1cm 10cm 2cm}]{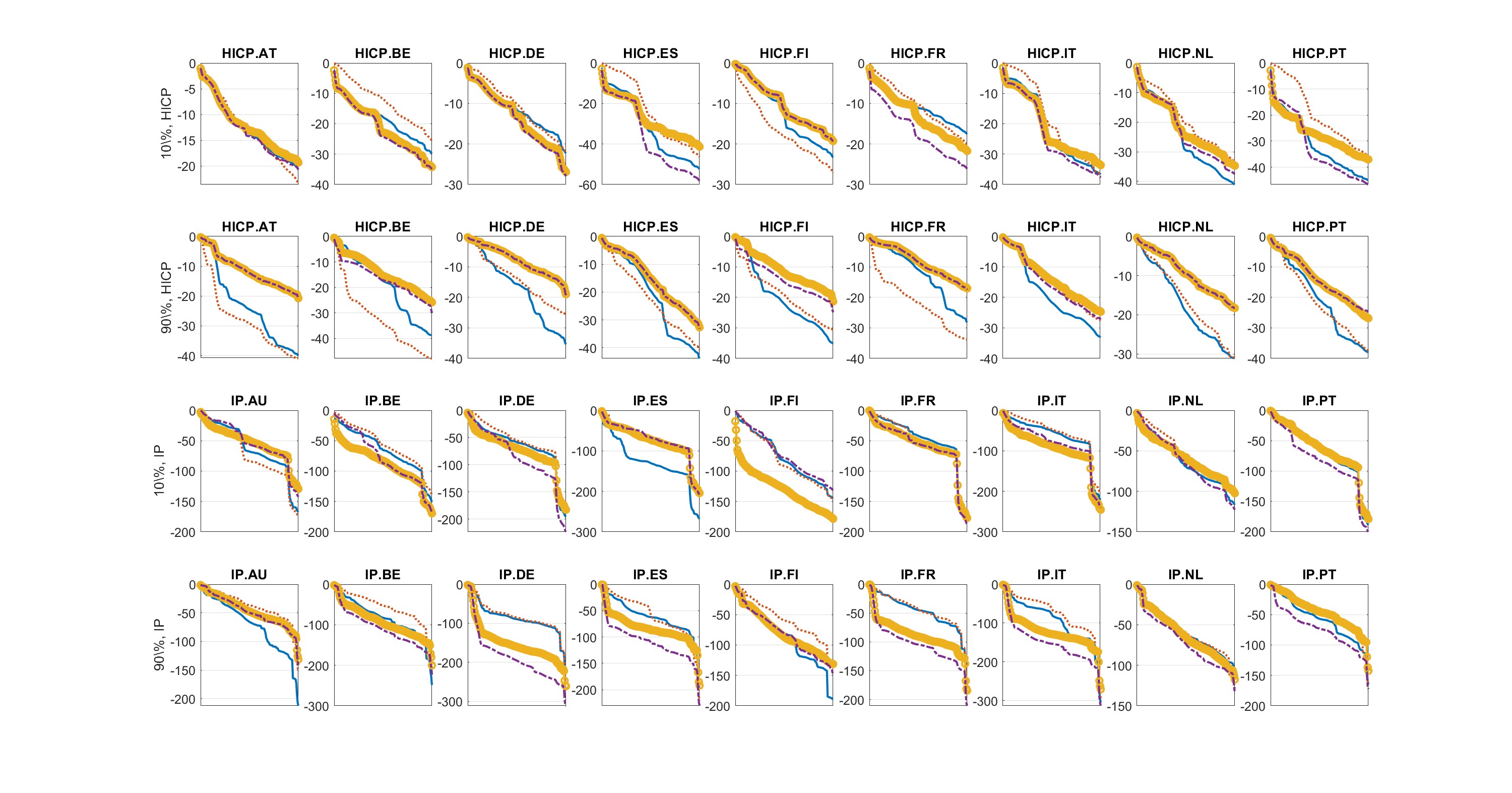}
\caption{Cumulative quantile score (QS) loss for forecast horizon $h=6$. Four models are compared, QAR (yellow circled line), QAR-X (purple dashed line), QDFM (red dotted line) and QFAVAR (blue solid line). The out-of-sample evaluation period shown on the x-axis is 2011Jan to 2022Dec-$h$. First two rows show quantile scores for the 10th and 90th percentiles of inflation and third and fourth rows show quantile scores for the 10th and 90th percentiles of industrial production.} \label{fig:h6QSCORES}
\end{figure}

\begin{figure}[H]
\centering
\captionsetup{width=\linewidth}
\includegraphics[width=0.95\linewidth, trim={15cm 1cm 10cm 2cm}]{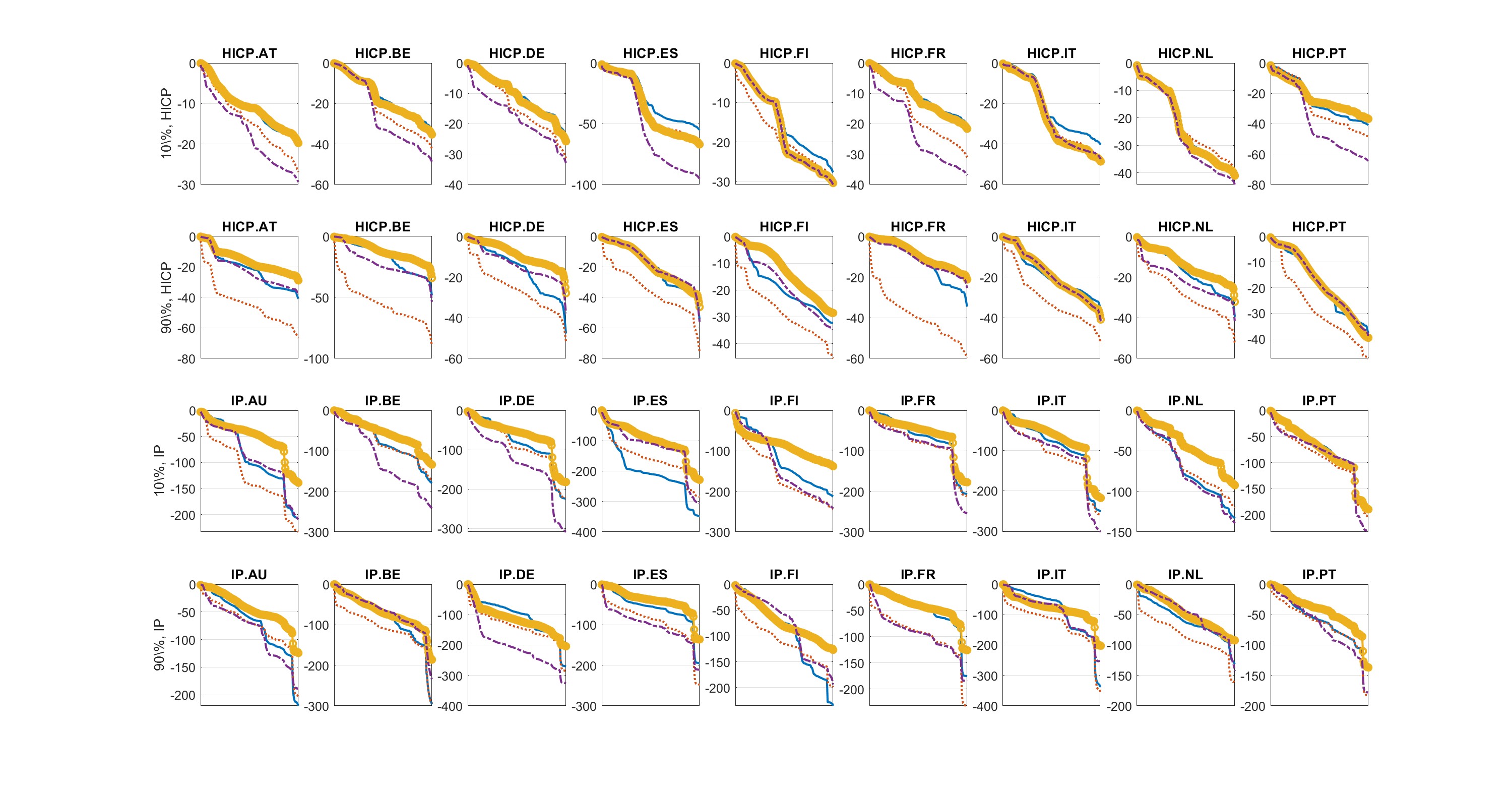}
\caption{Cumulative quantile score (QS) loss for forecast horizon $h=12$. Four models are compared, QAR (yellow circled line), QAR-X (purple dashed line), QDFM (red dotted line) and QFAVAR (blue solid line). The out-of-sample evaluation period shown on the x-axis is 2011Jan to 2022Dec-$h$. First two rows show quantile scores for the 10th and 90th percentiles of inflation and third and fourth rows show quantile scores for the 10th and 90th percentiles of industrial production.}\label{fig:h12QSCORES}
\end{figure}

\begin{figure}[H]
\centering
\captionsetup{width=\linewidth}
\includegraphics[width=0.95\linewidth, trim={15cm 1cm 10cm 2cm}]{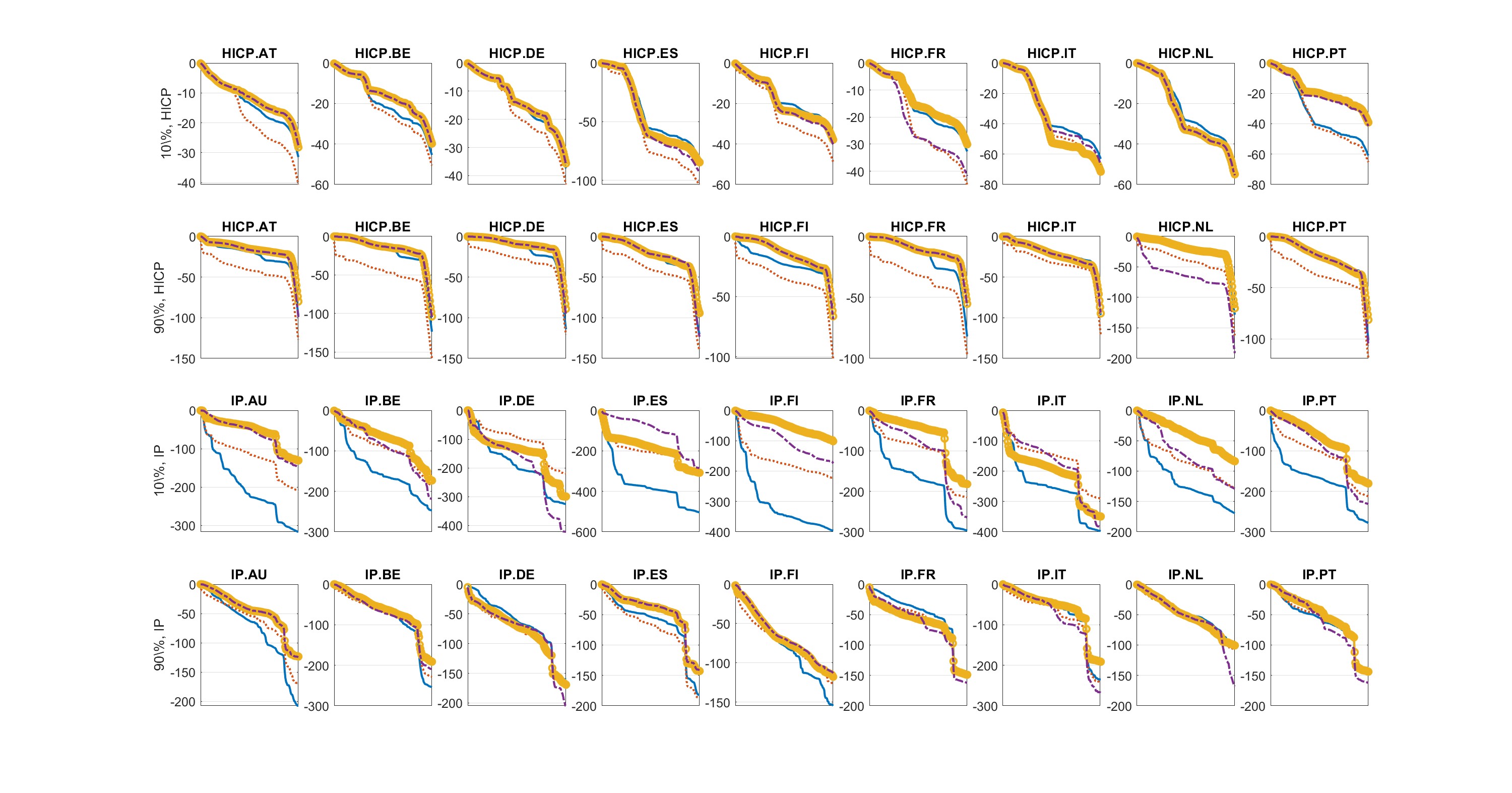}
\caption{Cumulative quantile score (QS) loss for forecast horizon $h=24$. Four models are compared, QAR (yellow circled line), QAR-X (purple dashed line), QDFM (red dotted line) and QFAVAR (blue solid line). The out-of-sample evaluation period shown on the x-axis is 2011Jan to 2022Dec-$h$. First two rows show quantile scores for the 10th and 90th percentiles of inflation and third and fourth rows show quantile scores for the 10th and 90th percentiles of industrial production.} \label{fig:h24QSCORES}
\end{figure}

\subsection{Detailed quantile IRFs}

\begin{landscape}
\begin{figure}[H]
\centering
\captionsetup{width=\linewidth}
\includegraphics[width=0.95\linewidth, trim={15cm 1cm 10cm 2cm}]{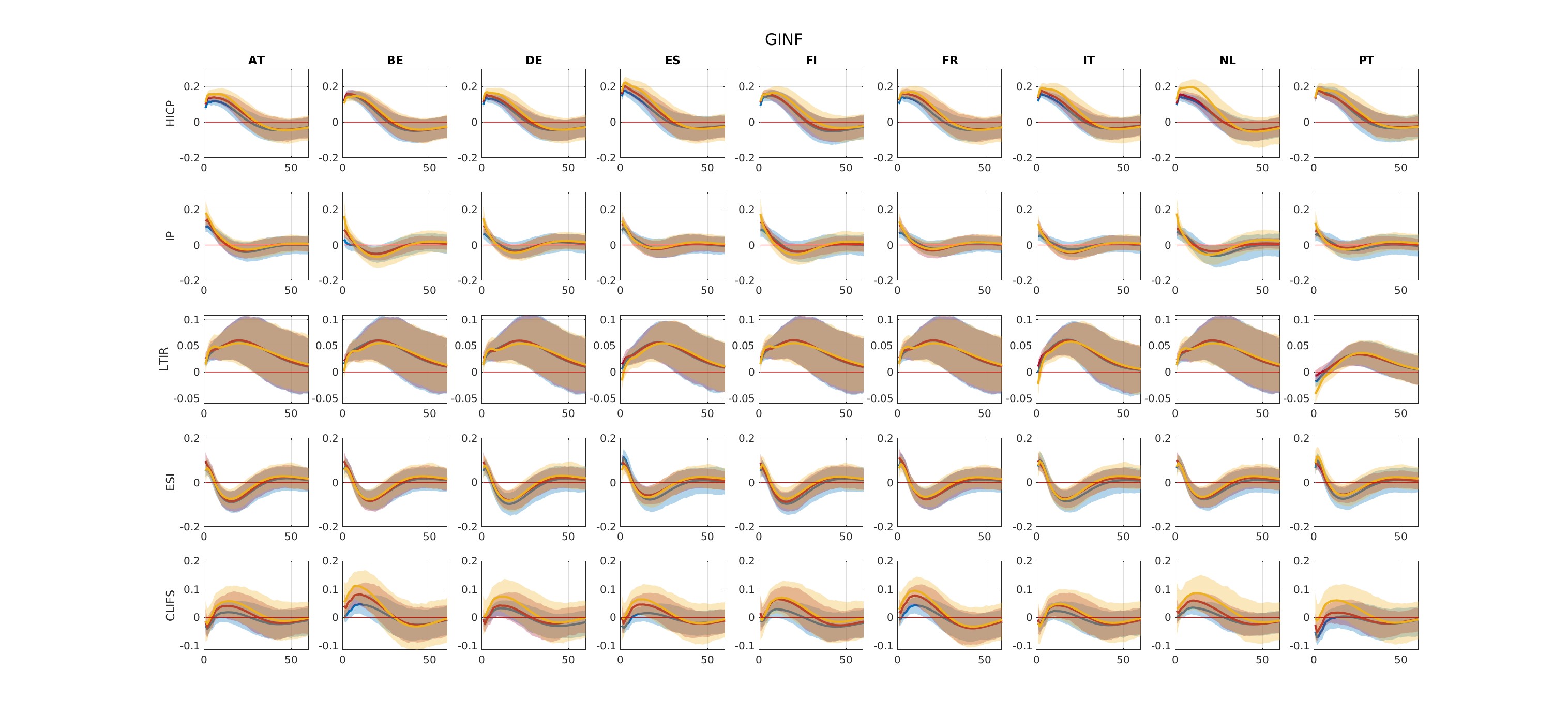}
\caption{Detailed quantile impulse response functions, shock to global inflation (GINF).} \label{fig:full_GINF}
\end{figure}
\end{landscape}

\begin{landscape}
\begin{figure}[H]
\centering
\captionsetup{width=\linewidth}
\includegraphics[width=0.95\linewidth, trim={15cm 1cm 10cm 2cm}]{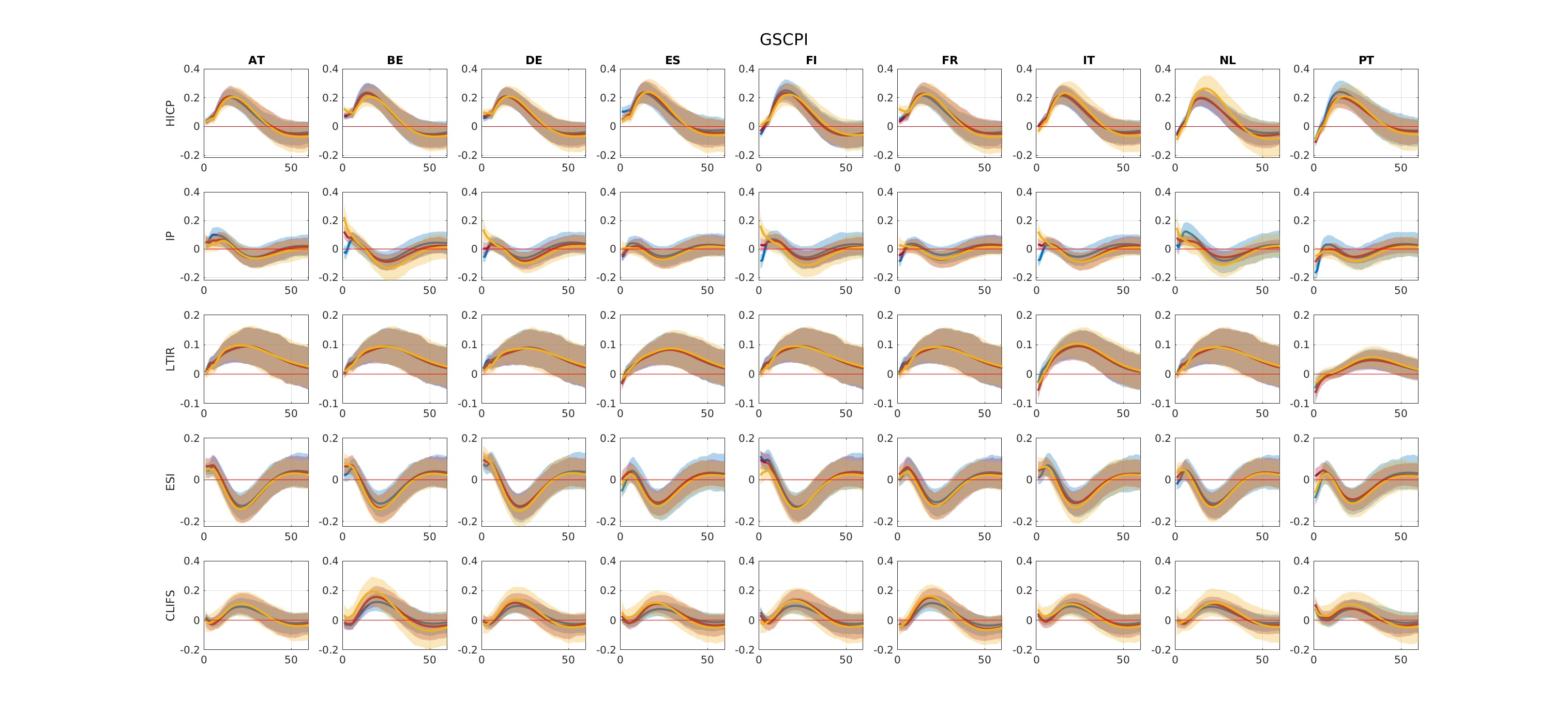}
\caption{Detailed quantile impulse response functions, shock to global supply chain pressure index (GSCPI).} \label{fig:full_GSCPI}
\end{figure}
\end{landscape}

\begin{landscape}
\begin{figure}[H]
\centering
\captionsetup{width=\linewidth}
\includegraphics[width=0.95\linewidth, trim={15cm 1cm 10cm 2cm}]{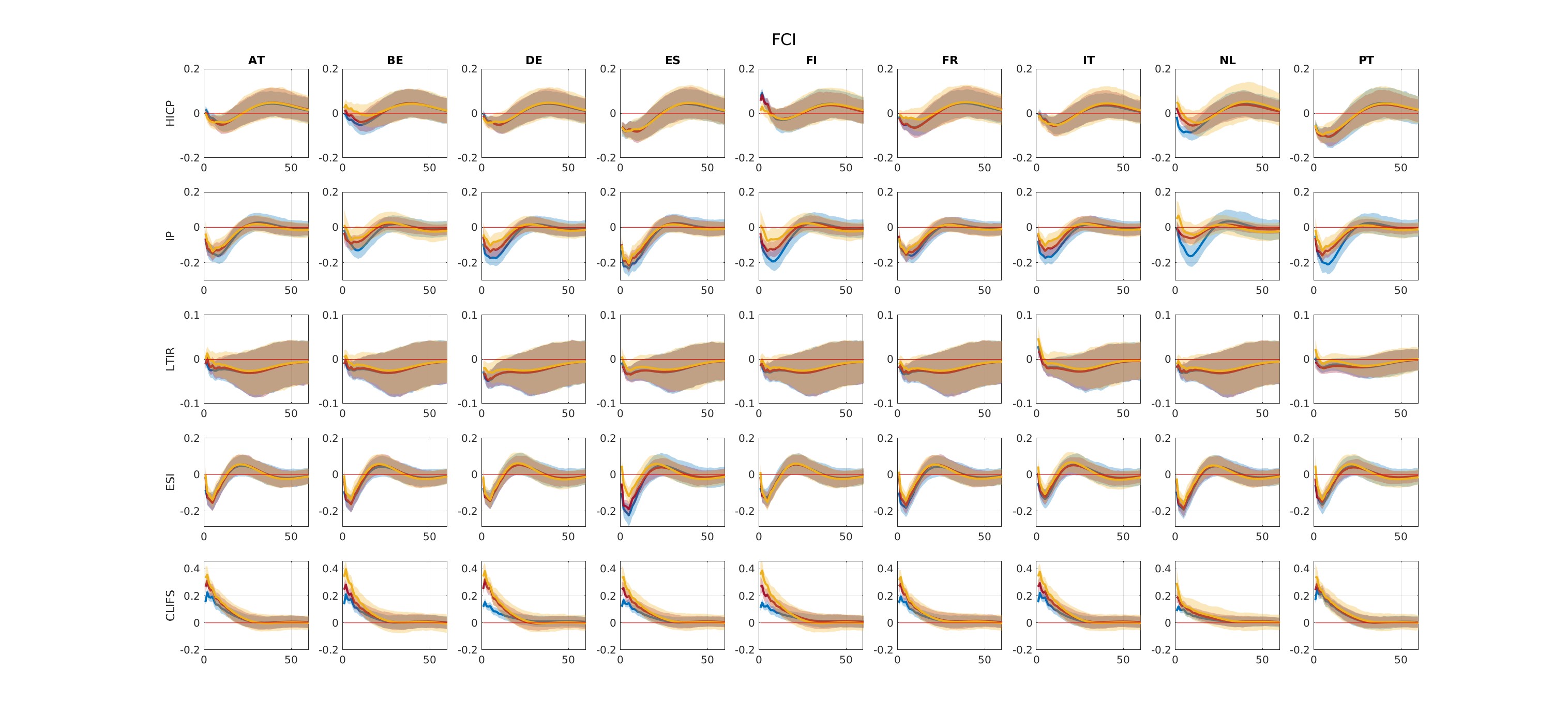}
\caption{Detailed quantile impulse response functions, shock to global (US) financial conditions (FCI).} \label{fig:full_FCI}
\end{figure}
\end{landscape}

\begin{landscape}
\begin{figure}[H]
\centering
\captionsetup{width=\linewidth}
\includegraphics[width=0.95\linewidth, trim={15cm 1cm 10cm 2cm}]{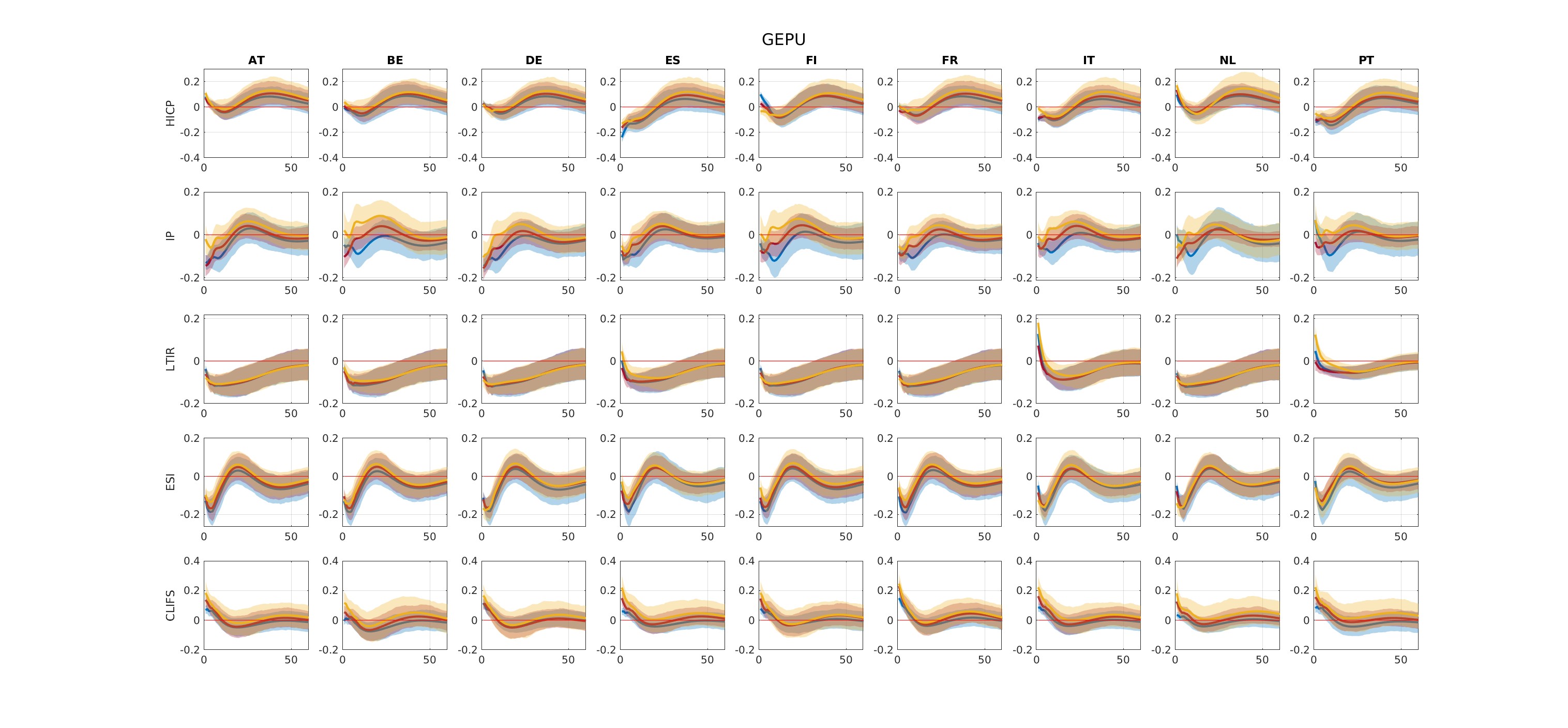}
\caption{Detailed quantile impulse response functions, shock to global economic policy uncertainty (GEPU).} \label{fig:full_GEPU}
\end{figure}
\end{landscape}
 
\end{document}